\documentclass[10pt, sigconf]{acmart}
\pdfoutput=1

\newif\ifdebugdoc\debugdocfalse
\newif\ifflow\flowfalse

\usepackage[labelfont={bf,small},textfont={bf,small},skip=0pt]{caption}
\usepackage[caption=false]{subfig}

\usepackage{graphicx}
\usepackage{epstopdf}
\usepackage{balance}
\usepackage{multirow}
\usepackage[normalem]{ulem}
\usepackage{url}  
\def\compactify{\itemsep=-3pt \topsep=-3pt \partopsep=0pt \parsep=0pt}
 \let\latexusecounter=\usecounter

\usepackage[procnames]{listings}
\usepackage{color}
\definecolor{keywords}{RGB}{255,0,90}
\definecolor{comments}{RGB}{0,0,113}
\definecolor{red}{RGB}{160,0,0}
\definecolor{green}{RGB}{0,150,0}
\definecolor{blue}{RGB}{0,0,150}
\definecolor{grey}{gray}{0.6}
\definecolor{mygray}{RGB}{153,153,153}
\definecolor{myblue}{RGB}{0,176,240}
\newcommand{\paragraphb}[1]{\vspace{1mm}\noindent{\bf #1}}

\usepackage{xspace}
\def\ie{i.e.\xspace}

\def\eg{e.g.\xspace}

\def\fig{Figure}
\def\tab{Table}

\usepackage{color, colortbl}
\definecolor{Gray}{gray}{0.9}
\definecolor{LightCyan}{rgb}{0.5,1,1}

\ifdebugdoc

\newcommand{\fyi}[1]{}

\newcommand{\add}[1]{\textcolor{red}{#1}}
\newcommand{\del}[1]{\textcolor{blue}{\sout{#1}}}
\newcommand{\outline}[1]{\textbf{\colorbox{yellow}{Outline:}\textcolor{red}{#1.}}}
\newcommand{\old}[1]{\large{\colorbox{blue}{Former: #1}}}
\newcommand{\todo}[1]{\textcolor{red}{\textbf{ToDo: #1}}}

\newcommand{\fdel}[1]{\textcolor{blue}{\sout{#1}}}
\newcommand{\tbd}[1]{\textcolor{green}{Discussion:#1}}

\newcommand{\save}[1]{}
\newcommand{\nosub}[1]{} 

\newcommand{\chunyi}[1]{\textcolor{magenta}{Chunyi:#1}}
\newcommand{\junpeng}[1]{\textcolor{blue}{Junpeng:#1}}

\else

\newcommand{\fyi}[1]{}

\newcommand{\add}[1]{#1}
\newcommand{\del}[1]{}
\newcommand{\outline}[1]{}
\newcommand{\old}[1]{}
\newcommand{\todo}[1]{}
\newcommand{\tbd}[1]{}

\newcommand{\fdel}[1]{}

\newcommand{\save}[1]{}
\newcommand{\nosub}[1]{} 

\newcommand{\chunyi}[1]{}
\newcommand{\junpeng}[1]{}
\fi

\ifflow
\newcommand{\p}[1]{\vskip 1ex\noindent\colorbox{yellow}{\parbox{\columnwidth}{\textbf{Point:} #1}}}
\newcommand{\paragraphpoint}[1]{\vskip 1ex\noindent\colorbox{yellow}{\parbox{\columnwidth}{\textbf{Point:} #1}}}
\newcommand{\q}[1]{\vskip 1ex\noindent\colorbox{cyan}{\parbox{\columnwidth}{\textbf{Question:} #1}}}
\newcommand{\z}[1]{}
\else
\newcommand{\p}[1]{}
\newcommand{\paragraphpoint}[1]{}
\newcommand{\q}[1]{}
\newcommand{\z}[1]{}
\fi

\def\name{\texttt{DCC}\xspace}
\usepackage{booktabs} 

\usepackage{tikz}
\usepackage{xcolor}
\newcommand*\circled[1]{\tikz[baseline=(char.base)]{
            \node[shape=circle,fill,inner sep=0.5pt] (char) {\textcolor{white}{#1}};}}

\newcommand*\circledblue[1]{\tikz[baseline=(char.base)]{
            \node[shape=circle,fill=myblue,inner sep=0.5pt] (char) {\textcolor{white}{#1}};}}

\usepackage{amssymb}
\usepackage{pifont}

\usepackage{centernot}

\setcopyright{none}

\renewcommand\footnotetextcopyrightpermission[1]{} 

\usepackage{siunitx}

\begin{document}

\title{Towards Live Video Analytics with On-Drone Deeper-yet-Compatible Compression}
\author{Junpeng Guo}
\affiliation{%
  \institution{Purdue University}
}
\email{ guo567@purdue.edu}

\author{Chunyi Peng}
\affiliation{%
  \institution{Purdue University}
}
\email{ chunyi@purdue.edu}

%

%


\begin{abstract}
In this work, we present \name (Deeper-yet-Compatible Compression), one enabling technique for real-time drone-sourced edge-assisted video analytics built on top of the existing codec. \name tackles an important technical problem to compress streamed video from the drone to the edge without scarifying accuracy and timeliness of video analytical tasks performed at the edge. 
\name is inspired by the fact that not every bit in streamed video is equally valuable to video analytics, which opens new compression room over the conventional analytics-oblivious video codec technology. 
We exploit drone-specific context and intermediate hints from object detection 
to pursue adaptive fidelity needed to retain analytical quality.
We have prototyped \name in one showcase application of vehicle detection and validated its efficiency in representative scenarios. 
\name has reduced transmission volume by 9.5-fold over the baseline approach and 19-683\% over~\cite{wang2018bandwidth} 
with comparable detection accuracy. 
\end{abstract}

%
%

%
%
%

\maketitle
\pagestyle{plain} 

\section{Introduction}\label{sect:intro}


Unmanned aerial vehicles (also known as drones) are gaining momentum to serve as moving eyes in the sky.
They fly in the low sky above a target area 
and perform real-time \save{ visual sensing and} video surveillance for a wide variety of authorized and even critical activities such as large event (\eg, sports, concert, festival) surveillance, infrastructure safety inspection, search and rescue, 
wildfire surveillance, law enforcement and public safety assistance, to name many~\cite{restas2015drone,ham2016visual,george2019towards,tsag19-drone-app-public-safety,sudhakar2020unmanned}.
Drones possess unique advantages over its counterparts -- cameras pre-deployed on the ground: 
a drone can cruise at a much higher altitude with a broader field of view (FoV) and an unparalleled perspective (birds' eye view) for surveillance;
It is \save{quite infrastructure-less and} always available to operate on demand. 
%
%


However, it is not easy to make drone-sourced video analytics {\em live}, particularly, to detect and track target objects in real-time on resource-constrained drones alone. 
Despite rapid advances in computer vision, object detection/tracking cannot 
meet the real-time requirement (the inference rate $>$ video source rate)
without any hardware accelerations\save{ on small drones}~\cite{cai2018cascade,zhang2017live,Kristan2019a}.
The state-of-the-art approaches are based on deep neural networks, precisely convolutional neural networks (CNNs), and most popular 
models (\eg, YOLO~\cite{redmon2018yolov3,bochkovskiy2020yolov4,jocher2020yolov5}, faster R-CNN~\cite{ren2015faster}, DRNet\cite{denton2017unsupervised}) are trained on ImageNet datasets and work with low-resolution images; However, high-definition videos (say, 4K or more) are often required for drone-sourced video analytics due to much larger distance from objects; 
Inference on such videos is not time efficient; In particular, YOLOv5~\cite{jocher2020yolov5}, one of the most recent studies, achieves real-time detection on only 1K videos even with GPU acceleration, which is unaffordable on drones.

\begin{figure}[t]
\includegraphics[width=0.95\columnwidth]{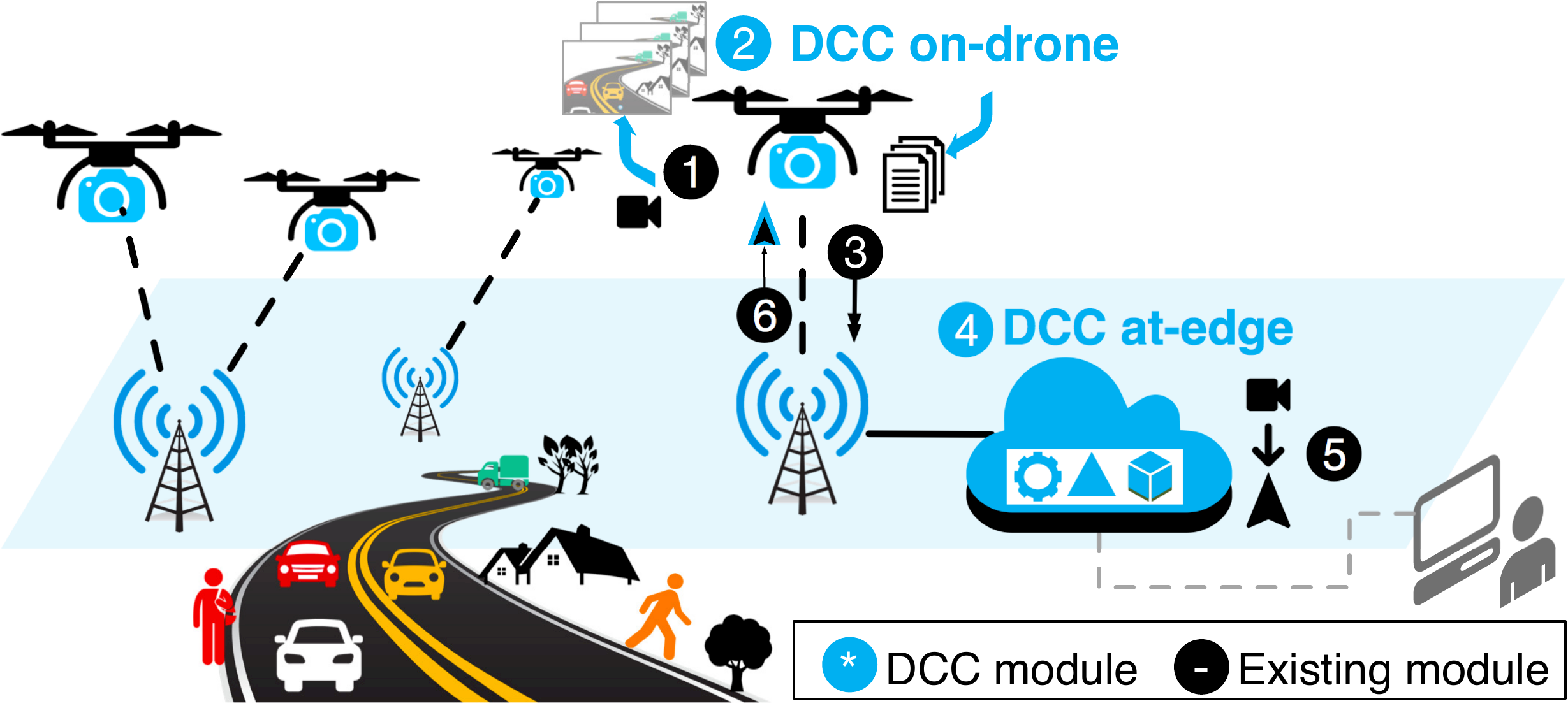}
\caption{\name for drone-sourced edge-assisted video analytics.}
\label{fig:drone-scene}
\vspace{-5mm}
\end{figure}

To this end, a promising paradigm is to offload heavy computing tasks to the edge~\cite{george2019towards,sudhakar2020unmanned,wang2018bandwidth,majid2018shuffledet}.
As illustrated in \fig~\ref{fig:drone-scene},
the drone is responsible for recording and streaming the captured video to the edge while the edge performs heavy analytical tasks\save{ (here, object detection)} over the received video frames. 
But it is still challenging because high bandwidth connectivity needed for high-definition video streaming is not available at all times~\cite{TR36.777,hayat2019experimental,lin2019mobile}.
\save{Due to distinct radio propagation and interferences,}
Instead, drones in the sky suffer from more failures and performance degradations at higher altitudes, thereby resulting in unpredictable and long delay in networking; 
Simply downgrading the quality of video sources (\eg, lower resolution) to reduce network traffic is hardly acceptable, which hurts accuracy of analytics. In a nutshell, center to live drone-sourced video analytics is to relieve tension between networking and local/edge computing.

In this work, we aim to tackle this\save{ core} problem, precisely, {\em how can we reduce the traffic volume of video frames to be delivered from the drone to the edge, while retaining high-quality surveillance accuracy and low on-drone computing load?} 


We design \name \footnote{This is our extension work of ADC ~\cite{guo2021towards} with a more comprehensive evaluation.} 
,deeper-yet-compatible compression, on top of the existing video codec to accelerate drone-sourced edge-assisted video analytics. \name is inspired by a simple fact that not every bit in streamed video frames is equally valuable to the analytical tasks, which opens new room over the conventional video codec technology which is analytics-oblivious; \name transfers less without scarifying accuracy and timeliness of video analytical tasks performed at the edge by compressing more over the existing video codec. 
By exploiting drone's sensory context and distinct quality demanded by drone-sourced video analytics, \name finds the most suitable codec parameters to tune a video codec, rather than directly change/upgrade a video codec. As a result, it is compatible to the existing video codec but 
looks for a larger compression space to minimize network transmission while guaranteeing quality of computing (here, object detection accuracy).  
In particular, \name explores three compression dimensions including resolution, inter-frame prediction and quantization, and devises three key technical components: {\bf adaptation resolution},{ \bf global motion compensated encoding} and {\bf RoPI-aware quantization}~(\S\ref{sect:design}).

We prototype and evaluate \name for a showcase application of vehicle detection from the sky~(\S\ref{sect:eval}). Our evaluation covers 100 mins of video \save{captured from drones}, two popular CNN-based detectors (tiny-YOLOv3 and YOLOv5s) in three representative use scenarios over residence, local roads and highways.
Overall, \name has achieved 9.5-fold compression on average over the baseline approach. Compared to \texttt{ED}~\cite{wang2018bandwidth},  it has  compressed by 19\% -- 683\%, without scarifying detection accuracy; $F_1$ score reduces no more than 1.5\% (mostly $<$ 0.5\%).
Moreover, \name is able to operate at 10.1 fps, which is promising to perform live object detection over video sources no larger than 10fps. 



\save{It turns out that the gap between these two camps 
misses huge potentials for deeper video compression
(elaborated in \fig~\ref{fig:related} and \S\ref{sect:motivation}). \add{We thus have two goals: {\bf First}, we want to bridge this gap by exploring {\em every} possible dimension spanning the whole compression space to to encode `{\em just-necessary}' information needed for video analytics. {\bf Second}, our solution is expected to be lightweight without requiring any special hardware to be equipped on drones. We aim to leverage functions in place and make it compatible with the existing architecture to 
lower additional on-drone computing as much as possible. }}

\section{Background and Related Work}
\label{sect:back-related}

We first introduce basic operations for edge-assisted video analytics and then present related work to make it live. 


\subsection{Background on Edge-assisted Video Analytics}
\label{sect:back}

\fig~\ref{fig:drone-scene} shows the basic operation flow for edge-assisted video analytics with six steps.
%
The drone is responsible for visual sensing (\circled{1}), local processing (\circledblue{2}) and then streaming the processed video frames to the edge (\circled{3}). 
The edge is responsible for retrieving video frames out of the received bit stream (\circledblue{4}) and performing analytical tasks (here, detecting objects, \circled{5}).
To further facilitate and optimize on-drone processing, recent solutions propose to ask the edge to offer intermediate results or feedback to the drone at runtime (\circled{6}).  
Next, we briefly present how theses existing steps work.

\paragraphb{Visual sensing (\circled{1}).} Visual sensing is to capture a 3D physical world into continuous 2D frames using a CMOS sensor in the camera. 
Its sensing capabilities is determined by three key factors: field of view (FoV), resolution and frame rate; 
All three are limited by camera hardware. 
FoV is the extent of the physical world captured in one frame, which can be adjustable by the camera's position and rotation, namely, yaw, pitch, and roll about the $Z_c, Y_c, X_c$ axises in the 3D coordinate system.
Resolution and frame rate determine how many pixels for each frame and how often to capture one frame. Both affect the data volume per second.
The off-the-shelf medium-end or small drones mostly support up to 90\textdegree\, FoV, 4K resolution (3840$\times$2160) and 120 frame per second (fps), which poses a huge pressure on network transmission for live video analytics.

%
%



\paragraphb{Video codec and pre-processing (\circled{2} \& \circled{4}).} Video encoding is an essential technology embedded on drones that compresses a raw captured video into a format that takes up less capacity for efficient transmission. It is one key operation in on-drone processing (\circled{2}). Today's commercial drones support for both H.264 and H.265, the most widely used coding formats~\cite{wiki-codec}.


All video coding techniques work with three standard procedures: (i) prediction (motion estimation and compensation), (ii) transformation and (iii) bitstream encoding.
Prediction is to estimate and compensate inter-frame motion.
It processes a video frame in units of {\em macroblock} ($m \times m$ pixels). 
It uses $n \times n$ (\eg, $16 \times 16$ or  $4 \times 4$) blocks to predict the macroblock in the current frame from its surrounding blocks in the previously-coded frames. 
In particular, live video frames are encoded as either I-frame or P-frame. An I-frame is a complete image and a P-frame is a delta-frame with only the changes from the previous frames to be encoded, which heavily relies on inter-frame prediction (motion estimation). 
{Transformation is to turn a block of pixels into a set of coefficients with rich redundancy, enabling more compression room for quantization.}
The sole standard transform is Discrete Cosine Transform (DCT).
Its output further quantized, namely, each coefficient is divided by an integer quantization parameter (QP). 
QP is used to balance compression and video quality. A larger value means more coefficients are set to zero, resulting in high compression but at the cost of poor decoded video quality. RoI encoding allows the user to adjust QP for each macroblock and apply different compression strategy for different regions. Finally, these quantized transform coefficients and metadata information about compressed data structure and video sequence will be encoded to the bitstream through a concise representation. Video decoding(\circled{4}) is a reverse process, reconstructing the video from the coded bitstream.

In addition, video frames can be pre-processed on drones (detailed next) to facilitate the analytical task at edge. It is first done at \circled{2} and its subsequent operations is further performed at \circled{4} before frames are loaded for video analytics.

\begin{figure}[t]
\centering
\includegraphics[width=0.92\columnwidth]{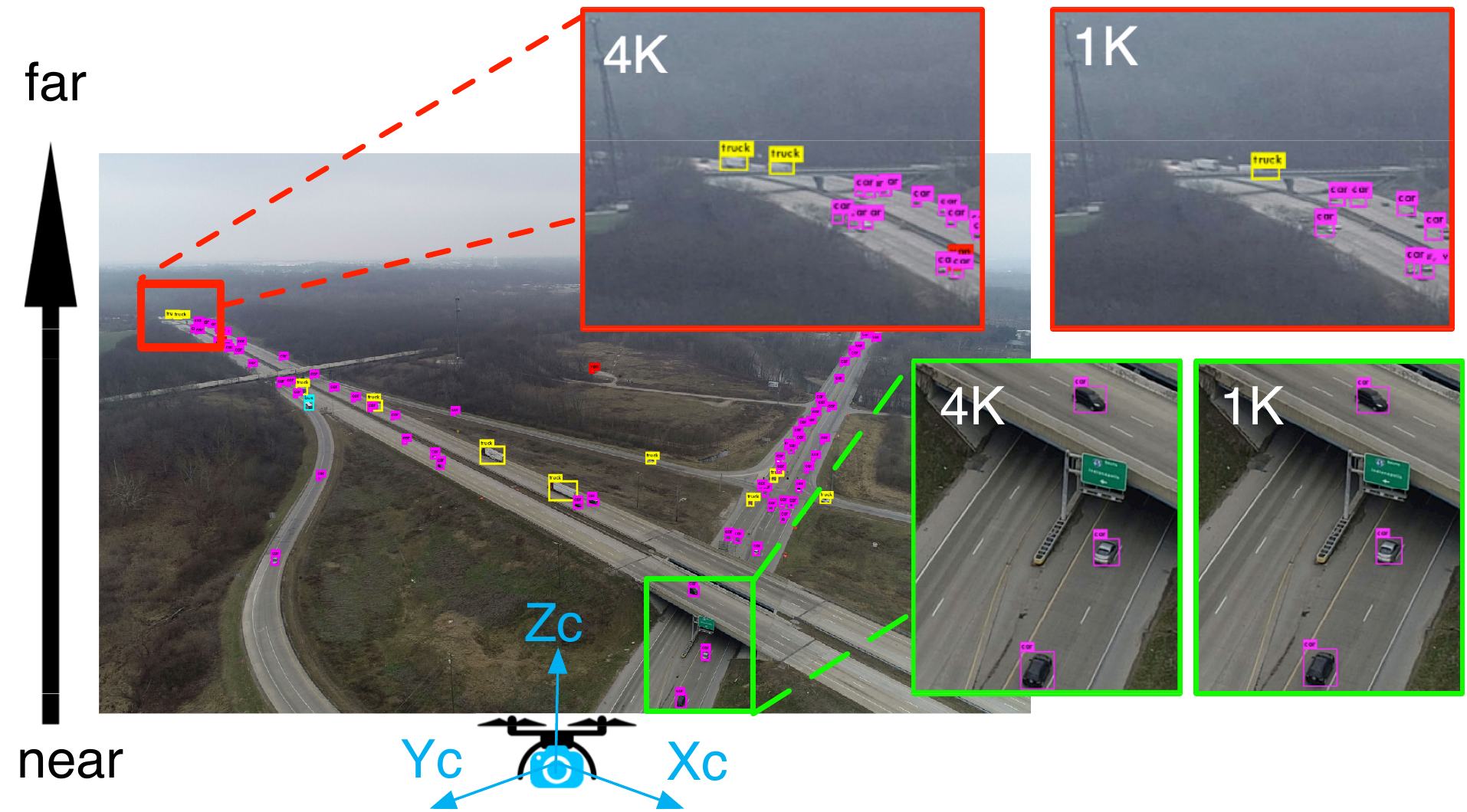}
\caption{An example of drone-sourced vehicle detection.}
\label{fig:drone-ex}
\vspace{-5mm}
\end{figure}

\begin{figure*}
\vspace{-2mm}
\includegraphics[width=1\textwidth]{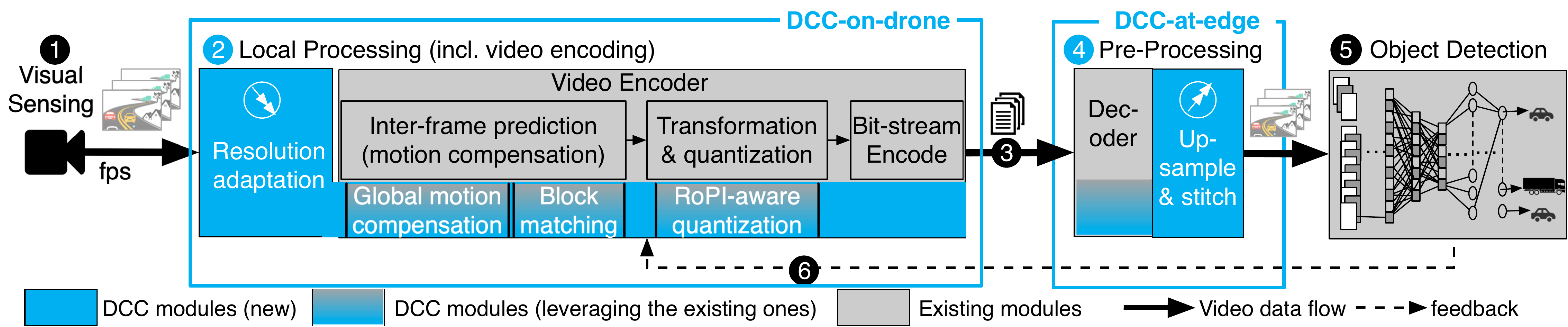}
\caption{\name's architecture and workflow.}
\label{fig:arch}
\vspace{-5mm}
\end{figure*}

\paragraphb{Video analytics/object detection (\circled{5}).} 
The decoded video frames will be applied to different video analytical tasks on demand. 
We choose object detection (particularly, vehicle detection) as our video analytical task in this paper while the proposed techniques are conceptually applicable to other detection tasks~\cite{Kristan2019a}. \fig~\ref{fig:drone-ex} gives an illustrative example, where a bounding box is used to locate and label the detected vehicle (here, a car in purple, a truck in yellow). It is referred to as a region-of-interest (RoI).
Detection typically applies conventional neural networks (CNNs) to extract features ~\cite{cai2018cascade,redmon2018yolov3,bochkovskiy2020yolov4,jocher2020yolov5,ren2015faster}.
Among them, the popular (and most efficient) technique is to use one-stage detection(e.g. YOLO~\cite{redmon2018yolov3,jocher2020yolov5}).  Specifically, they use a unified architecture for the input images to detect all objects simultaneously and the whole feature map is candidate region. 
In this work, we use YOLOv3~\cite{redmon2018yolov3} and YOLOv5~\cite{jocher2020yolov5} for object detection.
 
\subsection{Existing Approaches to Make it Live}
\label{sect:related}

There are numerous efforts to accelerate video analytics or even make it in live in the recent years. They are from two research camps: (1) computer vision and edge computing,
and (2) multimedia and video codec. 
The former focuses on image-based (frame-based) object detection but largely overlooks intrinsic relevance of consecutive frames, namely, the power of a video format; 
The latter is devoted to efficient video sensing and coding but typically targets at good viewing experience for humans, not tailored to video analytics by machines. 
We next briefly describe the representative and most relevant studies from each camp and their joint efforts.

\paragraphb{Faster object detection.} 
One most popular approach is to run lightweight DNNs at device to reduce the content needed to be transmitted while ensuring accuracy~\cite{wang2018bandwidth,han2016mcdnn,ananthanarayanan2020project,rocket}. 
The device runs weak detectors that are typically designed for mobile platforms or lightweight version of the full architecture, so that only interesting frames need to be transmitted.
Specifically, \cite{wang2018bandwidth} tiles the frames into a few sub-frames and discards uninteresting ones based on local tiny-YOLOv3 outcomes.  
Microsoft Rocket~\cite{ananthanarayanan2020project,rocket} proposes to run a light DNN first and invoke a heavy DNN only when required. Such a light DNN can be performed at the device though it is originally proposed at the edge~\cite{han2016mcdnn}.
Another popular approach is to transfer image-based detection networks to video sources for robust detection (\eg,~\cite{zhu2017deep,wu2019sequence, lu2017online,kang2017object}). 
They exploit spatial and temporal correlation between frames. Note that most solutions do not target for drones but can be conceptually applicable. There are a number of studies to optimize object detection algorithms for drones. For example, \cite{majid2018shuffledet} enhances CNNs with grouped convolution and less inference time; \cite{wu2019delving} develops cross-domain models for robust detection; ~\cite{benjdira2019car} validates that YOLOv3 outperforms faster R-CNN for drone-sourced detection. Given their heavy computation load, they must run at the edge, which is orthogonal to \name.

\paragraphb{Advanced video compression techniques.}
%
Video codec has made new progress to compress more, with the rapid development of computer vision and neural networks. 
One is to take both classical compression architecture and non-linear representation capabilities of neural networks to learn parameters for video compression~\cite{liu2019edge,ma2020video}; These solutions target at retaining high viewing quality for human vision. An alternative direction starts to consider new video codecs tailored to machines~\cite{duan2020video,duan2018compact}; 
However, they are limited to specific tasks and cannot be extended to object detection and generic video analytics for now.

\paragraphb{Exploit feedback from the edge.} 
The last direction attempts to leverage joint power from computer vision and video codec camps.
The basic idea is to exploit feedback from the edge to tune video coding or select what to transmit at the device~\cite{liu2019edge,chen2015glimpse,meuel2018region}.
In particular, \cite{liu2019edge} exploits intermediate feature map to adjust QP values and develop dynamic RoI-based encoding; 
\cite{chen2015glimpse} exploits stale hints to send only trigger frames to the edge; 
\cite{meuel2018region} leverages the outcome of detection results to locate the RoIs across frames and only transmit them later.
In a many-camera system, it can reduce more by leveraging cross-camera analytical results to determine the frames that best capture the scene and combine videos from many cameras~\cite{zhang2015design,cheng2018vitrack,jain2018rexcam,liu2019fusioneye,greengard2019drones}.

Our approach is aligned with the above efforts. Our difference is that the existing studies only explore a very limited compression space and we explore more. 
Actually, there are more tuning knobs (e.g., resolution, RoI, motion, QP, frame rate, etc.), and putting all together should compress more.
However, despite their potentials, it is hard to achieve the synergy effect due to the gap between computer vision and video compression. In particular, the goal of video codec is to preserve the high-fidelity video reconstruction without too much information loss. Video codec uses a fixed frame rate and resolution throughout all video frames to better leverage temporal and spatial correlation. However, computer vision only cares about the target objects. 
\save{All the optimizations are to reduce the useless and redundant information for fast responsiveness. Not all the frames even pixels need be treated equally, thus there are lots of rooms for compressions.}
There is no need for a fixed high frame rate and resolution. Moreover, background information (not RoIs) can be sacrificed. 
Little progress has been made to glue such gap, except for RoI encoding. 
Our work is inspired to examine more tuning parameters to pursue compression room which has been not explored yet. 
Moreover, we make \name compatible with the existing codec so as to make full use of hardware/software optimization in place.

\section{\name Design}\label{sect:overview}
\label{sect:design}


We design and develop \name, deeper-yet-compatible compression to leverage the compression dimensions exposed by the existing codec for drone-sourced video analytical tasks.
Its core idea is to stay adaptive in every possible dimension to encode ``{\em just-necessary}''\footnote{Information may be still over-provisioned as a design tradeoff for lightweight processing load and system compatibility.} information needed for object detection. To make it feasible operated on drones, we build \name on top of the well-established operation flow discussed in \S\ref{sect:back-related} with minor changes. 

\fig~\ref{fig:arch} shows \name's workflow. The modules added for \name are marked in blue, the modules modified on top of the existed ones are marked in hybrid and the rest unchanged (in black). 
\name changes only two modules: on-drone processing (\circledblue{2}) and at-edge preprocessing (\circledblue{4}) so as to work with any detector. In particular, \name adds software patches to video encoder/decoder in place, with no need to change their existing functions and APIs, making it feasible to work on commercial drones without hardware changes.
We explore three dimensions empowered by drones for deeper compression: adaptive resolutions, motion compensation (essential to inter-frame prediction and compression) and quantization and develop three key techniques:


{\bf 1. Resolution adaptation.} Drone-sourced videos exhibit strong far-near effects due to its higher altitude and wider FoV. We find that there is no need to use the same high resolution for the whole frame. Instead, video frames can be divided into smaller-granularity slices each using the needed resolution. Video slices are adaptively downsampled at the device and then upsampled and assembled at the edge for detection. 

{\bf 2. Global motion compensated encoding.}The changes between consecutive frames are introduced by movement of the drone and movement of objects in the physical world. The former introduces global motion across the frames which can be extracted and compensated ahead of time. We thus use sensory contexts of drone movement, say, flight direction and speed to compensate changes induced by drone movement. As a result, we encode changes induced by motions in the real world and reduce the volume that needs to be encoded.
 

{\bf 3. RoPI-aware quantization}. We extend RoIs to Region of Possible interests (RoPIs) as the edge may not quickly and accurately send the feedback to the drone. Moreover, we leverage task-specific context (e.g., vehicles on the roads or parking lots, not everywhere) to fast locate RoPIs with different confidence levels and apply various quantization strategies accordingly.


We next elaborate on each technique.

\subsection{Resolution Adaptation}\label{sect:resolution}

In \name, more than one resolutions can be assigned within the same frame.
At the start of \circledblue{2}, a raw frame tiles into several slices and then is downsampled to different resolutions for transmission. 
At the end of \circledblue{4}, they will be finally upsampled back to one high resolution (here, 4K) that suits computing for object detection. 
We first show two motivations of this idea.

\begin{figure}[t]
\vspace{-5mm}
\centering
\centering
\subfloat[With far-near effects]{
\includegraphics[width=0.5\columnwidth]{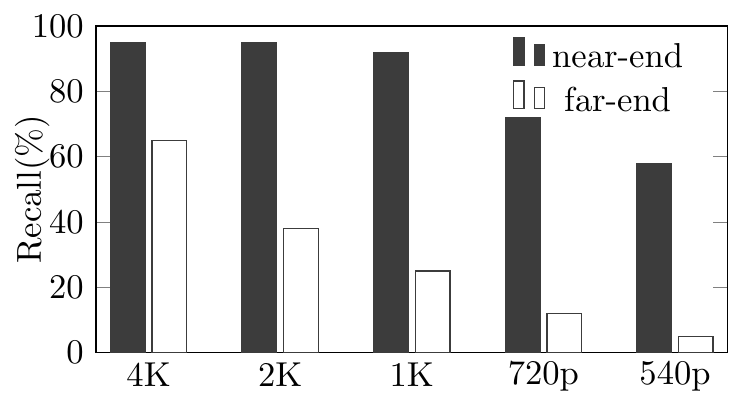}
\label{fig:minimumReso} 
}         
\subfloat[Via downsampling-upsampling]{
\includegraphics[width=0.5\columnwidth]{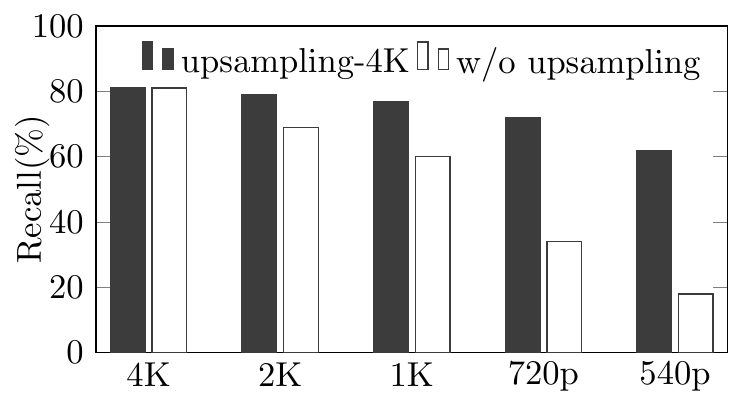}
\label{fig:upsample-overall}
}
\caption{Detection accuracy (recall) results with adaptive resolution.}
\vspace{-3mm}
\end{figure}

{\em $\circ$ Lower resolution tolerated with upsampling prior to object detection.} The video captured by the drone exhibits the strong {\em far-near effect} and there is no need to assign the uniform high resolution for the whole frame. The simple idea is that we can consider the processing at the granularity of slice and assign only needed resolution for detection for each slice. 
As shown in \fig~\ref{fig:drone-ex}, a vehicle at the near-end occupies 200x more pixels than the one at the farest end. We clearly see that object detection can tolerate 1K resolution at the near-end but demands for 4K at the far-end. 
We extend this example to a 10-minute video study in \fig~\ref{fig:minimumReso} and the result show that different resolution requirement for detection within a single frame. This can be exploited to pursue {\em just-necessary} resolution and thus reduce number of unnecessary pixels for closer sub-frames. 

{\em $\circ$ Lower resolution tolerated with upsampling prior to object detection.} High-definition videos are still needed to ensure that the objects that are not near-end occupy enough pixels and have clear features to be detected . However, we find that it is unnecessary to have every pixel identical to the original one as long as the object feature is roughly retained. This makes it possible to transmit at a lower resolution and then upsample at the edge for computing. To validate its effectiveness, we use 500 randomly selected video frames covering various altitudes and perspective angles. All the video frames are originally 4K and then they are first downsampled to 2K, 1K, 720p, 540p and then upsampled back to 4K.
\fig~\ref{fig:upsample-overall} shows upsampling can make up and help detection objection tolerate lower resolution. Specifically, upsampled 2K video can reach a similar recall to that of 4K video, but directly using 2K does not work. 

The key technical challenges are how to slice the video and how to assign a proper resolution for each slice at a small cost.  

\begin{figure}[t]
\includegraphics[width=1\columnwidth]{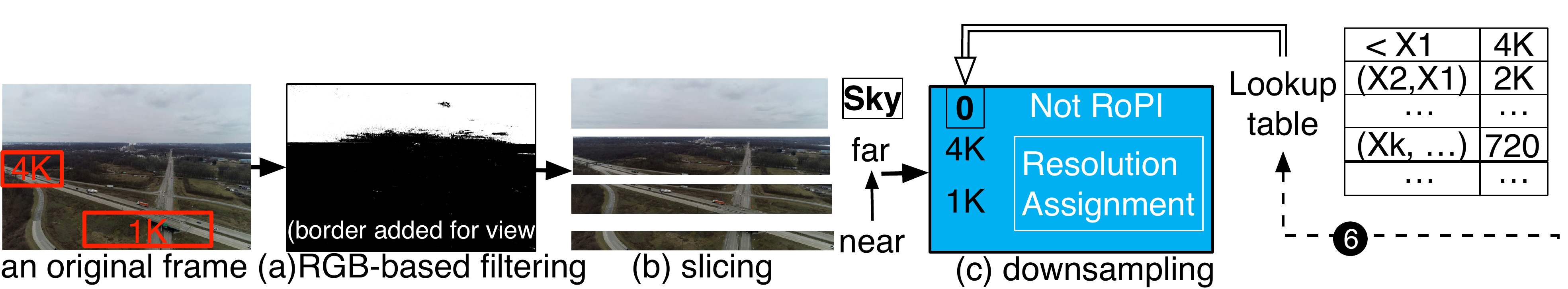}
\caption{Illustration of resolution adaptation.}
\label{fig:ex-resolution}
\vspace{-5mm}
\end{figure}

In our design, we exploit RGB pattern to segment sky and tile the rest part based on the {\em far-near effect}. Then we look up a hash-table to determine the necessary resolution based on the estimation of the pixel occupied for target objects via camera projection model. The process is illustrated in \fig~\ref{fig:ex-resolution}. 

\paragraphb{Camera projection model.}
We first introduce the projection model used to map the physical world coordinate system into the frame coordinate system. 
Note that there are three coordinate systems for the physical world ($O_P-X_PY_PZ_P$), the camera ($O_C-X_CY_CZ_C$) and the frame ($O_F-X_FY_F$). 
Given drone's current location ($x_P$,$y_P$,$z_P$) in the world coordinate system, rotation angles ($\alpha$: yaw, $\beta$: pitch, $\gamma$: roll), and other camera parameters, 
we are able to derive the (approximate) projection from the 3D physical world to the 2D video frame 
using the perspective transformation matrix based on the famous pinhole sensing model in any computer vision textbook~\cite{szeliski2010computer}. 
\vspace{-2mm}
\begin{equation}
P_F = K [ I_{3 \times 3}  \quad 0_{3 \times 1} ]P_C , P_C \approx \left [\begin{array}{ll} R(\alpha,\beta,\gamma) & D_{3 \times 1} \\ 0_{1\times3} & 1\\ \end{array} \right ]P_P.
\end{equation}

Here, $P_P = [x_P, y_P, z_P,1]^T$, $P_C = [x_C, y_C, z_C,1]^T$ and $P_F = [x_F, y_F,1]^T$ are the homogeneous coordinates of the same point in the physical, camera and frame coordinate systems. Moreover, 
\begin{equation}
K = \left [\begin{array}{lll} S_x  & 0 & L/2 \\ 0 & S_y & W/2 \\ 0 & 0 & 1 \\ \end{array} \right ], R(\alpha,\beta,\gamma)  = R_Z(\alpha) \cdot R_Y(\beta), \cdot R_X(\gamma), 
\end{equation}
$K$ is a constant matrix with $(S_x,S_y)$ being the focal length in terms of pixels, $L$ and $W$ as the length and width of the video frame. $T$ is the displacement between $O_P$ and $O_F$; $R(\alpha,\beta,\gamma)$ is a combined rotation matrix determined by the yaw, pitch and roll angles in use;  where $R_Z(\alpha)$ is a standard rotation matrix with $\alpha$ rotation along the Z-axis. $D_{3 \times 1}=[d_x,d_y,d_z ]^T$ is the displacement between the origins $O_P$ and $O_F$ in the physical world coordinate system.  
 
 \begin{figure}[t]
\centering
\includegraphics[width=0.32\columnwidth]{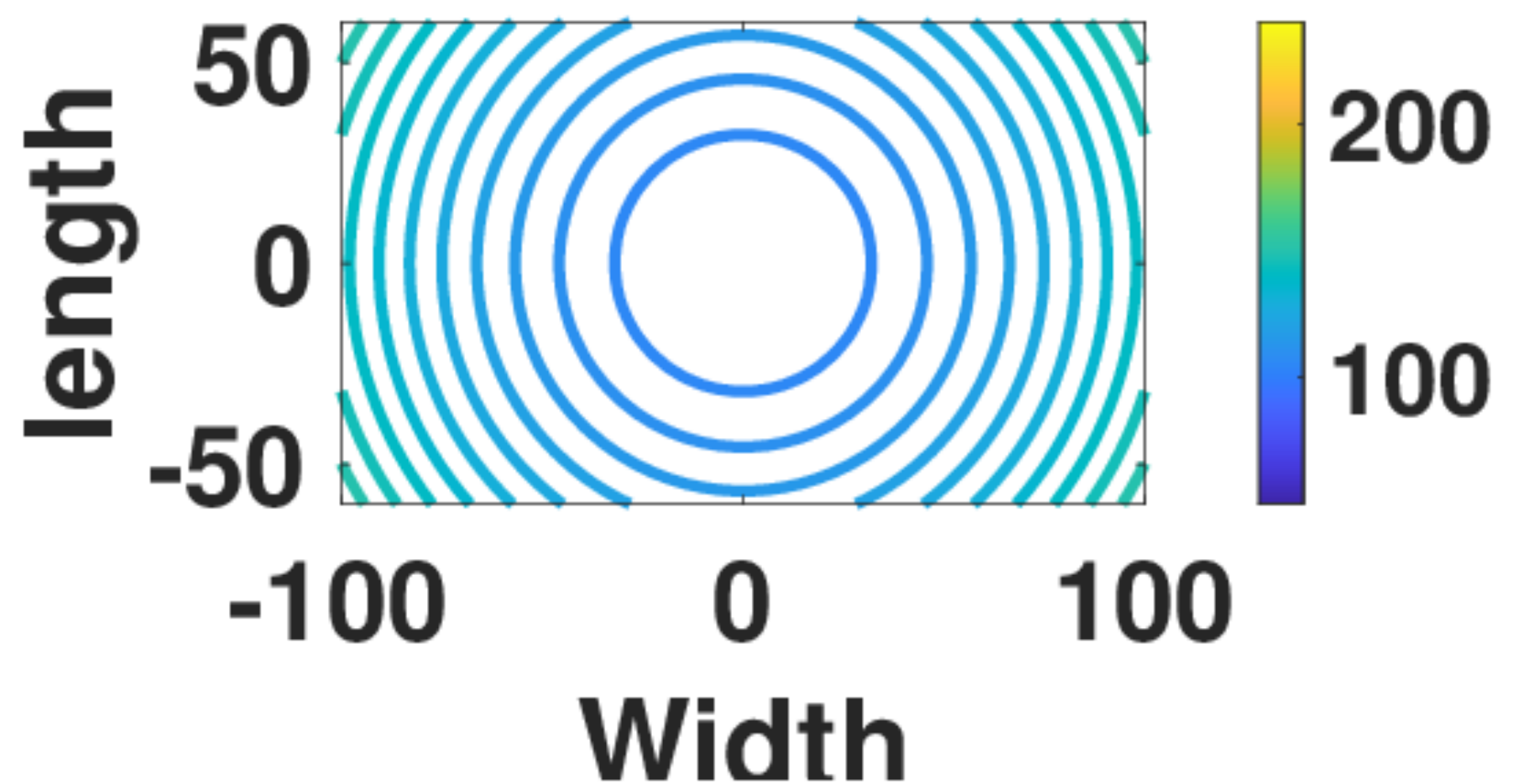}
\label{fig:0degree}
\includegraphics[width=0.32\columnwidth]{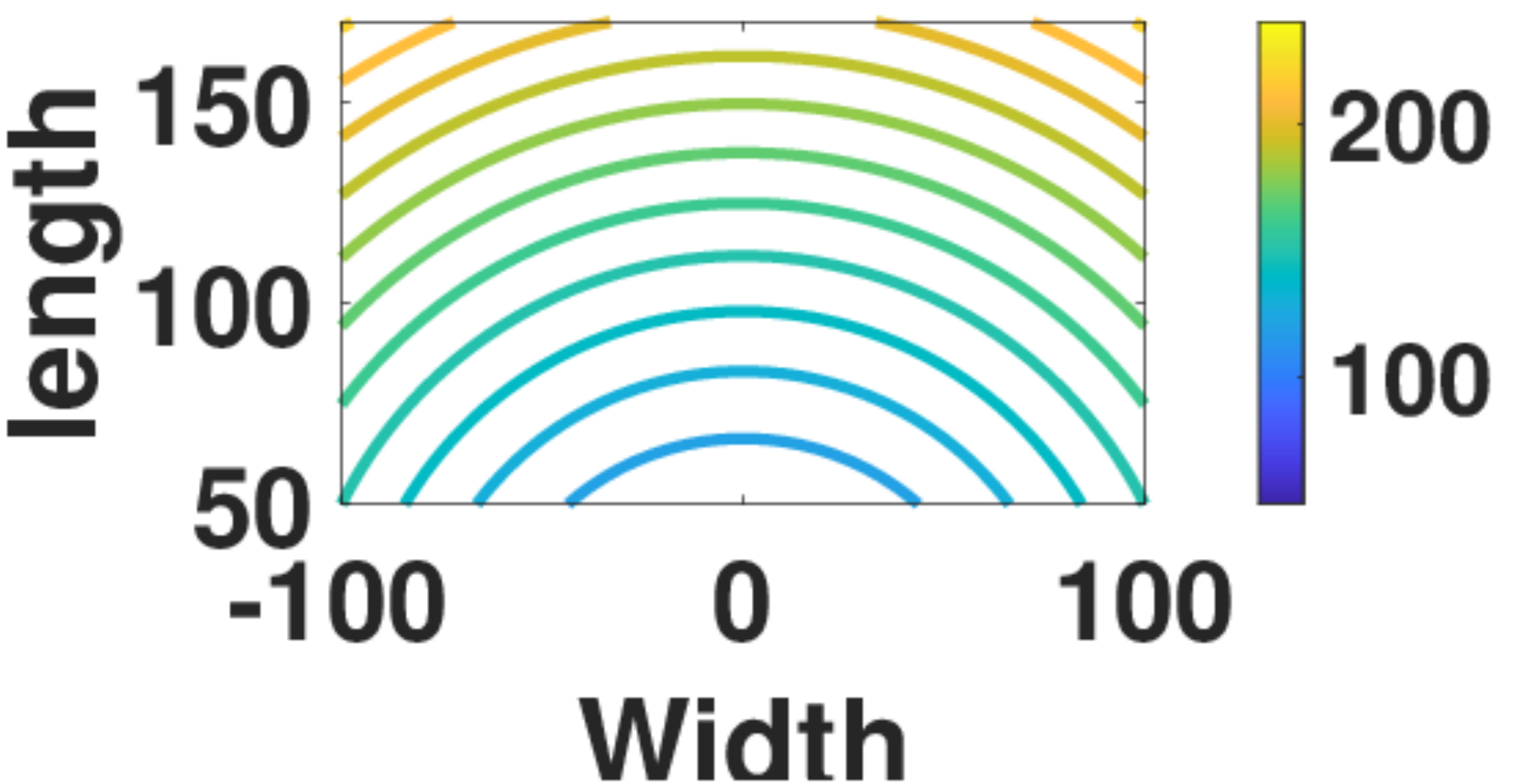}
\label{fig:30degree}          
\includegraphics[width=0.32\columnwidth]{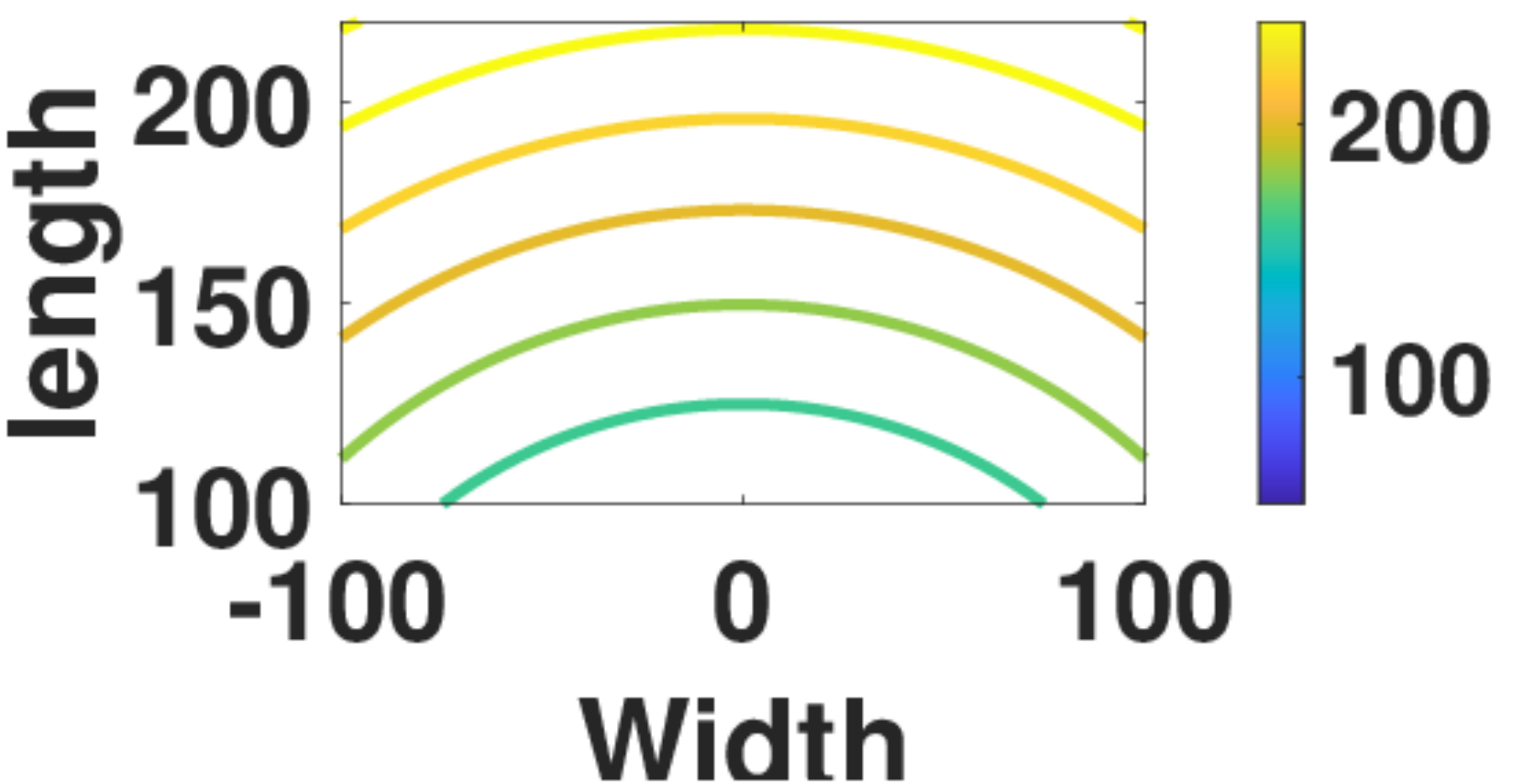}
\label{fig:60degree}
\caption{Working distance map in a frame captured at 100m with various pitch angles (from left to right:  
90 \textdegree, 60 \textdegree, 30 \textdegree).}
\label{fig:degree}
\vspace{-5mm}
\end{figure}

In theory, we can use the above formula to estimate the working distance from the ground to the camera for every pixel. Then tile the video and assign a proper resolution for each slice accordingly.
In practice, there is no need to do such fine-grained calculation for two reasons. On the one hand, we need to retain the shape of slice to be regular(say rectangular) to make it compatible with the current on-device processing. 
Since the contours under different working distances are irregular, in order to make the slices finally rectangular, approximate cropping is necessary.
On the other hand, we can exploit common patterns (say, sky) and projections associated with sensor data (angles) to quickly locate far and near ends, which can give lots of hints. Thus, we do a coarse-grained scene segmentation in \name and just do the horizontal slicing. We next show how we leverage the drone's context to do slicing.

\paragraphb{Sky segmentation.} 
Sky is the salient object in the drone-sourced video analytics, which is apparently not RoPI (at least in case of vehicle detection). We first check if the sky exists via a simple filtering (a). This filter is based on mature computer vision algorithms to do semantic segementation (here, sky). It converts the blocks (pixels) to 0 (non-sky, in black) or 1 (sky, in white). 
Note that the commercial gimbal can set the \textit{Follow} option to maintain the camera stable during the flight. In other words, the angle between the gimbal's orientation and drone' remains constant (almost) at all times. As a result, once captured in the frame, the sky hardly tilts. In the rare case with a sloping sky, it can be recognized from the drone's posture and the camera's relative rotation and these frames are discarded (frames can be reserved if such shooting angle is allowed and reserve rotation is performed, which is our future work). 
We then use color segmentation to locate the sky and identify a horizontal skyline based on a filter in RGB color space. The segmented sky only needs to be sent once to reduce the transmission volume, if there is no scene change. 
After we crop the sky part, the rest is the ground we are interested in.


\paragraphb{On-device slicing.} 
We next tile the subframe into multiple slices based on its far-near effect (b). We use two heuristic rules.
First, when the sky is detected (usually at a small pitch angle), 
the ground directly below it is more distant.
The working distance monotonically decreases from the ground closer to the sky to the ground to the camera.  
Second,
we notice that the far-near effect is related to the rotation angle.
For simplicity, we use a simple and common setting where other two rotation angles except the pitch angle are zero and the pitch angle is used to adjust the shooting perspective. 
To be specific, there is almost no far-near effect for vertical shoot as shown in \fig~\ref{fig:degree} (left). When the pitch angle decrease, the far-near effect grows. It is worth noting that when the angle continually decrease and the sky appear, the distant world scene is greatly distorted even converged as a vanishing point as shown in \fig~\ref{fig:drone-ex}.  
More working distances are involved in a frame when the pitch angle is small.  Moreover,
the smaller angle flats the contour at distinct working distance (see the changes from left to right in \fig~\ref{fig:degree}.
Hence, we can horizontally tile the subframe and determine the number of slice based on its far-near effect. 
In the implementation, we tile the subframe into two, three slices and remain one slice, corresponding to the pitch angle (60\textdegree,75\textdegree) $\cup$ [0\textdegree, 30\textdegree), [30\textdegree,60\textdegree) and  [75 \textdegree,90\textdegree) respectively. Note that this process is only need to be done once if there is no scene change and the processing is lightweight due to the rule-based mechanism. 

\paragraphb{Adaptive resolution assignment.}
We determine the resolution for each slice by a table lookup. We observe that the minimal resolution required for detection is highly correlated with the number of pixel occupied by the target object and we can construct a table to store such mapping ahead of time. The key question at this point is how to know the number of pixel occupied. The commodity drones have limited compute resources that even preclude compressed network(tiny-yolo $<$ 1fps). Thus, we are not able to locate the object on drone and calculate the occupied pixel directly. 

To lower the computational overhead, we use projection model to do the approximate pixel size estimation. We leverage the fact the working distance monotonically decreases from the top to the bottom. For each slice, the smallest size of object may appear in the upper center, where we defined as the origin of world coordinate. Then we can calculate $D_{3 \times 1}$ and $R$ accordingly. Since we know the true size of the object(3m x 1.8m) in the world coordinate and we can estimate the number of pixel through(1). Then we can look up a hash-table to determine the needed resolution and downsample each slice to the assigned value(c). Note that these slices are independent with each other and can be processed in parallel. In order to resemble them at edge, we generate metadata using 2 bits for the number the slice and 2 bits for the resolution to be assigned. Such metadata is integrated into the bitstream and sent to the edge. 

\paragraphb{Construct a look-up table.}
We create a multinomial Logistic Regression model on the server, which is essentially a lookup table. During training, we use different images as supervision to find the smallest acceptable size of object to be detected in different resolutions. The model can be learned offline using historical data. Note that we can also do online training based on the frames transmitted from the drone to tune the trained weight in real-time if needed.  Specifically, if the recall obtained from CNN are continuously low (say, below $70\%$) for more than 5 frames, we will run such online training. The result will be encoded as a hash table and send to the drone to lower on-drone processing load at runtime. Since there are a small number of standard resolution options in total, we can just do a coarse-grained classification. Correspondingly, the size of table should be small and it is easy to transmitted back to device.

\subsection{Global Motion Compensated Encoding}\label{sect:motion}
\begin{figure}[t]
\includegraphics[height=2.75cm]{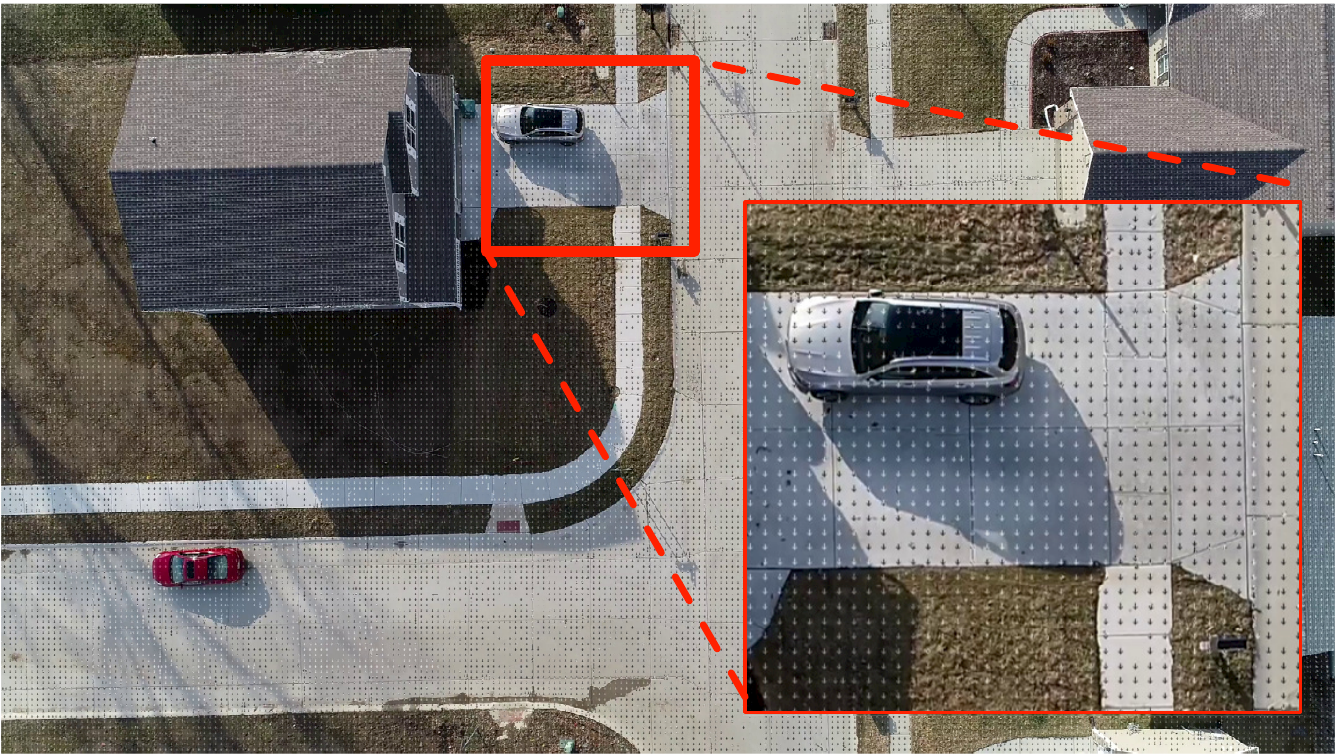}
\includegraphics[height=2.75cm]{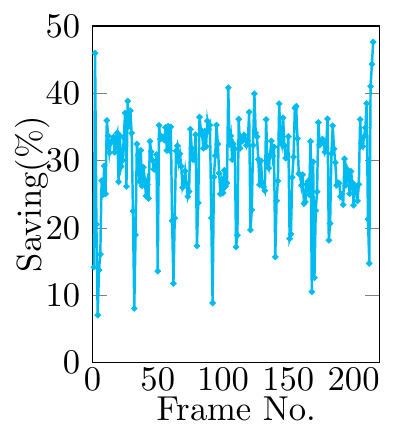}
\raisebox{0.12cm}{
\includegraphics[height=2.5cm]{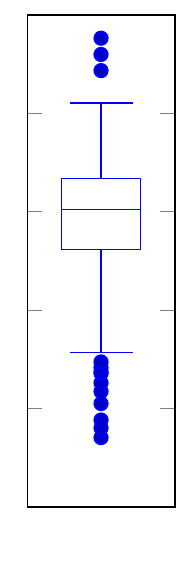}
}
\caption{Illustration of a global motion vector induced by drone's movement and its possible saving gains.}
\label{fig:common_mv}
\vspace{-3mm}
\end{figure}

The changes between two adjacent frames are caused by both movement of the drone and real movement of objects in the physical world. Clearly, we only want to encode the latter one since the more changes involved, the larger the amount of data after encoding. The most intuitive idea is to compensate for the change introduced by drones. 

To validate this idea, we dig into the intermediate result of inter-frame prediction.
 \fig~\ref{fig:common_mv} (left) gives an illustrative example of motion vectors over all the macroblocks at one frame, using traditional video codec (H.264). We clearly see that all the blocks have non-zero motion vectors.  
But when we look closer, most blocks share a common motion vector because the drone is flying southward.
As the drone movement can be known from the sensor data even ahead of the time, we can extract this common motion vector to do motion compensation. 
We evaluate its potential saving gain in \fig~\ref{fig:common_mv} (right). 
For a stream of frames (here, 220 frames), we assume that they are encoded as P-frames in reference of its previous frame. 
We achieve $32\%$ bit saving on average for these P frames by compensating the global motion. 

\begin{figure}
\vspace{-3mm}
\subfloat[2 consecutive frames]{
\includegraphics[width=0.32\columnwidth]{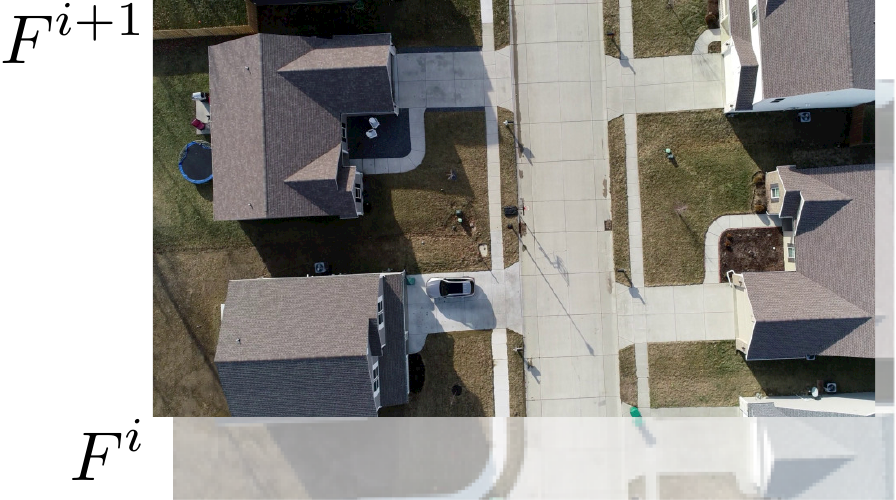}
\label{fig:ex-motion-1}
}
\subfloat[Global motion offsets]{
\includegraphics[width=0.32\columnwidth]{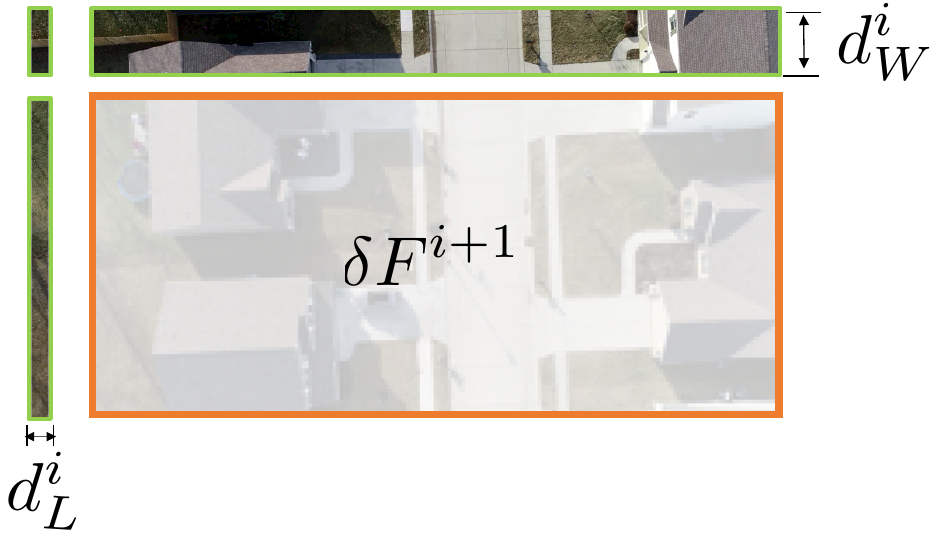}
\label{fig:ex-motion-2}
}
\subfloat[Inter-frame changes]{
\includegraphics[width=0.33\columnwidth]{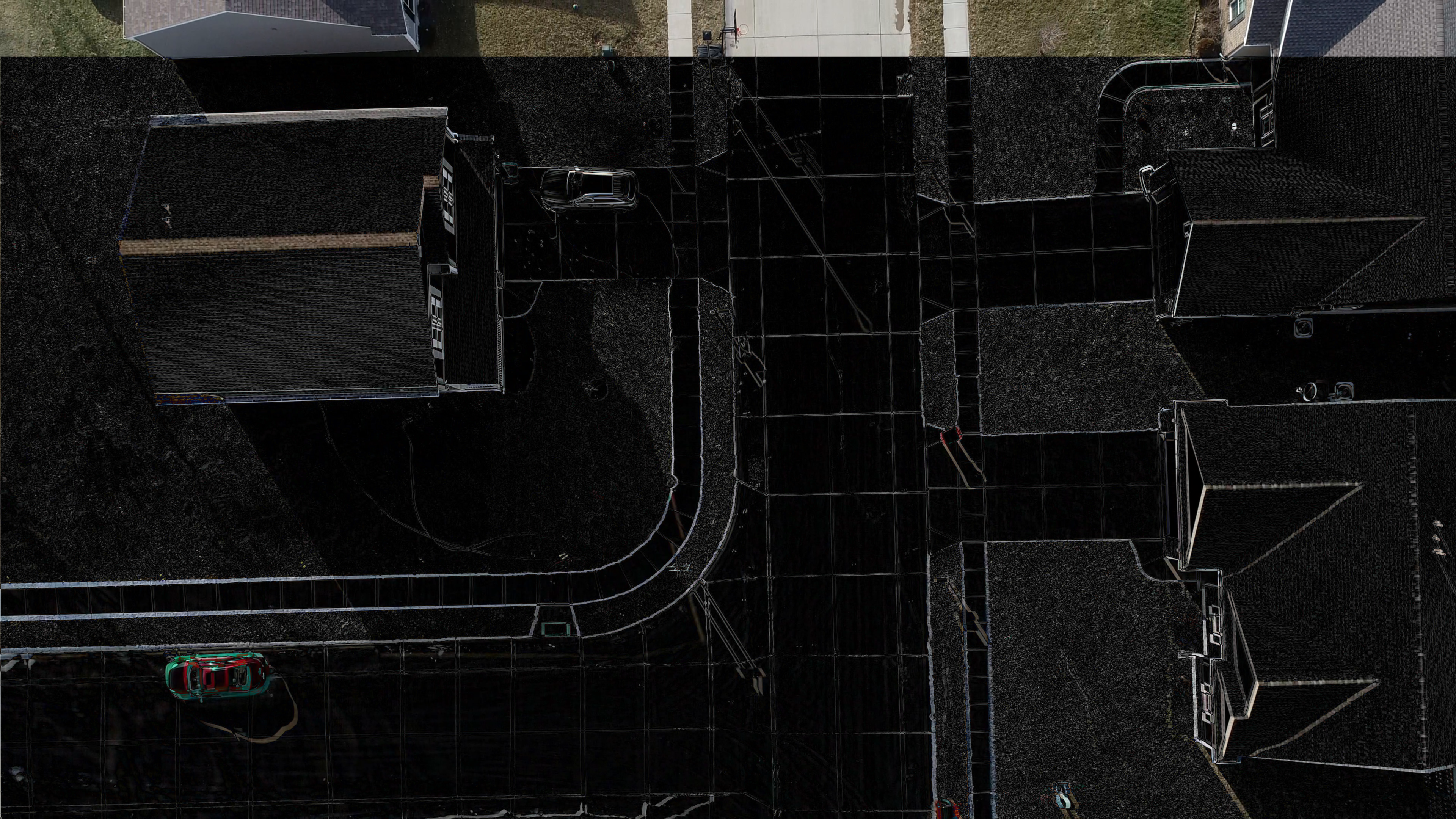}
\label{fig:mv_residuals}
}
\caption{Intermediate results of global motion-compensated encoding (example: \fig~\ref{fig:common_mv}).
In (c), we visualize the residuals (inter-frame changes) after motion-compensation (no change in black).
}
\label{fig:ex-motion}
\vspace{-5mm}
\end{figure}

Moreover, we later show that the residuals after motion compensation can be further compressed. As illustrated in white in \fig~\ref{fig:mv_residuals}, the non-zero parts are mostly tiny and will not impact object detection. These non-zero residual coefficients are also useful for us to track the movement of real objects in the physical world. They can be used as hints to locate or mutually check RoPIs (detailed in \S\ref{sect:region}). 

\begin{figure}
\vspace{-3mm}
\centering
\includegraphics[width=1\columnwidth]{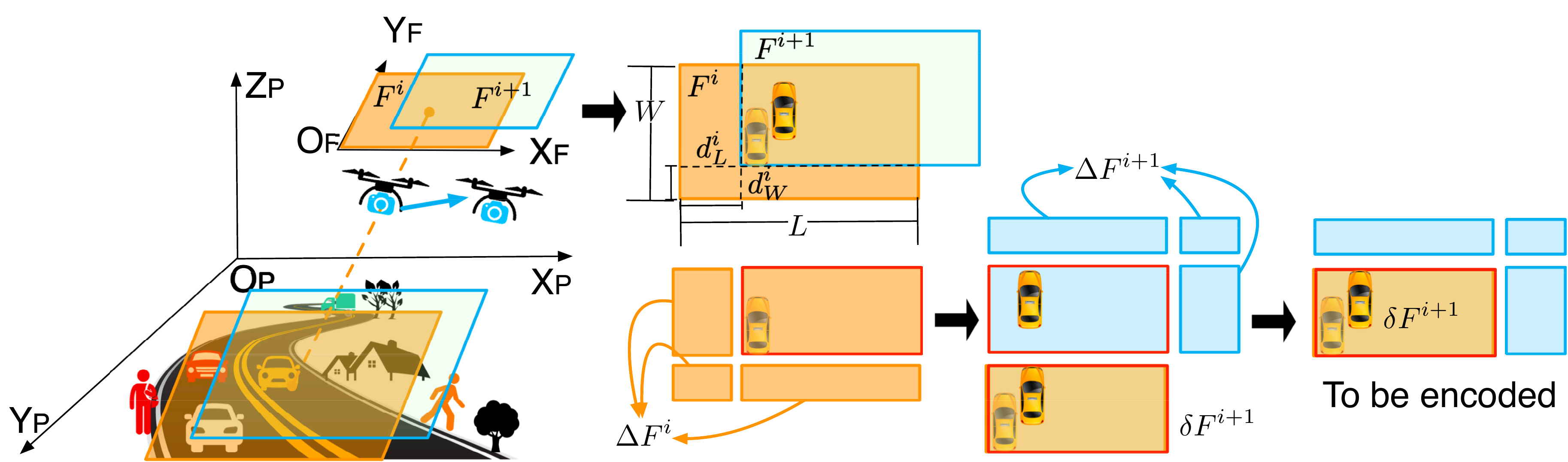}
\caption{Illustration of global motion-compensated encoding.}
\label{fig:motion}
\vspace{-3mm}
\end{figure}

To make it compatible with the existing codec components (\ie, inter-frame prediction), we consider two steps: global motion extraction and motion-compensated encoding. We will leverage \fig~\ref{fig:motion} to illustrate the whole process. 

\paragraphb{Global motion extraction.}
In order to compensate the changes introduced by the drone, we need to extract them based on the projection model given in Eq. (1) and Eq. (2). 
We examine two continuous frames $F^i$ and $F^{i+1}$ to showcase how to get $d_W^i$ and $d_L^i$\footnote{The frame here is the output of {\bf resolution adaptation}, not the original frame.}.
$d_W^i$ and $d_L^i$ are the offset in width ($X_F$-axis) and length ($Y_F$-axis) in the frame coordinate system. Collecting the sensor information from the drone, we know the velocity vector $V = [v_X, v_Y,  v_Z]^T$ and its current position at the i-th frame, as well as its position at the $i+1$-frame. 
We can derive $d_W^i$ and $d_L^i$ as
\begin{equation}
 [d_W^i, d_L^i, 0]^T = P_{F}^{i+1} - P_{F}^i = K' \cdot ( P^{i+1}_{P} - P^i_{P})
 \end{equation}
 $K'$ is derived from Eq. (1) and Eq. (2). 
 
While \fig~\ref{fig:motion} plots relative large offsets for a better view, the actual offsets are quite small.  
In the above example (\fig~\ref{fig:common_mv}), we get {$d_W=28$, $d_L \approx 6$} as illustrated in \fig~\ref{fig:ex-motion-2}. Here, $v_X=0.5m/s$, $v_Y=4m/s$, $v_Z=0m/s$, $d_Z = 50m$ (altitude), $\beta=90$\textdegree (drone flies in the horizontal plane above the ground with a nadir-view), the used resolution is 720p for this slice and the interval between two frames are 200 ms (5 fps). 
Other cases can be similarly extracted. 
Note that our global motion extraction is inspired by \cite{mi2019sensor} which proposes a sensor-assisted global motion estimation method. We share the same idea to estimate motion based on the projection model but our work focuses on working with codec for video compression and handle imperfect motion compensation. 
 


\paragraphb{Motion-compensated encoding.}
We want to integrate global motion compensation into the current codec (here, H.264). 
By manually shifting $F^i$ with the offsets $d_W$,$d_L$, we get a new reference frame $F^{*}_i$. 
The object's dynamics can be retrieved out of the successive substitutions between the overlapping $F^{i+1}$ and $F^{*}_{i}$. \fig~\ref{fig:mv_residuals} gives an illustrative example. We mark this pixel as white (black) if change (no change) is observed.
Clearly, motion compensation is effective to reduce unnecessary motion vectors introduced by drone's flying.

We next describe how to encode these frame changes over the existing codec. 
All the popular encoders including H.264 implement different mechanisms for I-frames and P-frames. 
We need to first determine the frame type: P-frame or I-Frame and then prepare streams for their codec\add{s}. 
We calculate $\delta F^{i+1}$ as the changes between the overlapping $F^{i+1}$ and $F^{*}_{i}$. 
If $\delta F^{i+1}$ is below a given threshold, $F_{i+1}$ will be treated as P-frame, otherwise I-frame.
If P-frame, we divide it into four blocks; Three of blocks (bottom-right, upper-right and upper-left) are the newly appearing regions and there are no reference, they will be treated as I-slice (\ie, I-subframe). 
For the overlapped blocks, we reuse the existing codec modules (\ie, inter-frame prediction) but replace the input; We use the reference frame $F^{*}_i$ instead of $F^i$ to find the best coding mode with the minimum rate-distortion cost. This way, we only encode the changes $\Delta$ and a small amount of motion vectors and other blocks can select \texttt{SKIP} mode and no information need to be encoded. Correspondingly, the frame $\hat{F}^{i+1}$ can be reconstructed from $\Delta,d_W,d_L ,\hat{F}_i$ at the decoder. We also would like to highlight that both $d_{W}$ and $d_{L}$ are small, compared to W and L. Thus, very few newly introduced parts without the reference need to be encoded. 


\paragraphb{Handling imperfect compensation.} 
We know that there is no need to keep the pixels exactly the same as these sensed by the camera for the sake of object detection. 
This gives us extra room to further compress video frames after motion compensation. 
We carefully examine the residuals (as illustrated in \fig~\ref{fig:mv_residuals}) and still a portion of them are non-zero though they are still in the physical world. This is because distortion is introduced by global motion compensation.
There are two reasons: First, the barrel distortion caused by the camera varies by frames. 
Second, the offsets we calculate cannot achieve a perfect compensation due to imperfect sensing data (speed and angles).  
Moreover, the pixel offsets must be integers, different from the real displacement. 
To resolve these problems, we round all the small residuals to 0 as they are negligible to object detection. 
To this end, we use the parameter available in the codec, which is called \texttt{decimate-mb}.
When residual coefficients are below this threshold, they will be round to 0 and skip the frame prediction. 
In \name, we increase this \texttt{decimate-mb} threshold to eliminate the distortion introduced by imperfect motion compensation. 
This also handles inter-frame distortions inherent in an imperfect visual sensing. 
We rinse off these tiny changes which are of no impact on object detection, although desired for high-fidelity viewing.

\subsection{RoPI-aware Quantization}\label{sect:region}

\begin{figure*}
\vspace{-5mm}
\subfloat[RoPI: Road segmentation]{
\includegraphics[width=0.25\textwidth]{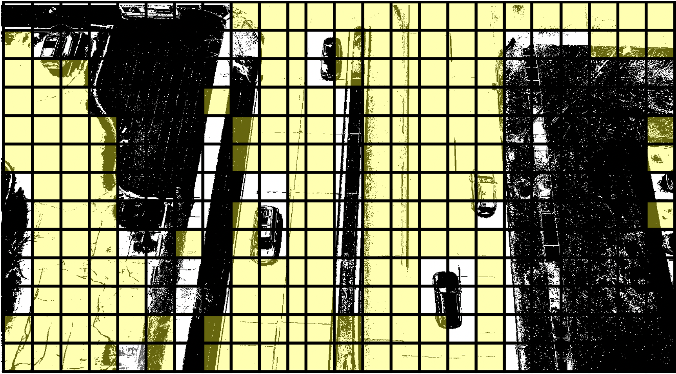}
\label{fig:segement}
}
\subfloat[RoPI: Nonzero residuals hints]{
\includegraphics[width=0.25\textwidth]{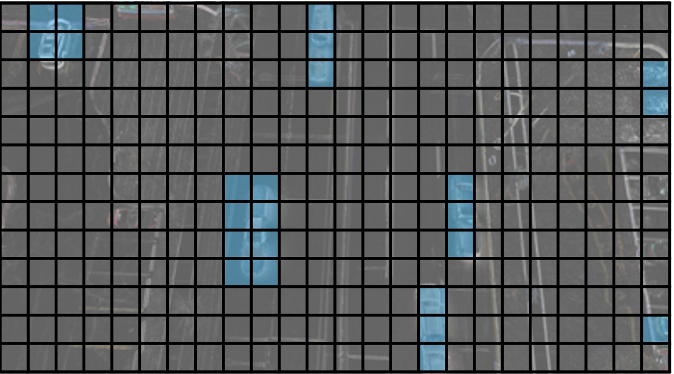}
\label{fig:non-zero}
}
\subfloat[RoPI: Extend detection results]{
\includegraphics[width=0.25\textwidth]{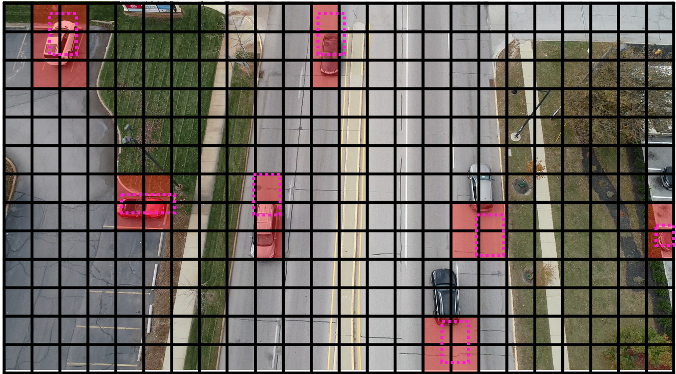}
\label{fig:detection}
}
\subfloat[Finalized RoPI: Union set ]{
\includegraphics[width=0.25\textwidth]{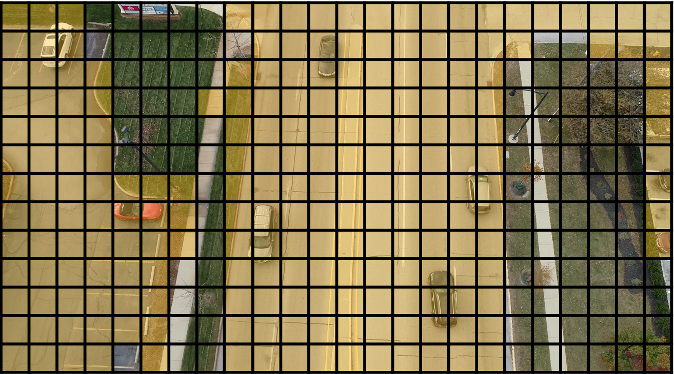}
\label{fig:union}
}
\caption{Intermediate results of RoPI-aware quantization (better view in color).
}
\label{fig:ropi}
\vspace{-3mm}
\end{figure*}


Encoding is almost done after the blocks of residuals and motion vectors are transformed to DCT coefficients by the existing codec. Quantization is the last step to control (actually, reduce) information carried by 
these DCT coefficients. As introduced in the background (\S\ref{sect:back}), QP is used to balance compression and video quality. Each coefficient is divided by a QP and thus more are set to zero with a larger QP value. More compression is achieved at cost of lower video quality. 
In \name, we use this exposed parameter to realize RoPI-aware encoding.   

The idea of RoI-aware encoding is not new~\cite{liu2019edge}. By selectively applying a higher degree of compression to parts of the frame that are unlikely to contain objects of interest can reduce volume to transfer while ensuring high quality for the regions with candidate objects. The key challenge is how to determine the RoI frame by frame.
\cite{liu2019edge} leverages the intermediate results of CNN for last processed frame to define the RoI dynamically for mobile virtual reality (VR) detection.
However, such design may not work well in the drone scenario for three reasons. 
First, high-altitude object detection is less accurate and suffers from more false negatives compared with object detection for VR. 
These false negatives are propagated since lots of RoIs are missed, resulting in severe performance degradation.
Second, the target object location (its bounding box) in the current frame significantly deviates from the one detected in the previous frame due to high mobility of both drone and target objects. 
The displacement is hard to accommodate by slightly enlarging each detected RoI by one macroblock. 
Third, the network condition is varying and cannot assure in-time feedback to assist RoI encoding. 
In one word, simply counting on the feedback from the edge is not sufficient to handle the videos captured by a flying drone. 

To resolve above challenges, we exploit drone's application context to efficiently and reliably avoid missing regions of interests. We thus enlarge RoI to RoPI (of {\em probably} interests) to tolerate the possible false negatives in object detection. Moreover, we define region of no interests (RoNI) to exclude the regions that are surely irrelevant, for example, sky and lawn. Instead of only waiting for the feedback from the edge, the drone uses heuristic rules (similar to sky segmentation in \S\ref{sect:resolution}) to roughly locate RoPIs, in a progressive way and use previous results of global motion compensation to further tune the regions. We define RoPI as a unit set of three sets to guarantee no missing areas. To fully utilize the current codec, we use macroblocks of 16x16 pixels as the smallest granularity to define regions. We mark the macroblocks as RoPI if the majority of the pixels are of interests. We next elaborate on the process. 


The first set includes the street part, we leverage a fact that all the vehicles are on the road to locate RoPIs. We segment the road by color segmentation and mark the macroblocks as yellow if the majority of the pixels are of interest (in white).  Note that this process is only done once if drone is hovering, otherwise we can do it for every 2 seconds. The frequency could be altered based on the drone's speed. 
The sample output of road segmentation is given in \fig~\ref{fig:segement}. In this example, the video frame is captured above the 40m at a speed of 5m/s. 720p resolution is in use. There are  80x45 macroblocks in total and we just show a grid of 24x13 macroblocks for illustration. 
While most of the streets are covered, there are lots of vehicles that are filtered out due to different color. We also notice that some roads will not be included because of the shadow caused by the obstruction.

To address the above problems, we further tune the regions and patch to a complete road view. 
We make use of both on-device processing outcomes and likely-stale feedback from the edge. 
We would like to note that it is easy to track moving vehicles using the intermediate results of global motion compensation; Non-zero residuals for the current frame $F^{i+1}$ with the reference frame $F^{*}_i$ are likely moving objects. They are marked in blue in \fig~\ref{fig:non-zero}. This strategy ensures there are no false negatives for moving objects and mutually check RoPIs. For still vehicles, we thus leverage the object detection results by uniformly enlarging the bounding box according to the speed of the drone. Note that it can tolerate stale feedback because the vehicles are still. As seen in \fig~\ref{fig:detection}, we slightly enlarge the bounding box with one blocks by virtue of the offset $d_W$ and $d_L$ and mark these blocks in red. 
We see that all the detected still vehicles are included although the moving vehicles are not completely covered (because the speed of the vehicles is not considered but they are covered above). 
We use their union set as RoPIs (orange blocks in \fig~\ref{fig:union}), and the rest parts are RoNIs. 

Clearly, using a fixed QP can not realize adaptive compression quality for identified RoPIs and RoNIs. 
To remain compatible with the existing codec, we do quantization twice. The first is the default QP which is dynamically set by the existing codec submodule, and the second is to set  QP$=0$ ( lossless compression) for RoPIs and QP$=20$ for RoNIs.


All the parameters (metadata) generated and used by \name will be encoded into bit stream by {\em \name encode} submodule. Afterward, it is really ready to go. At edge, {\em \name decode} first extracts the metadata and then perform regular decoding. The decoded frames will be upsampled to one resolution and switched to restore the original frame. To reduce the effect of uneven seams on object detection during stitching, we smooth the transition using Gaussian pyramid.



\section{Evaluation}\label{sect:eval}
\newcommand{\tabincell}[2]{\begin{tabular}{@{}#1@{}}#2\end{tabular}}  

In this section, we first describe our evaluation methodology. Then we use representative examples to show benefits of our design. Finally, we show the statistical results given different impact factors. Our key findings are:

{\bf $\bullet$ Compression gain.}  \name has significantly compressed videos to be transferred. It has achieved 9.5-fold compression on average(up to 31.8-fold) over the baseline approach. Compared to \texttt{ED}(image-based state-of-the-art), it achieved 19\% -- 683\% compression gains.  

{\bf $\bullet$ Accuracy loss.} We confirm that \name have small impact on object detection accuracy.  In terms of  $F_1$ score (both precision and recall), accuracy loss is $<$1.5\% in most cases. 

{\bf $\bullet$ On-device overhead.}  \name is able to operate at 10.1 fps on average for original(4K) video without any hardware acceleration, highlighting the ability to detect video sources no larger than 10 fps in real time. 

{\bf $\bullet$ Detector independence. } \name can apply for different CNN-based detectors, and the more advanced the detector, the less accuracy loss.
 
 \begin{table}
\vspace{-3mm}
\resizebox{1\columnwidth}{!}{{
\begin{tabular}{ l r r r | r}
\hline
 & {\bf Residence}  & {\bf Local} & {\bf Highway} & {\bf Total} \\
 \hline
All run length  &  23min13s & 13min1s & 64min20s  & {\bf 100min34s}\\
Altitude Setting(m) & 50, 75, 100 & 50, 75,100 &40, 50, 75, 100 \\
Pitch angle Setting(\textdegree) &  90 & 90 & 15, 30, 45, 60, 90 \\
Speed Setting(m/s) & 5, 10 & 0, 5, 10 & 0, 5, 10 \\
\hline
\end{tabular}
 }}
\caption{Dataset of original videos \save{captured by drones }for the evaluation.}
\label{tab:dataset}
\vspace{-3mm}
\end{table}

\begin{table*}
\vspace{-3mm}
\normalsize
\resizebox{1.0\textwidth}{!}{
    \begin{tabular}{lr|rrr|rrr|rrr|rrr|rrr|rrr|}
\cline{3-20}
&&
\multicolumn{3}{c}{(a) R, 50m, 5m/s,  90\textdegree} & 
\multicolumn{3}{c}{(b) L, 70m, 5m/s, 90\textdegree} &
\multicolumn{3}{c}{(c) H, 50m, 0m/s,  90\textdegree} &
\multicolumn{3}{c}{(d) H, 100m, 5m/s, 90\textdegree} &
\multicolumn{3}{c}{(e) H, 100m, 0m/s, 90\textdegree} &
\multicolumn{3}{c|}{(f) H, 100m, 0m/s, 30\textdegree} \\
&&
\multicolumn{3}{c}{\includegraphics[width=0.17\textwidth]{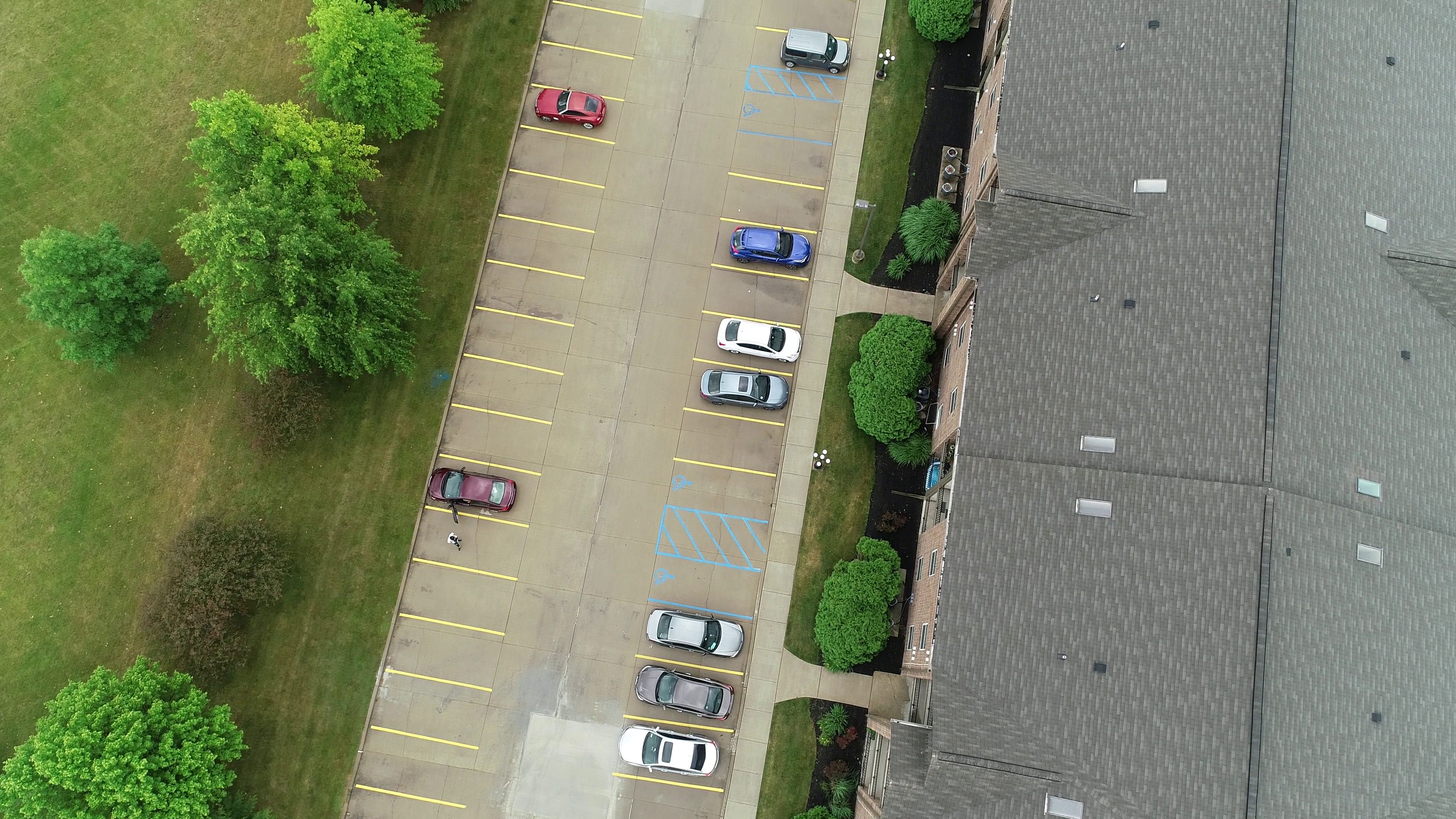}} &
\multicolumn{3}{c}{\includegraphics[width=0.17\textwidth]{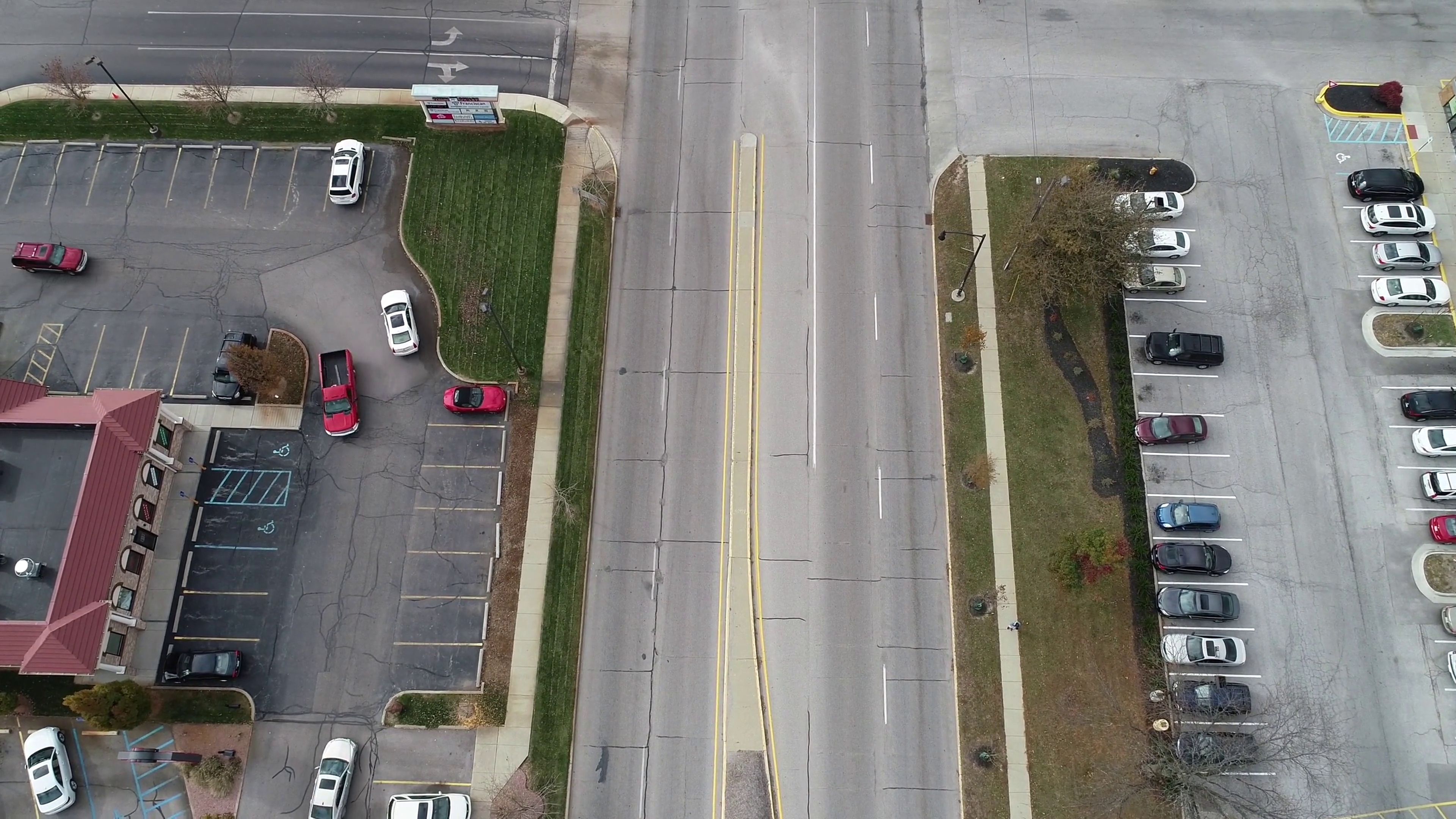}} &
\multicolumn{3}{c}{\includegraphics[width=0.17\textwidth]{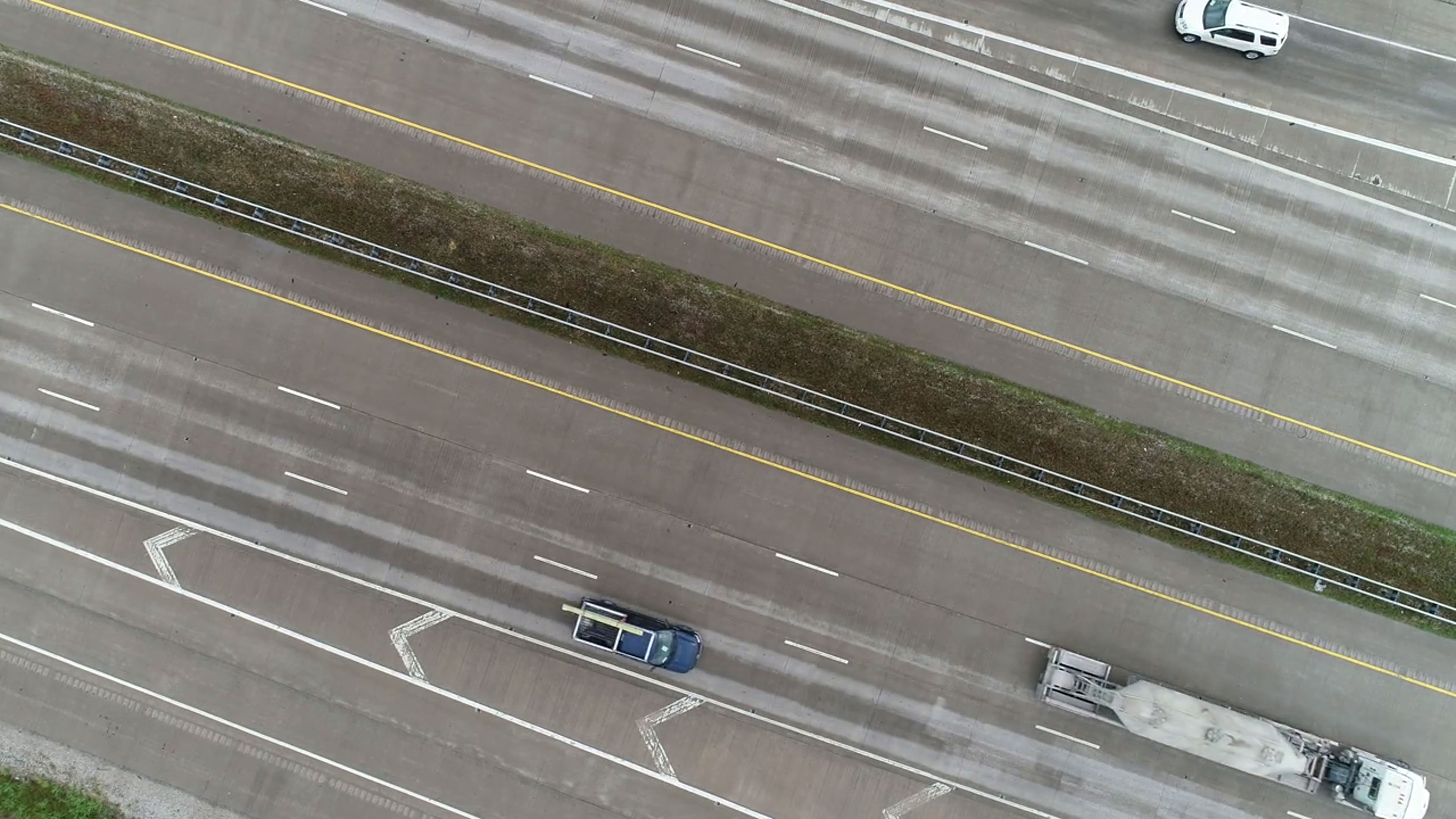}} &
\multicolumn{3}{c}{\includegraphics[width=0.17\textwidth]{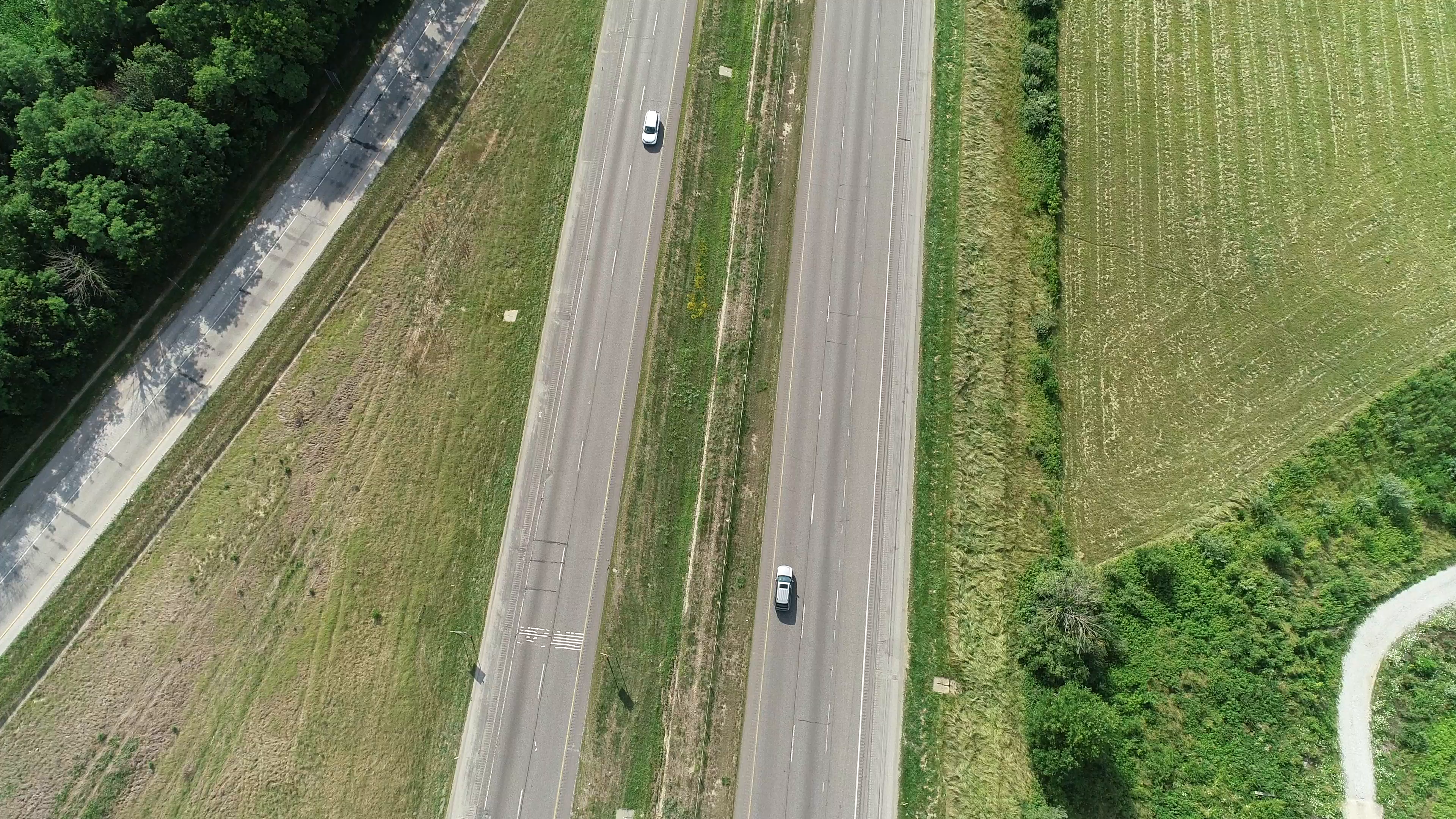}} &
\multicolumn{3}{c}{\includegraphics[width=0.17\textwidth]{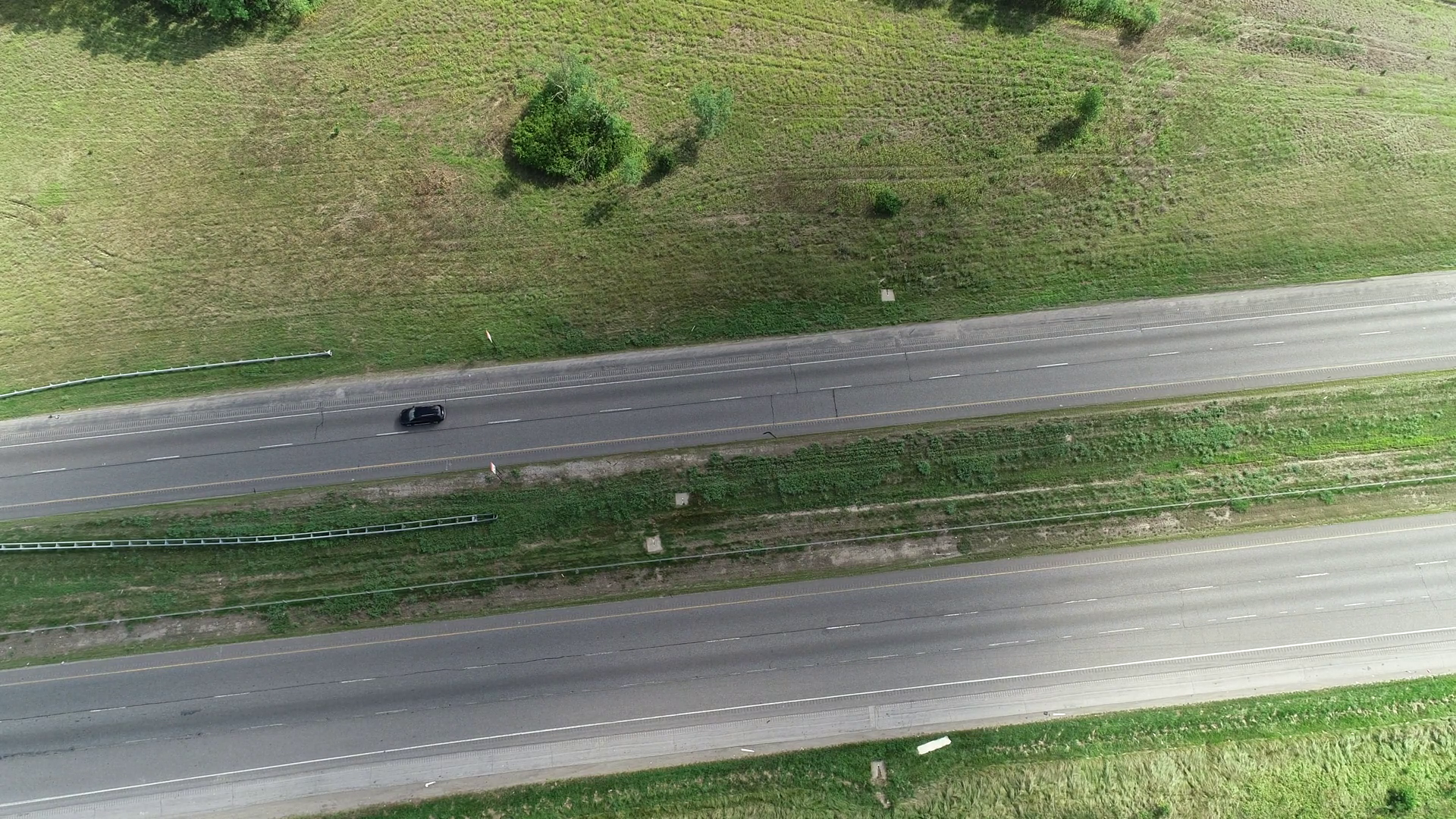}}&
\multicolumn{3}{c|}{\includegraphics[width=0.17\textwidth]{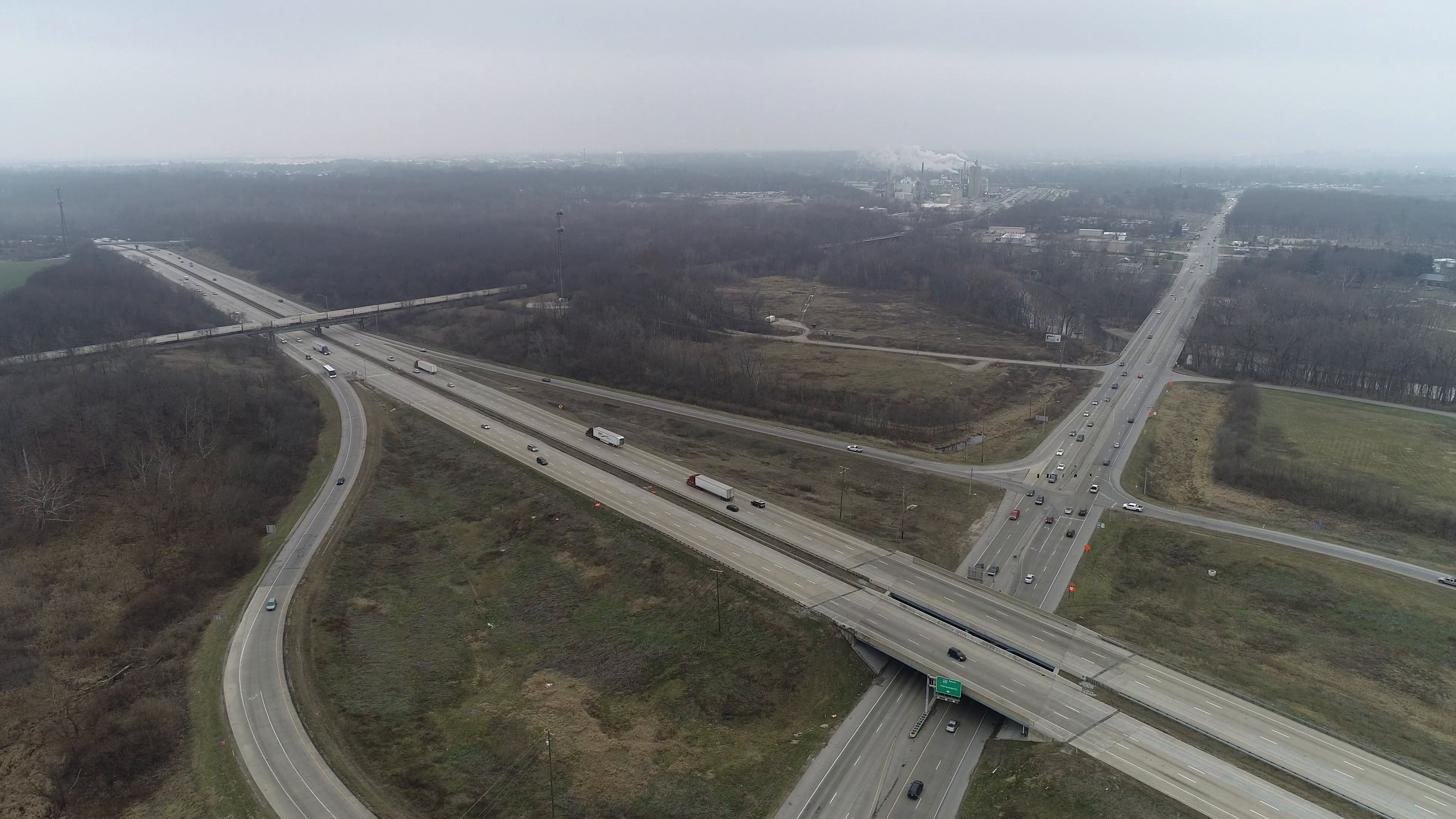}}
\\
&&
\multicolumn{3}{c}{\includegraphics[width=0.17\textwidth]{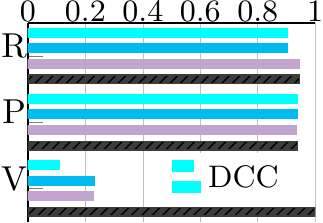}} &
\multicolumn{3}{c}{\includegraphics[width=0.17\textwidth]{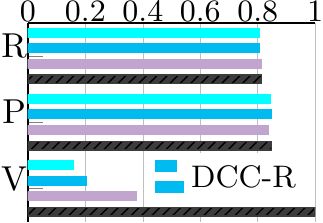}} &
\multicolumn{3}{c}{\includegraphics[width=0.17\textwidth]{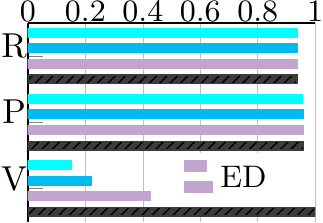}} &
\multicolumn{3}{c}{\includegraphics[width=0.17\textwidth]{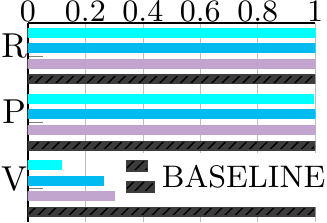}} &
\multicolumn{3}{c}{\includegraphics[width=0.17\textwidth]{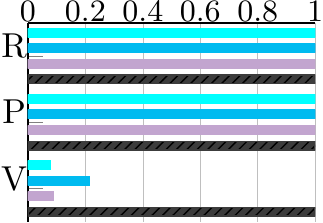}} &
\multicolumn{3}{c|}{\includegraphics[width=0.17\textwidth]{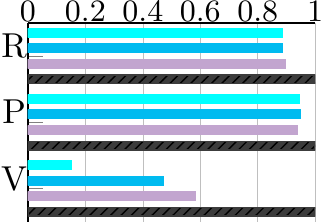}} 
\\
\cline{3-20}
& & V (MB) & P (\%) & R (\%) &
V (MB) & P (\%) & R (\%) &
V (MB) & P (\%) & R (\%) &
V (MB) & P (\%) & R (\%) &
V (MB) & P (\%) & R (\%) &
V (MB) & P (\%) & R (\%) \\
\hline
\multirow{2}*{\texttt{BASELINE}}
 & {\texttt{V3}}&
\multirow{2}*{33.1} & 93.9\%& 94.5\% &
\multirow{2}*{33.2} & 84.8\% & 81.4\% &
\multirow{2}*{31}& 95.9\% & 93.7\% &  
\multirow{2}*{47.7}& 94.6\%& 94.5\% &
\multirow{2}*{37.8}& 100\%  & 100\% &
\multirow{2}*{23.9} &  94.9\% & 90.7\%
\\ 
& {\texttt{V5}}
 & & 95.9\%& 98.1\% &
 & 88.6\% & 82.7\% &
& 95.9\% & 95.8\% &  
& 100\%& 100\% &
& 100\%  & 100\% &
 &  95.2\% & 94.3\%
\\ 
\hline
\hline

\multirow{2}*{\texttt{{ED}~\cite{wang2018bandwidth}}}
 & {\texttt{V3}}&
\multirow{2}*{7.58} & 93.3\%& 94.5\% &
\multirow{2}*{12.5} & 83.8\% & 81.4\% &
\multirow{2}*{13.2}& 95.9\% & 93.7\% &  
\multirow{2}*{20.2}& 94.5\%& 94.5\% &
\multirow{2}*{3.4}& 100\%  & 100\% &
\multirow{2}*{13.9} &  93.8\% & 89.5\%
\\ 
& {\texttt{V5}}
 & & 95.2\%& 97.7\% &
 & 88.8\% & 81.6\% &
& 95.9\% & 95.8\% &  
& 100\%& 100\% &
& 100\%  & 100\% &
 &  95.2\% & 93\%
\\
\hline
\multirow{2}*{\name-R }
 & {\texttt{V3}}&
\multirow{2}*{7.7} & 93.9\%& 90.2\% &
\multirow{2}*{6.7} & 84.7\% & 80.7\% &
\multirow{2}*{6.8}& 95.9\% & 93.7\% &  
\multirow{2}*{11.3}& 94.6\%& 94.2\% &
\multirow{2}*{8.1}& 100\%  & 100\% &
\multirow{2}*{11.3} &  94.9\% & 88.4\%
\\ 
& {\texttt{V5}}
 & & 95.5\%& 97.7\% &
 & 88.3\% & 82.4\% &
& 95.9\% & 95.8\% &  
& 99.8\%& 100\% &
& 100\%  & 100\% &
 &  95.2\% & 92.1\%
\\ 
\hline
\multirow{2}*{\name}
 & {\texttt{V3}}&
\multirow{2}*{3.6} & 93.8\%& 90.2\% &
\multirow{2}*{5.3} & 84.4\% & 80.7\% &
\multirow{2}*{4.7}& 95.4\% & 93.7\% &  
\multirow{2}*{5}& 94.4\%& 93.6\% &
\multirow{2}*{2.9}& 100\%  & 100\% &
\multirow{2}*{3.6} &  94.6\% & 88.4\%
\\ 
& {\texttt{V5}}
 & & 95.9\%& 97.7\% &
 & 88.3\% & 82.3\% &
& 95.4\% & 95.7\% &  
& 99.2\%& 100\% &
& 100\%  & 100\% &
 &  95.1\% & 92.1\%
 \\
\hline
\hline
\multirow{2}*{$\gamma$-\texttt{ED}~\cite{wang2018bandwidth}}
 & {\texttt{V3}}&
\multirow{2}*{ 337\%} & 0.6\%& 0\% &
\multirow{2}*{166\%} & 1.2\% & 0\% &
\multirow{2}*{135\%}& 0\% & 0\% &  
\multirow{2}*{136\%}& 0\%& 0\% &
\multirow{2}*{1012\%}& 0\%  & 0\% &
\multirow{2}*{72\%} &  1.1\% & 1.3\%
\\ 
& {\texttt{V5}}
 & & 0.4\%& 0.4\% &
 & -0.2\% & 1.3\% &
& 0\% & 0\% &  
& 0\%& 0\% &
& 0\%  & 0\% &
 &  0\% & 1.4\%
 \\
\hline
\multirow{2}*{$\gamma$-\name}
 & {\texttt{V3}}&
\multirow{2}*{ 819\%} & 0.1\%& 4.8\% &
\multirow{2}*{526\%} & 0.5\% & 0.9\% &
\multirow{2}*{560\%}& 0.5\% & 0\% &  
\multirow{2}*{754\%}& 0.2\%& 1.0\% &
\multirow{2}*{1203\%}& 0\%  & 0\% &
\multirow{2}*{564\%} &  0.3\% & 2.6\%
\\ 
& {\texttt{V5}}
 & & -0.4\%& 0.4\% &
 & 0.4\% & 0.5\% &
& 0.5\% & 0.1\% &  
& 0.4\%& 0\% &
& 0\%  & 0\% &
 &  0.1\% & 2.4\%
 \\ 
\hline
%
%
\end{tabular}
}
\caption{Compression and detection accuracy results in six representative instances. V: volume, P: precision, R: recall, V3: tiny-YOLOv3. V5: YOLOv5s}
\label{tab:example}
\vspace{-3mm}
\end{table*}

\subsection{\name Prototyping \& Evaluation Methodology}\label{sect:eval:method}


Because commercial off-the-shelf drones do not expose good programming interfaces to implement \name on-drone, we build a strawman prototype with all the original videos captured from the drone flying in the wild (\circled{1}) and their on-drone processing (\circledblue{2}) implemented at a MacBook Pro(2.8 GHz dual-core). All the codes for edge (\circledblue{4} and \circled{5}) are implemented at our lab server. 
Our objective is to reduce the total transmission volume, not to address how to transmit it. Network transmission (\circled{3} and \circled{6}) is done over the laptop-server connection. We choose a showcase application of vehicle detection from the drones to evaluate \name, as~\cite{benjdira2019car} does. Note that a detected vehicle is eventually classified as a car or a track.

%

\paragraphb{Video dataset.}
Due to lack of publicly available datasets with known extrinsic parameters of the flights(heights, rotation angle, speed), we use two DJI Phantom 4 Pro v2 drones to collect our own traffic surveillance video at 4K/30fps at different times sporadically from Nov 2019 to July 2020 in West Lafayette. A variety of experiment settings are considered as shown in \tab~\ref{tab:dataset}. We want to point out that the actual pitch angle is not 100\% precise within up to 5\textdegree error since the used drone are not equipped with precision control. We manually crop the videos into 10s segments as test instances. Our evaluation is based on replay of a suite of these clips.
%

\paragraphb{CNN models.} 
In this work, we consider two one-stage CNN detectors: YOLOv5 and YOLOv3. In particular, we use the small scale network architecture among all of the candidates, YOLOv5s and tiny-YOLOv3 respectively. 
Note that our result can be extended to the complete model with more layers. 
Since the backbone of YOLO to extract features, \texttt{Darknet} for tiny-YOLOv3, \texttt{CSPNet} for YOLOv5s, has been trained on the images that were captured on the ground(e.g. COCO Dataset) which has poor detection performance for the aerial view. To improve detection accuracy, we make two changes. First, we use transfer learning to fine-tune the pre-trained classifiers on small training sets of images that were captured from the drones. 
Specifically, we manually label a number of frames (images) under several typical scenarios for training. We temporally subsample the selected video frames by a factor of 150 to avoid the same vehicles constantly appear in the consecutive frames. We crop the frames into the subframes that contains objects for training. Each subframe is 512 $\times$512 pixels with 100 pixels overlapping to ensure each vehicle appeared as a complete object.  In total, about 7,100 subframes are randomly selected for training. All the ground truth are manually labelled by us.
Second, we improve the current image-based object detector by aggregating their results over consecutive frames. In particular, we define a detection window $W$ and the object is detected as long as it is detected out in $W$. This is a practical consideration, as for most latency-critical tasks there should have a time window constraint otherwise it is meaningless.  At the same time, the classifier uses a majority voting over all the classified results for the same object in these frames. This is taking the fairness into consideration to avoid the case that detecting many true negative\add{s}. We set $W=5$ in our evaluation. 

\paragraphb{Metrics}.
%
We use three measures of precision, recall and $F_1$ score to assess detection accuracy. 
Precision=$\frac{TP}{TP+FP}$, Recall = $\frac{TP}{TP+FN}$, where TP, FP and FN are the number of true positives, false positives and false negatives.  $F_1$ score, the harmonic mean of precision and recall, $F_1 = \frac{2*\textnormal{Precision} \times \textnormal{Recall}}{\textnormal{Precision}+ \textnormal{Recall}}$. For representative cases, we mainly use precision and recall to better understanding how accuracy is affected. Statistical results only report $F_1$ score.

To better understand the contributions of \name's components, we compare performance with two \name variants: \name (a full version) and \name-R (which employs adaptive resolution only). 
\name is compared with the \texttt{BASELINE} strategy, which just temporally downsamples the video to a lower frame rate (here, 5 fps), the same used by \name and then directly encodes via \texttt{H.264}.
We quantify \texttt{METHOD}'s compression gain (or accuracy loss) as $\gamma = \frac{[\texttt{BASELINE}] - [\texttt{METHOD}]}{[\texttt{METHOD}]}$. 
We also implement CMU's EarlyDiscard (\texttt{ED}) for comparison. \texttt{ED} is a typically image-based mechanism and proposed for CCTV surveillance. Specifically, it first tiles the frame into several subframes and uses a weak detector to determine whether the subframes contain target objects. Only subframes of interest are transmitted accordingly.

\subsection{Representative Instances}\label{sect:eval:example}

We first use six instances (\tab~\ref{tab:example}) to demonstrate benefits of our \name solution in representative scenarios where 
\vspace{-1mm}
\begin{itemize}
\itemsep 0pt
\item[(a)] the drone flies at 5~m/s, at height of 50~m above one residential  area with pitch $\approx$ 90\textdegree,
\item[(b)] the drone flies at 5~m/s, at height of 70~m above one local road with pitch $\approx$ 90\textdegree, 
\item[(c)] the drone hovers (at 0~m/s) at height of 50~m above one highway with pitch $\approx$ 90\textdegree,
\item[(d)] the flies at 5~m/s, at height of 100~m above one highway with pitch $\approx$ 90\textdegree, 
\item[(e)] the drone hovers (at 0~m/s) at height of 100~m above one highway with pitch $\approx$ 90\textdegree, and
\item[(f)] the drone hovers (at 0~m/s) at height of 100~m above one highway with pitch $\approx$ 30\textdegree.
\end{itemize}

\noindent These representative instances cover typical ground areas (residence, local and highway), speeds, altitudes and pitch angles. 

\tab~\ref{tab:example} shows compression and accuracy results,
along with one illustrative frame per instance.
Histogram of transmission volume is normalized by its \texttt{BASELINE} one.
We would like to highlight that some instances are selected to demonstrate the ``{\em lower-bound}'' of \name. 
We observe the lowest compression gain for \name (19\%) in (e) compared to the \texttt{ED} out of all experiments. Instance (b) has less satisfactory accuracy (in terms of both precision and recall). We have four main observations. 

First, the compression gains for \name reach 526\%--1203\%, greatly relieving network delivery pressure.  Moreover, \name outperforms \texttt{ED} in most of the case, and achieve a comparable performance in the case with the sporadic traffic(\ie case(e)), containing only three vehicles in a 10s video clip.  Clearly, \texttt{ED} can do a really good job especially when there is no car, like at night. However, drone flies on demands, which limits the benefits of \texttt{ED}. Our statistical results validate this and we will have a compete comparison in \S~\ref{sect:eval_sta}.
\begin{figure*}[t]
\centering
\includegraphics[width=0.195\textwidth]{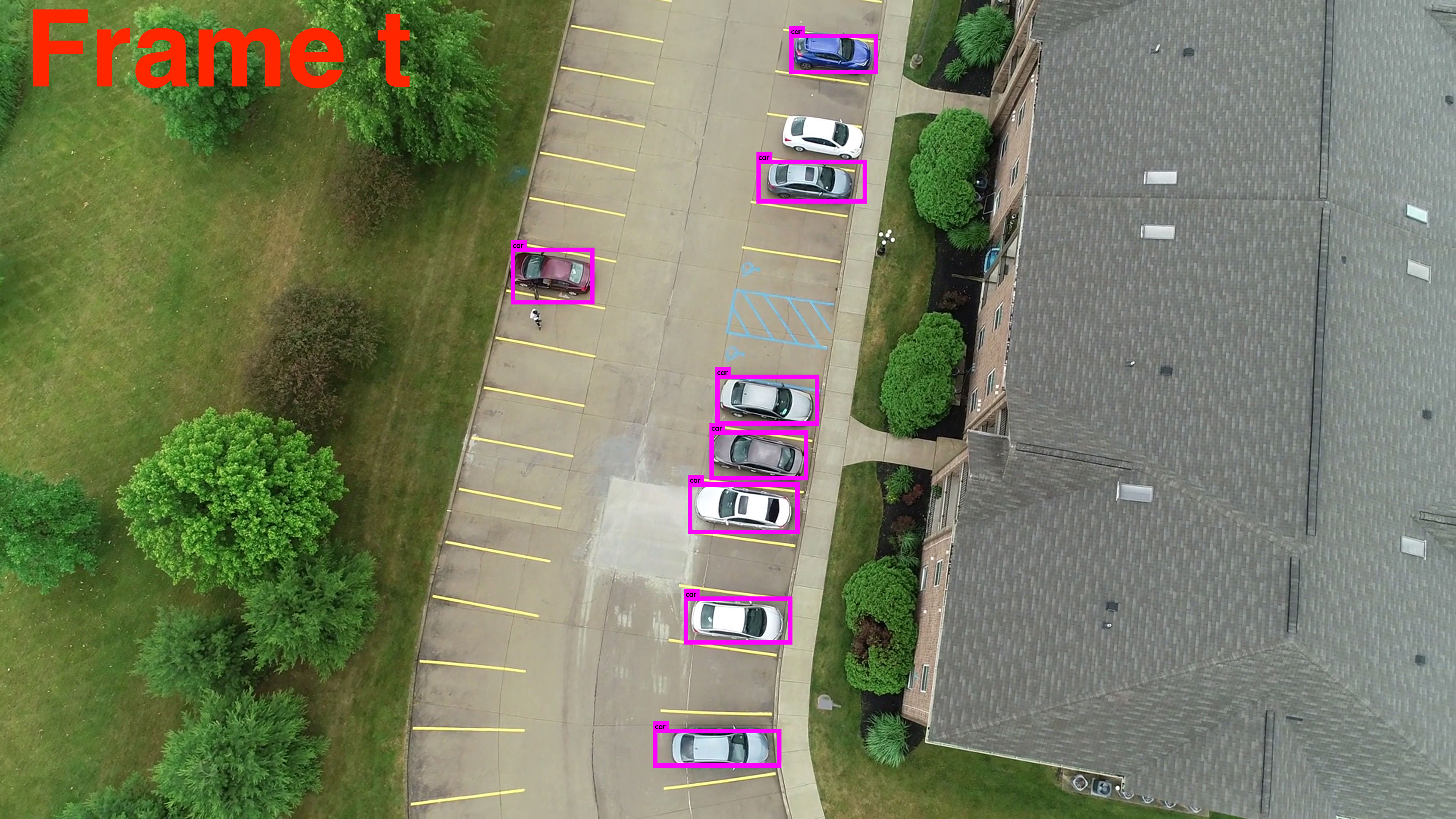}
\includegraphics[width=0.195\textwidth]{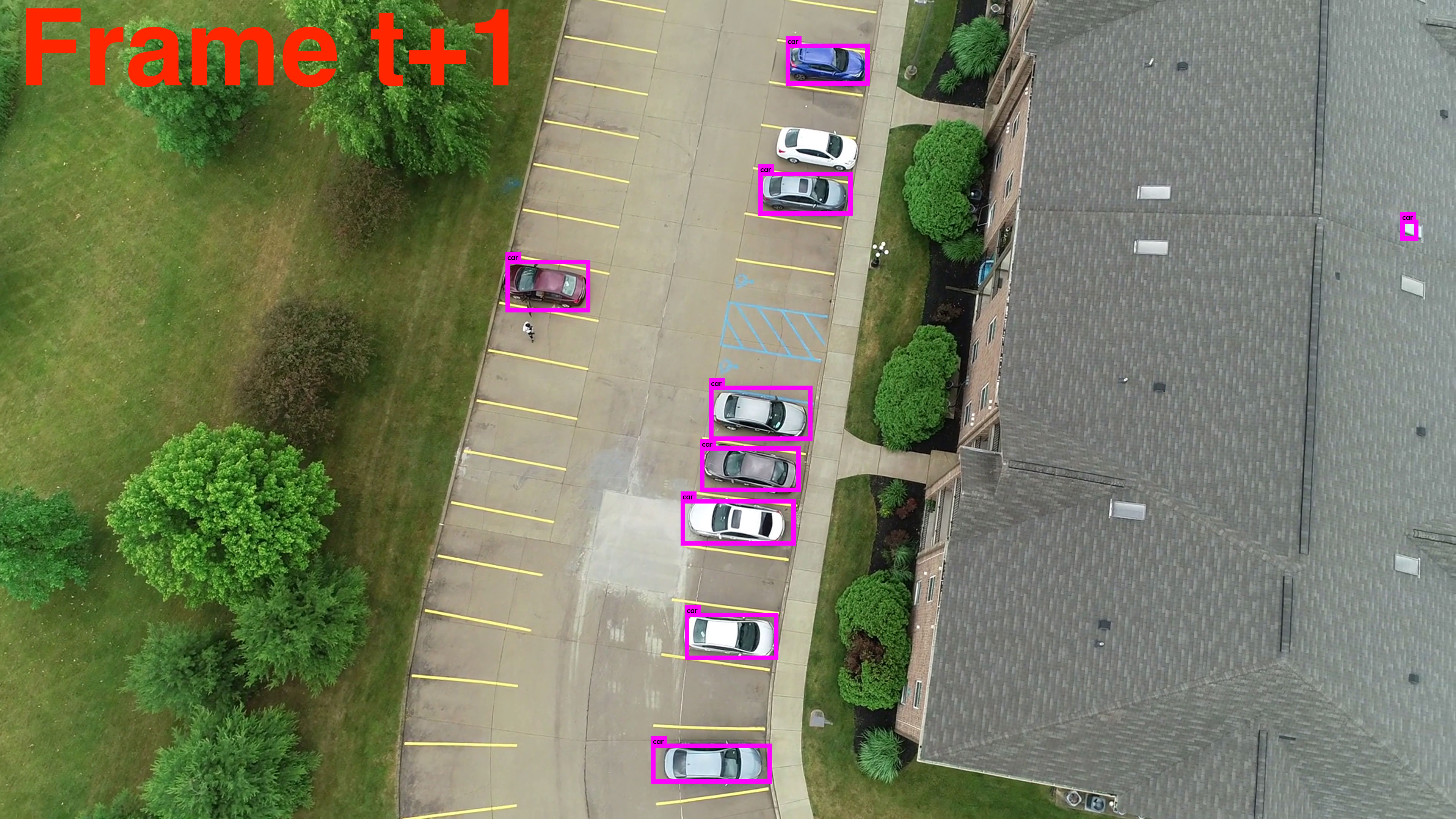}
\includegraphics[width=0.195\textwidth]{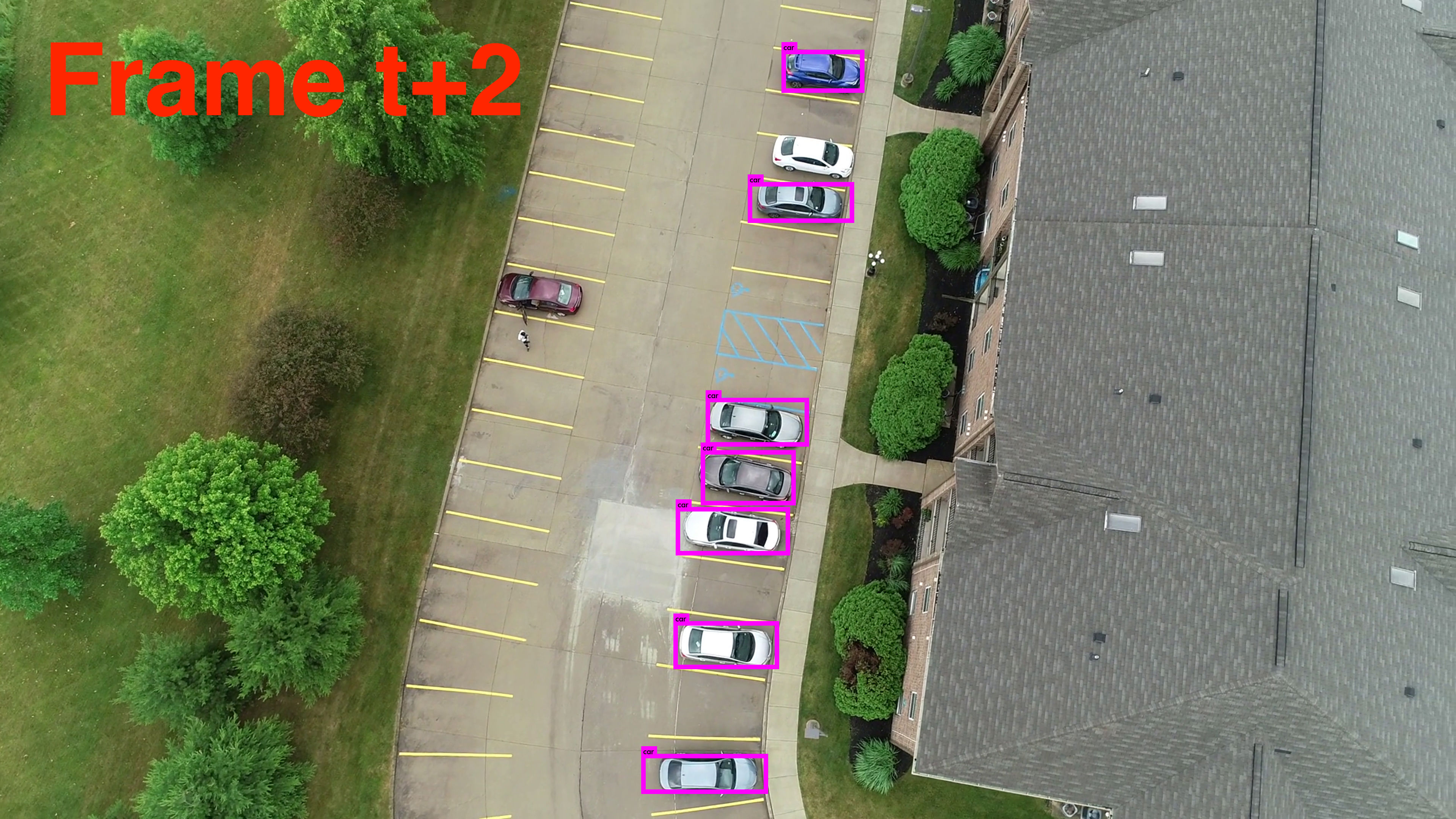}
\includegraphics[width=0.195\textwidth]{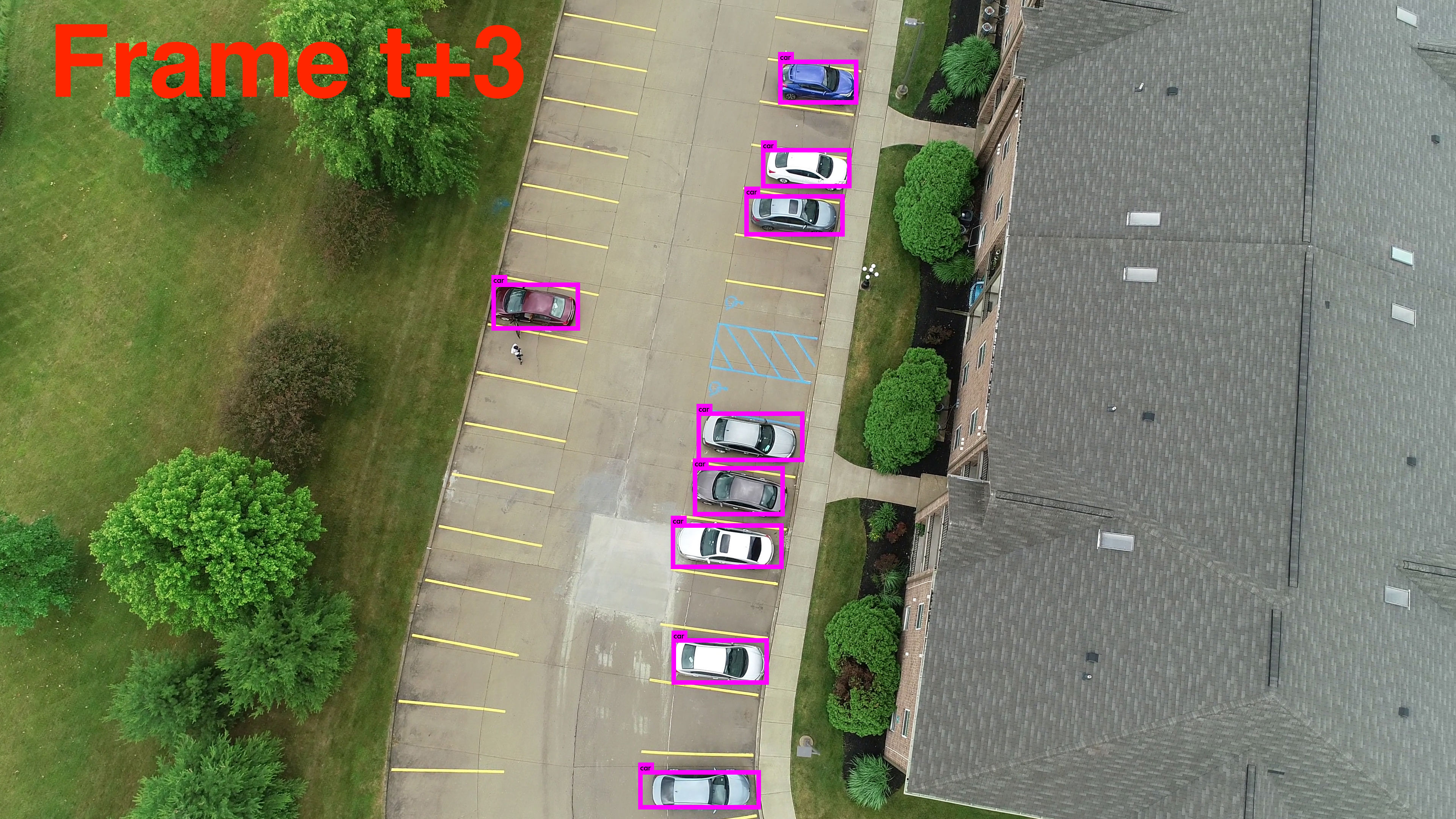}
\includegraphics[width=0.195\textwidth]{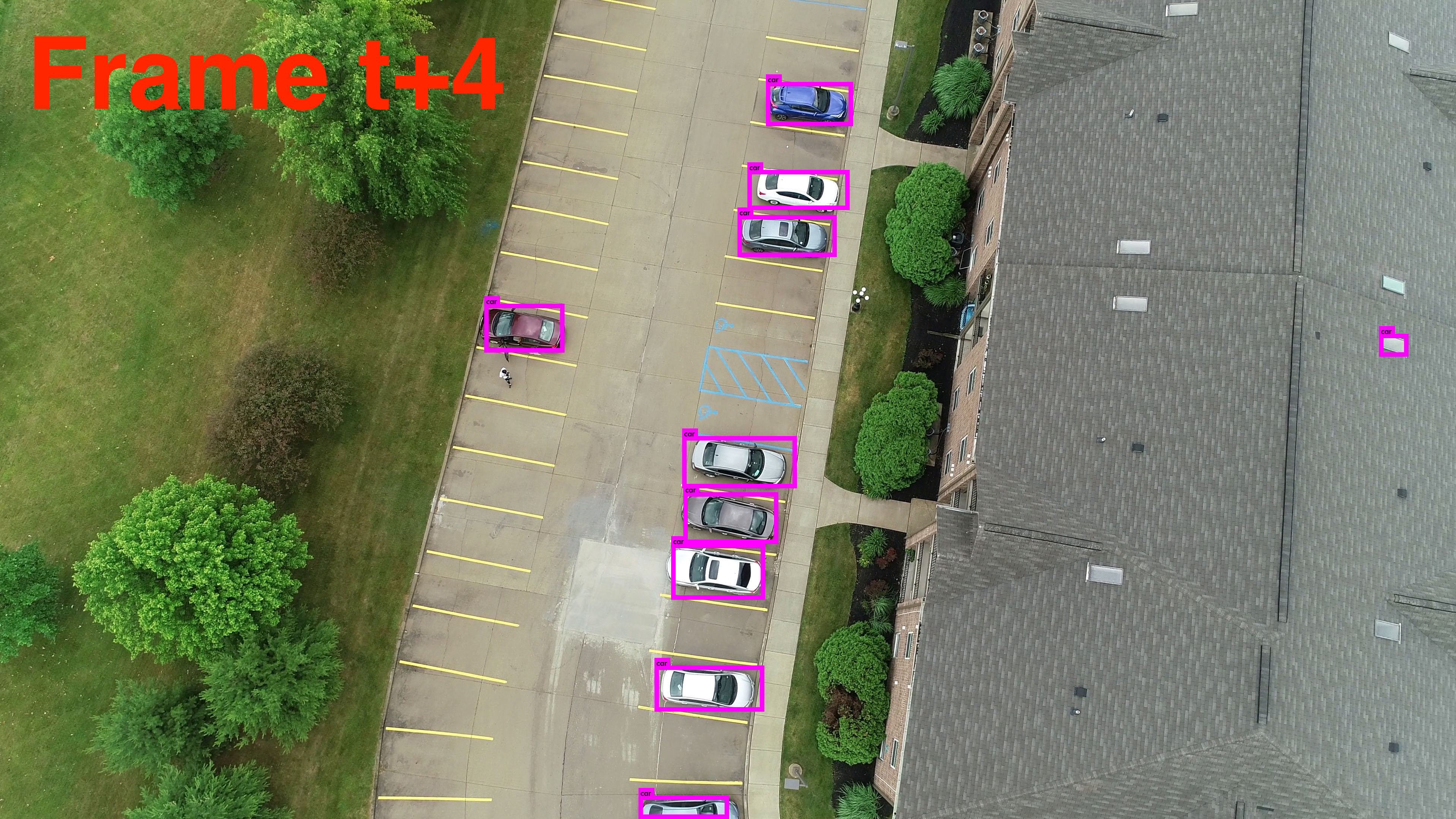}
\\
\includegraphics[width=0.195\textwidth]{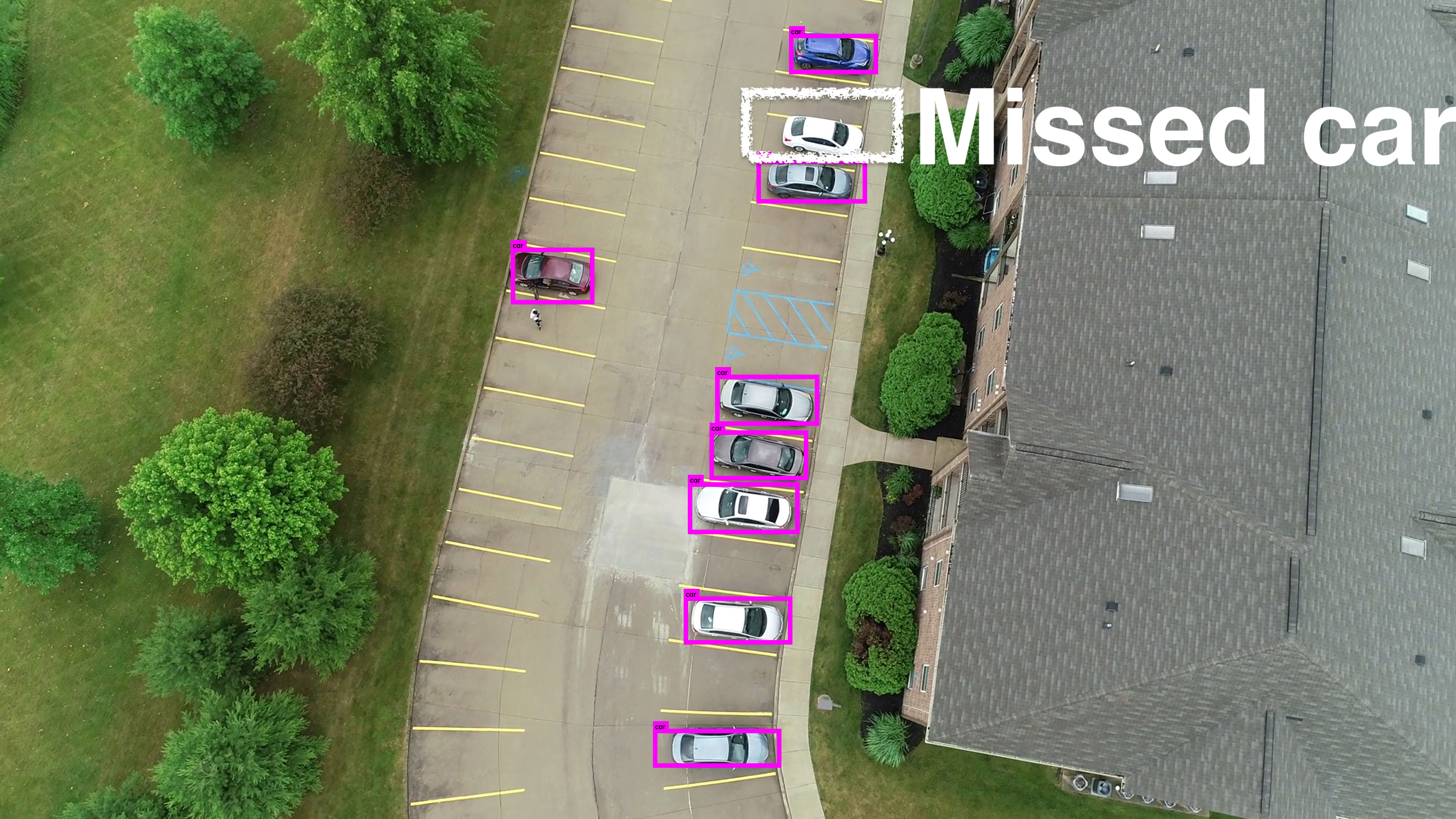}
\includegraphics[width=0.195\textwidth]{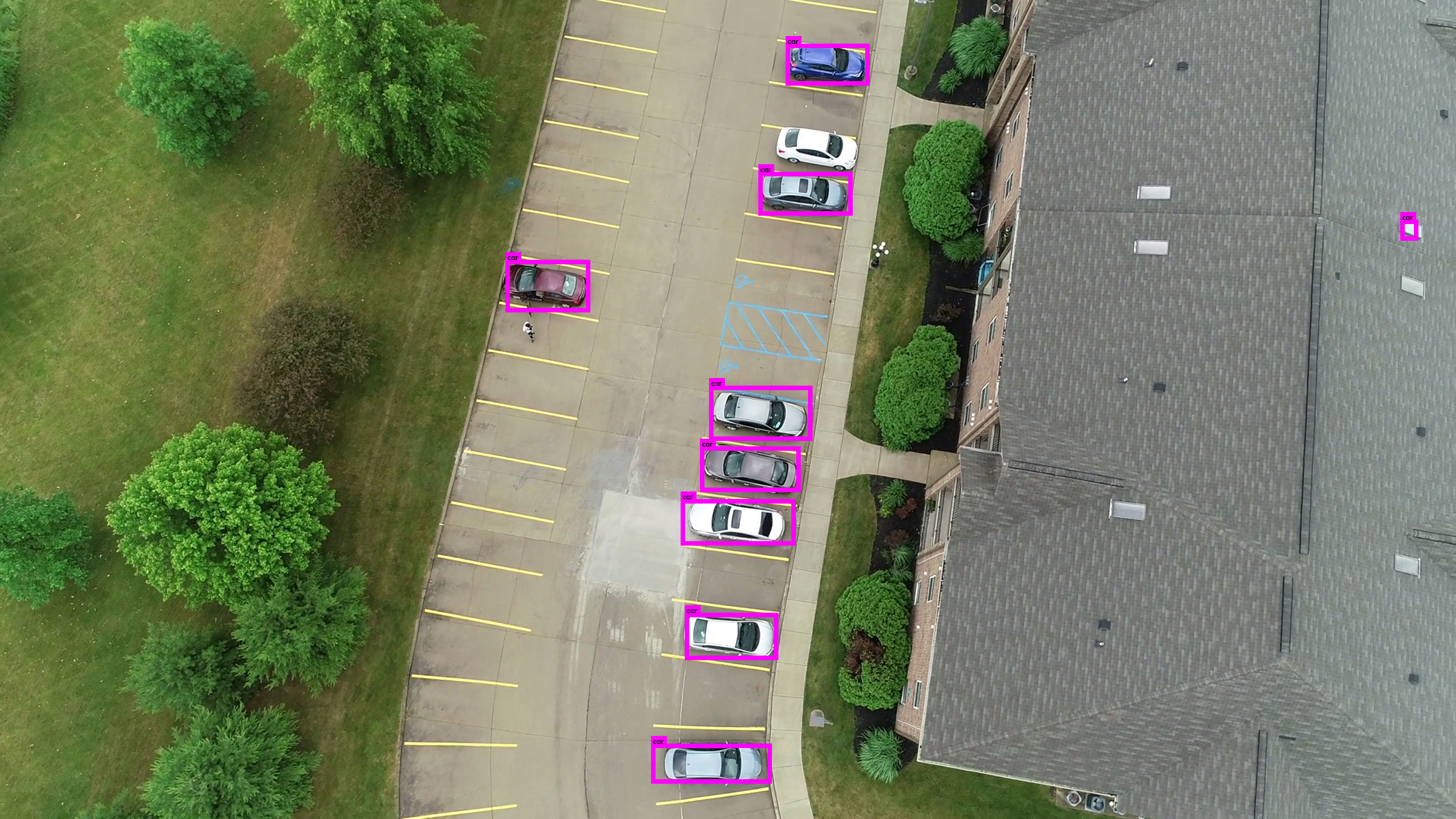}
\includegraphics[width=0.195\textwidth]{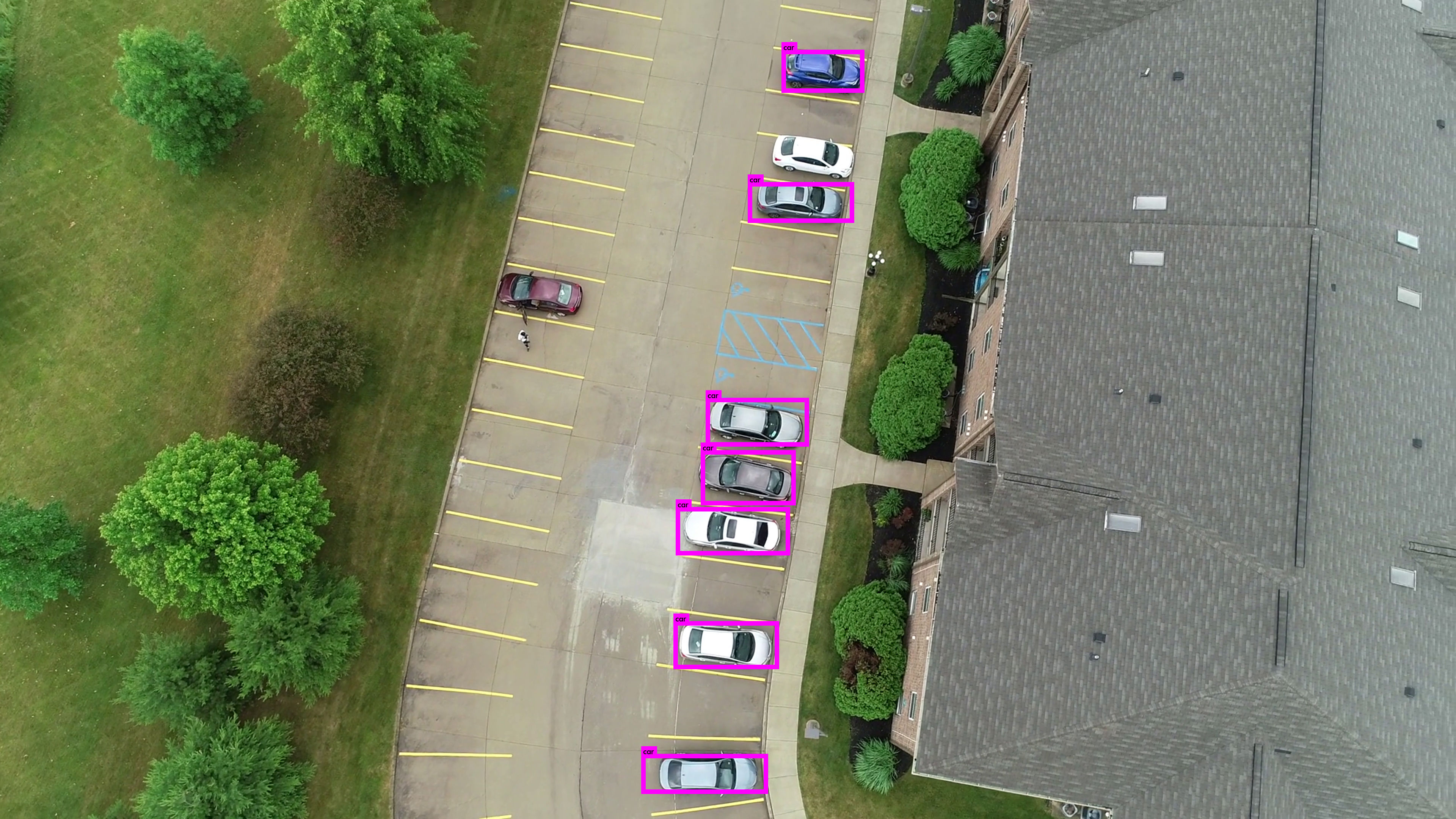}
\includegraphics[width=0.195\textwidth]{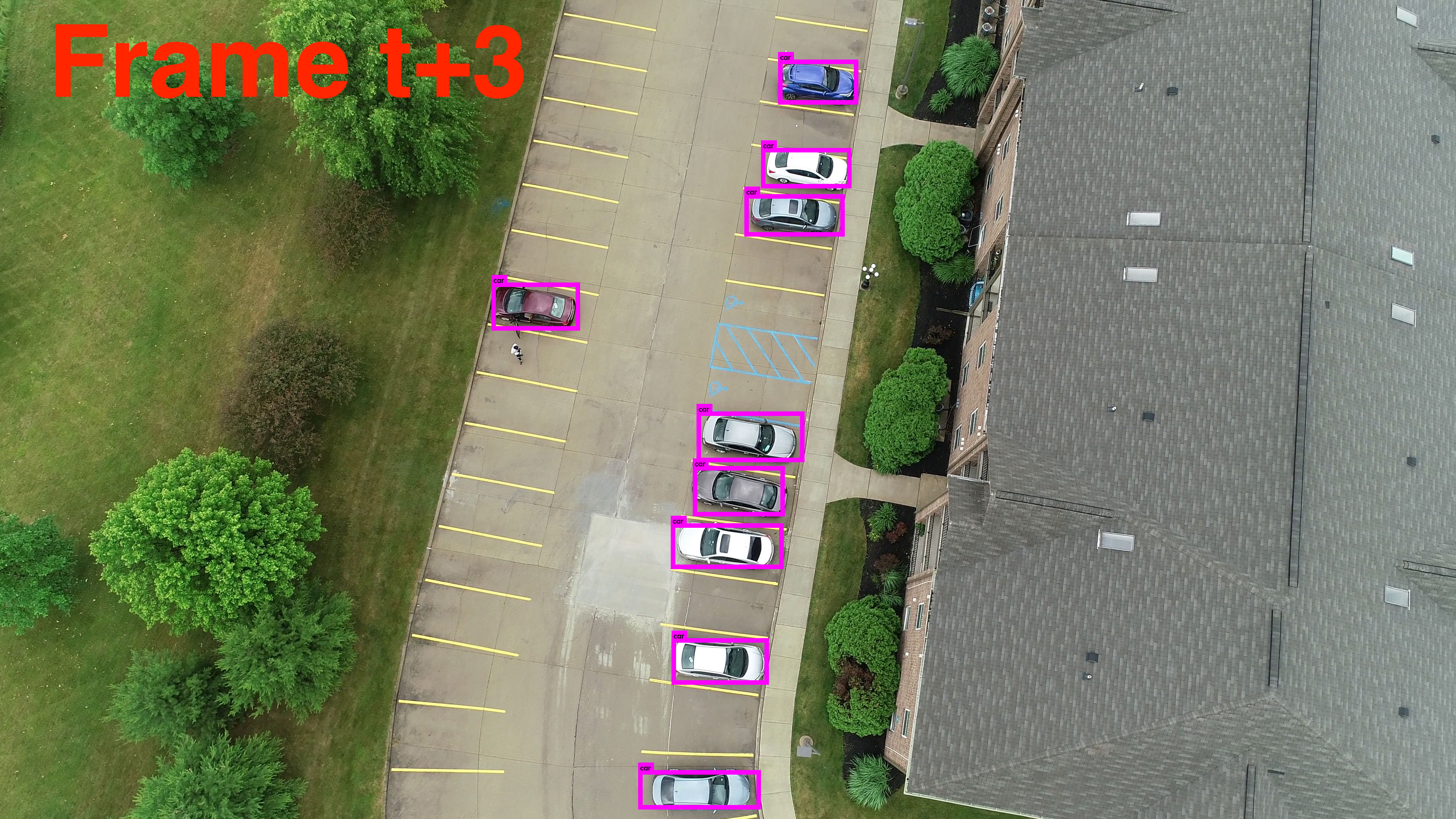}
\includegraphics[width=0.195\textwidth]{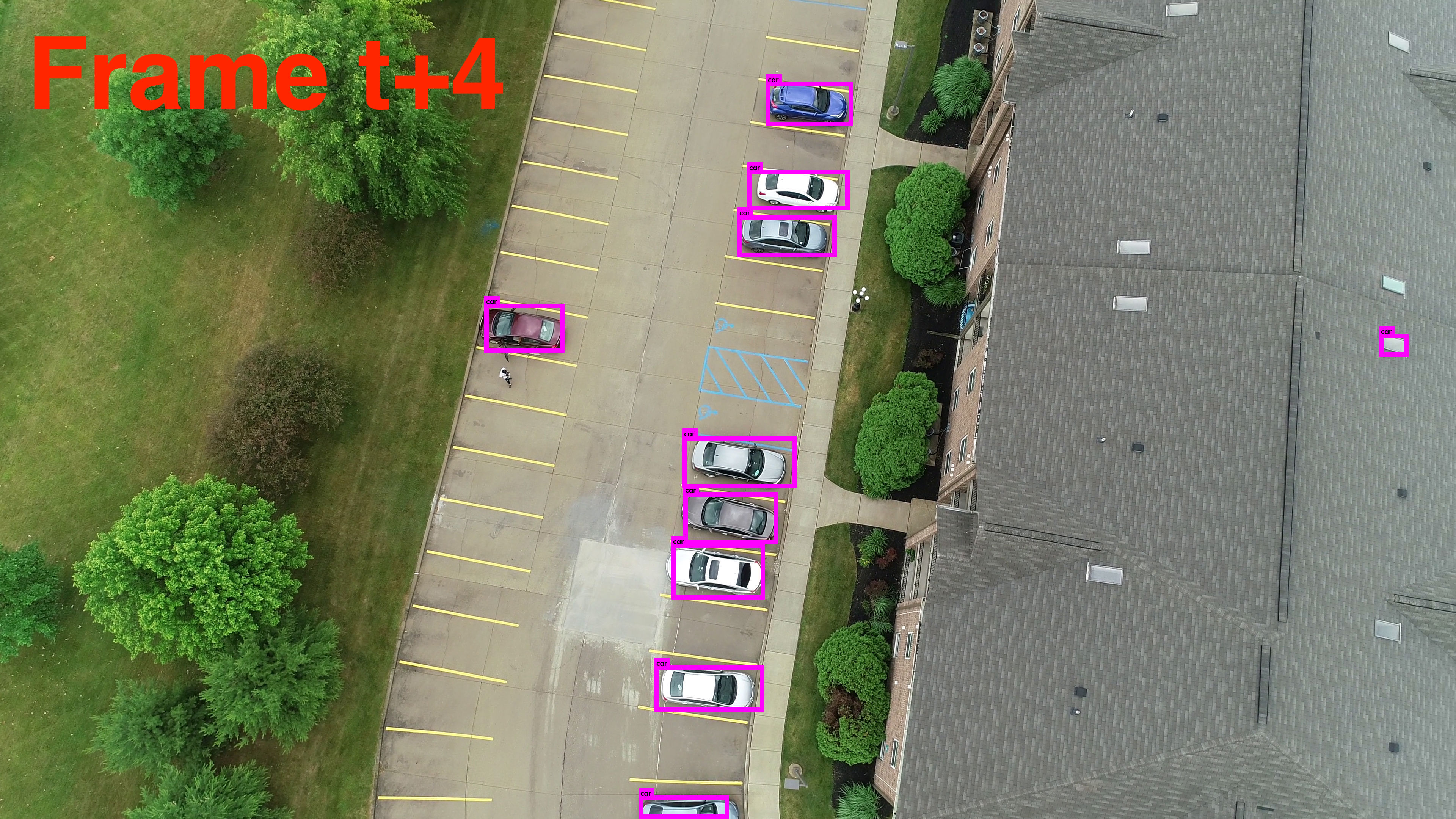}
\caption{ tiny-YOLOv3 intermediate detection results for consecutive frame in case (a): top (original) and bottom (\name). 
In this example, 
\name misses one car (the second) as labeled by white boundingbox.
}
\label{fig:ex-accuracy-a-yolov3}
\vspace{-3mm}
\end{figure*}
\begin{figure*}[t]
\centering
\includegraphics[width=0.195\textwidth]{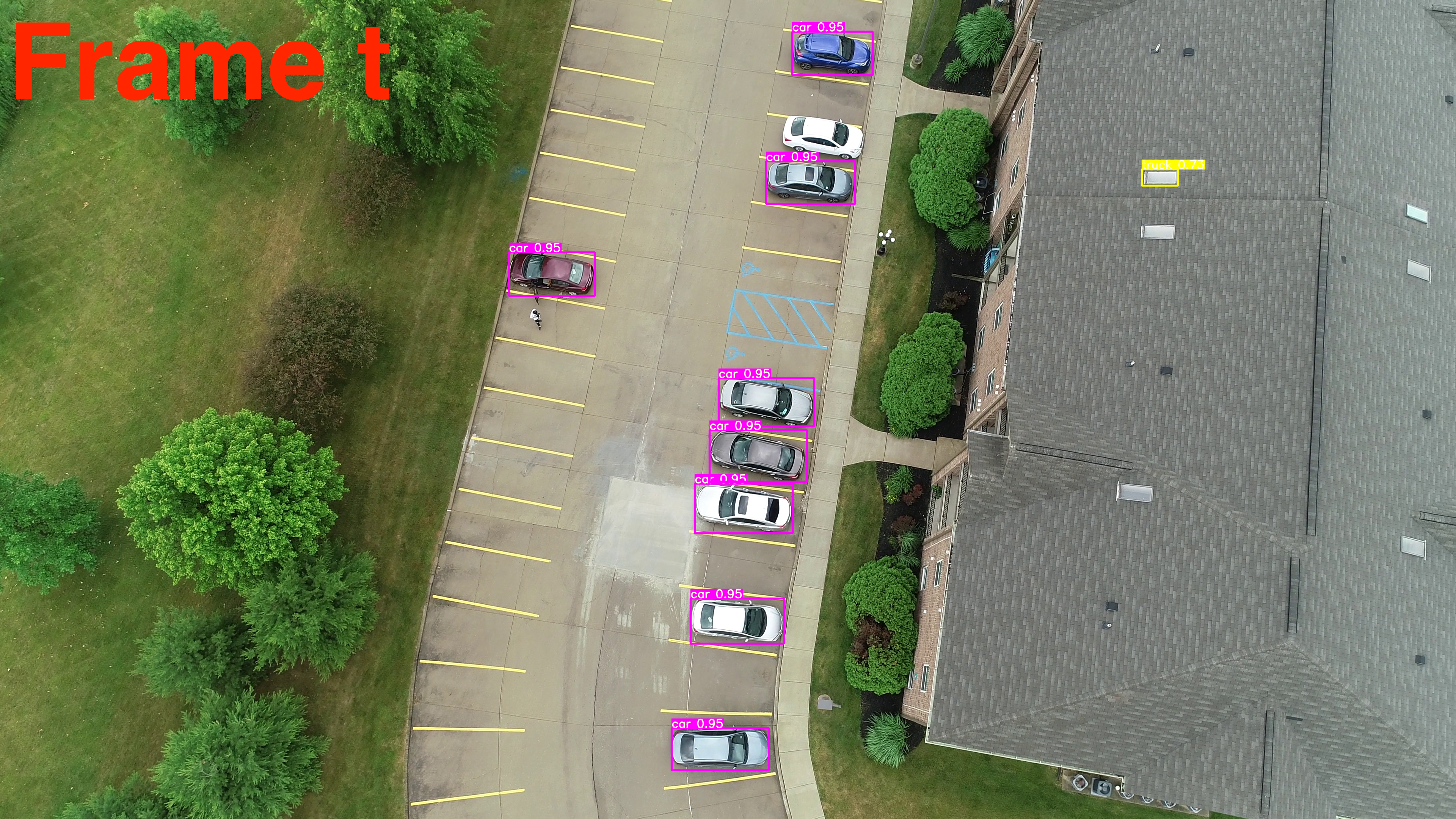}
\includegraphics[width=0.195\textwidth]{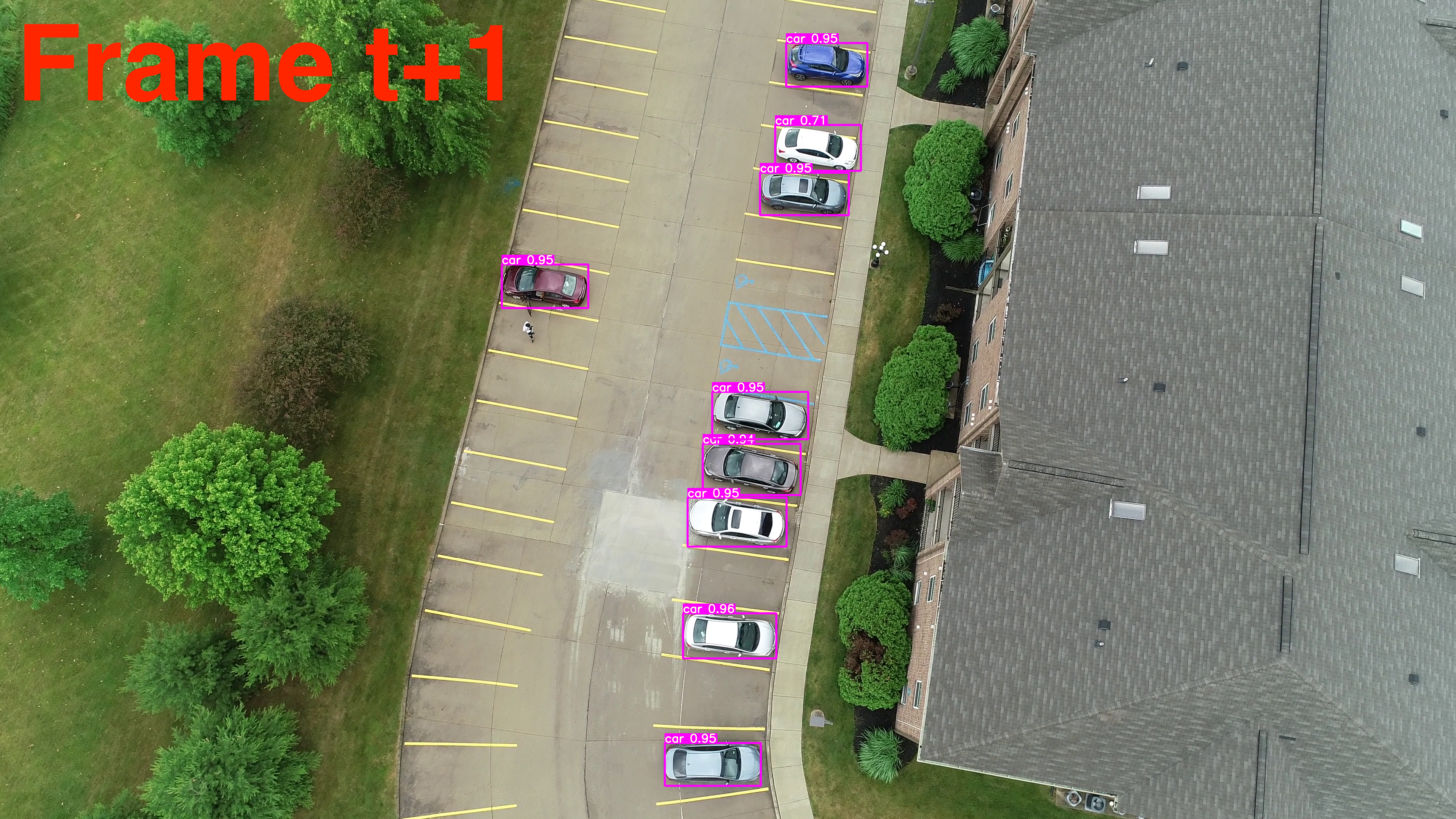}
\includegraphics[width=0.195\textwidth]{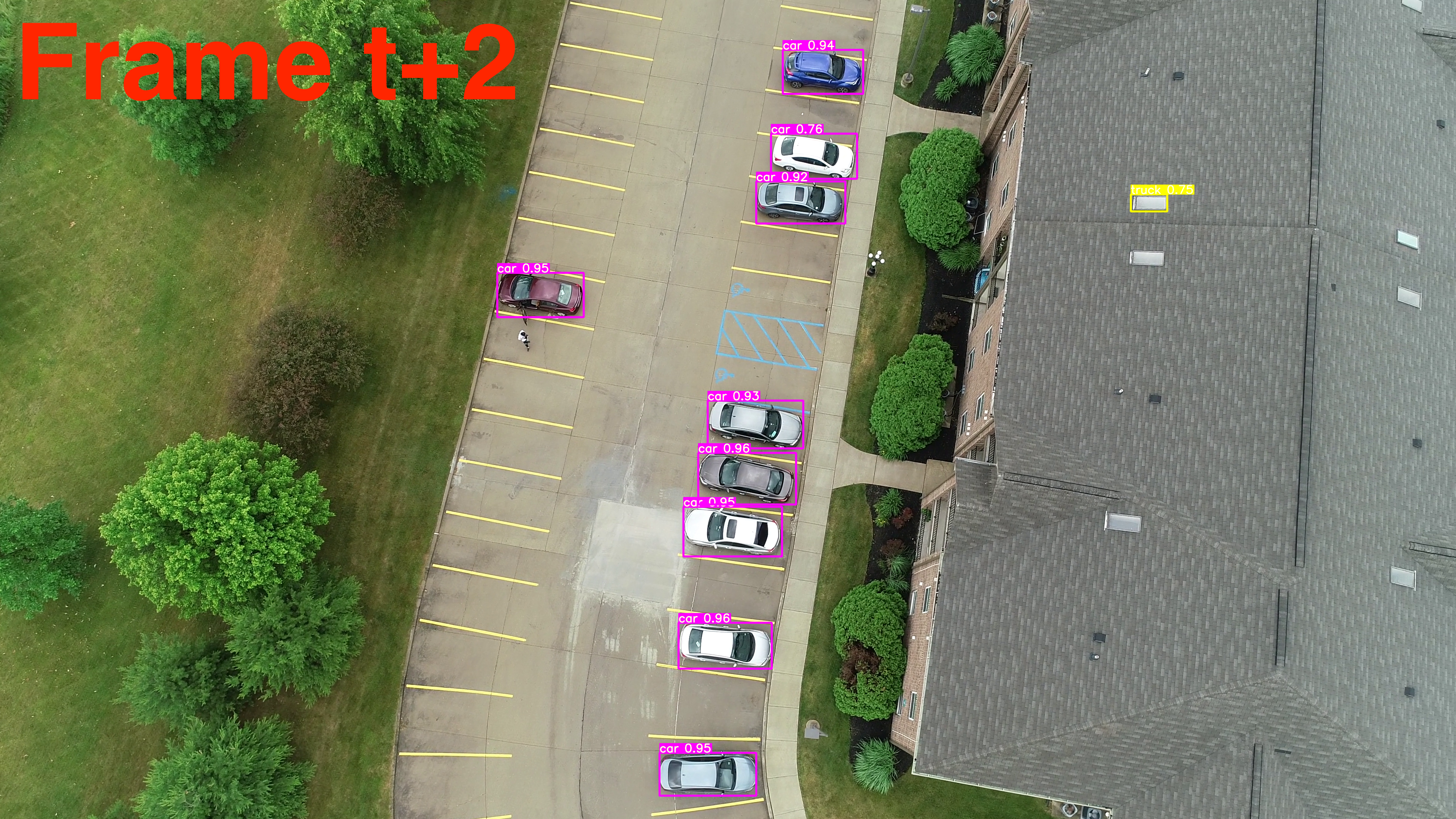}
\includegraphics[width=0.195\textwidth]{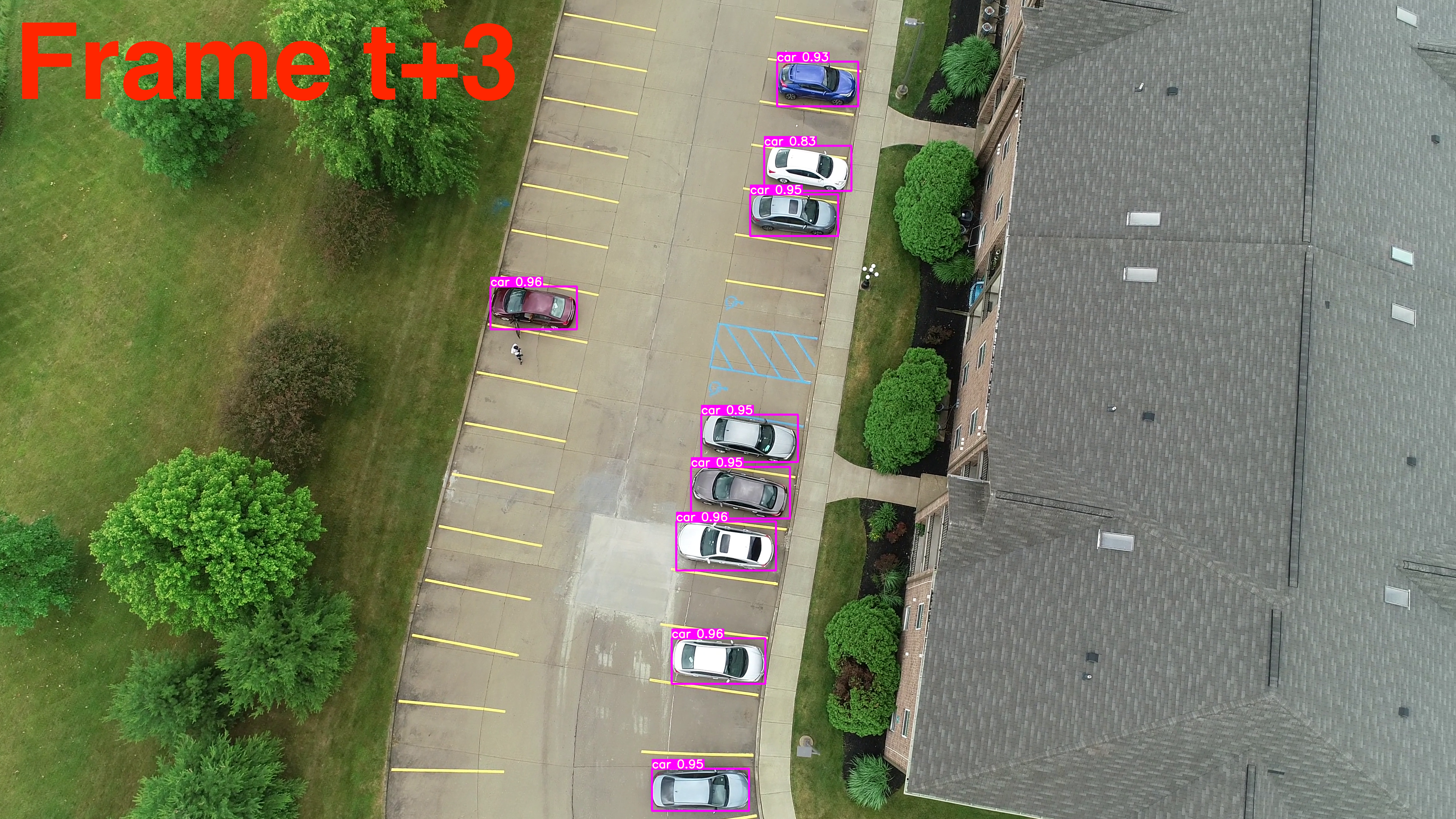}
\includegraphics[width=0.195\textwidth]{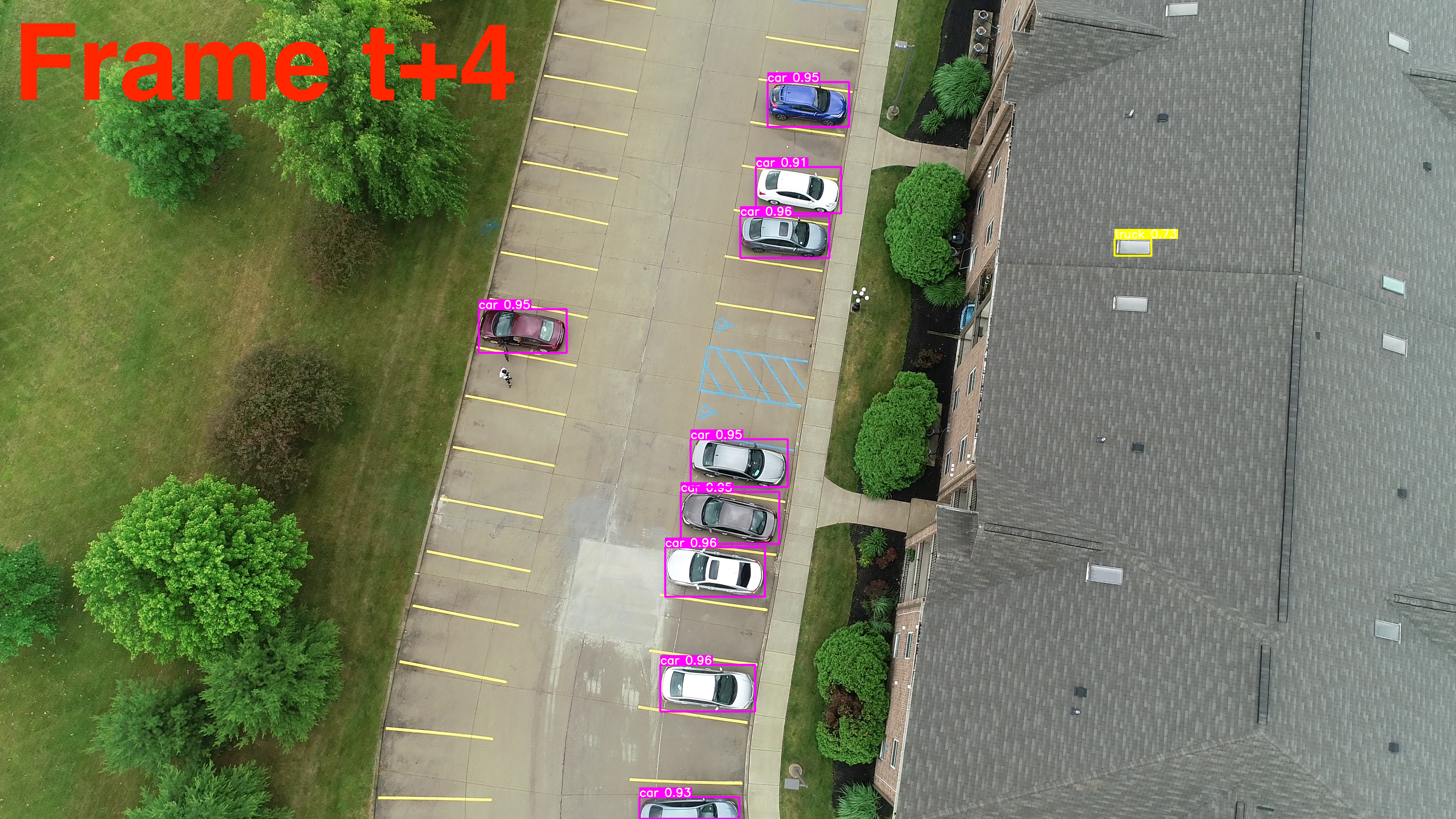}
\\
\includegraphics[width=0.195\textwidth]{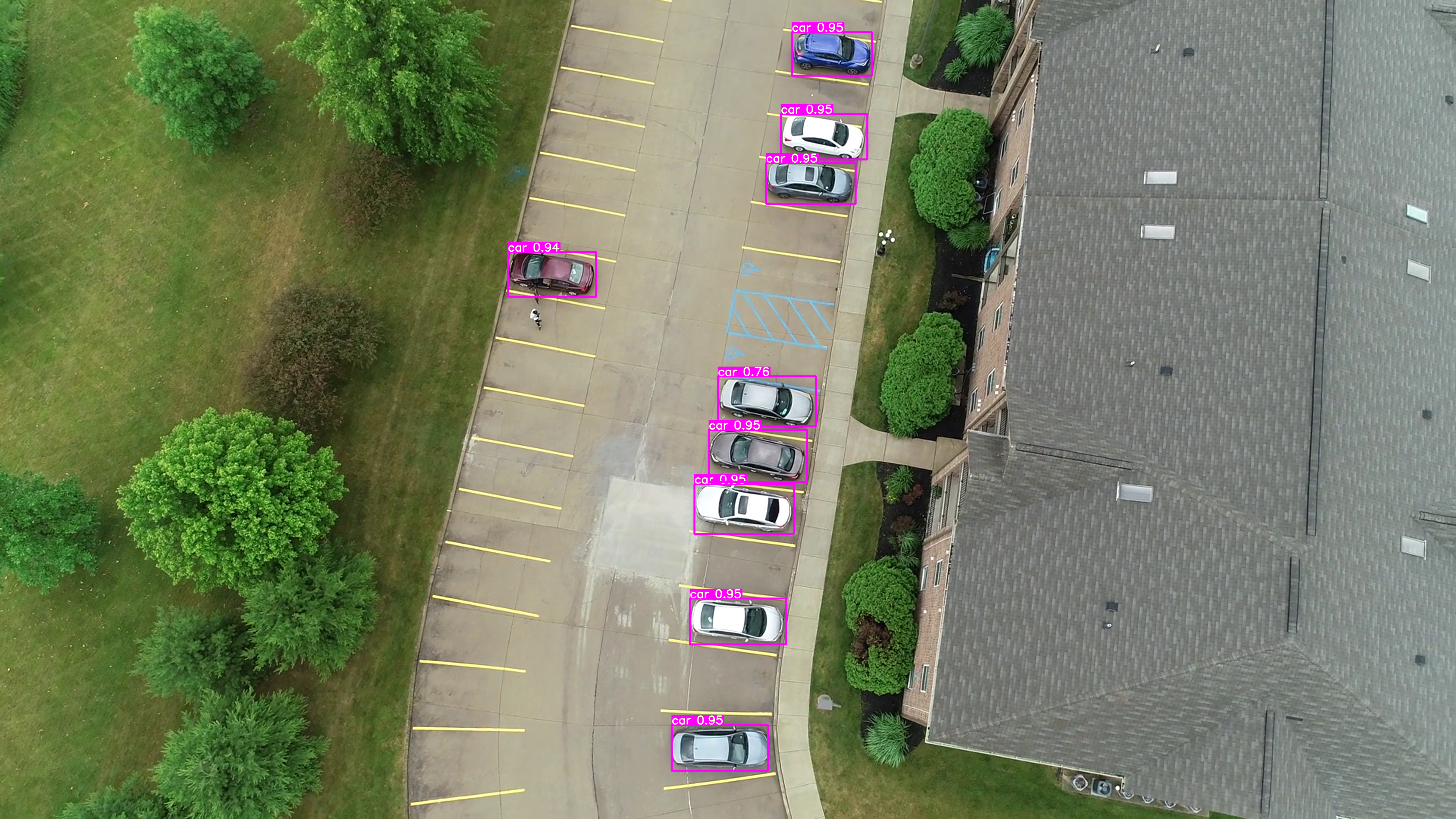}
\includegraphics[width=0.195\textwidth]{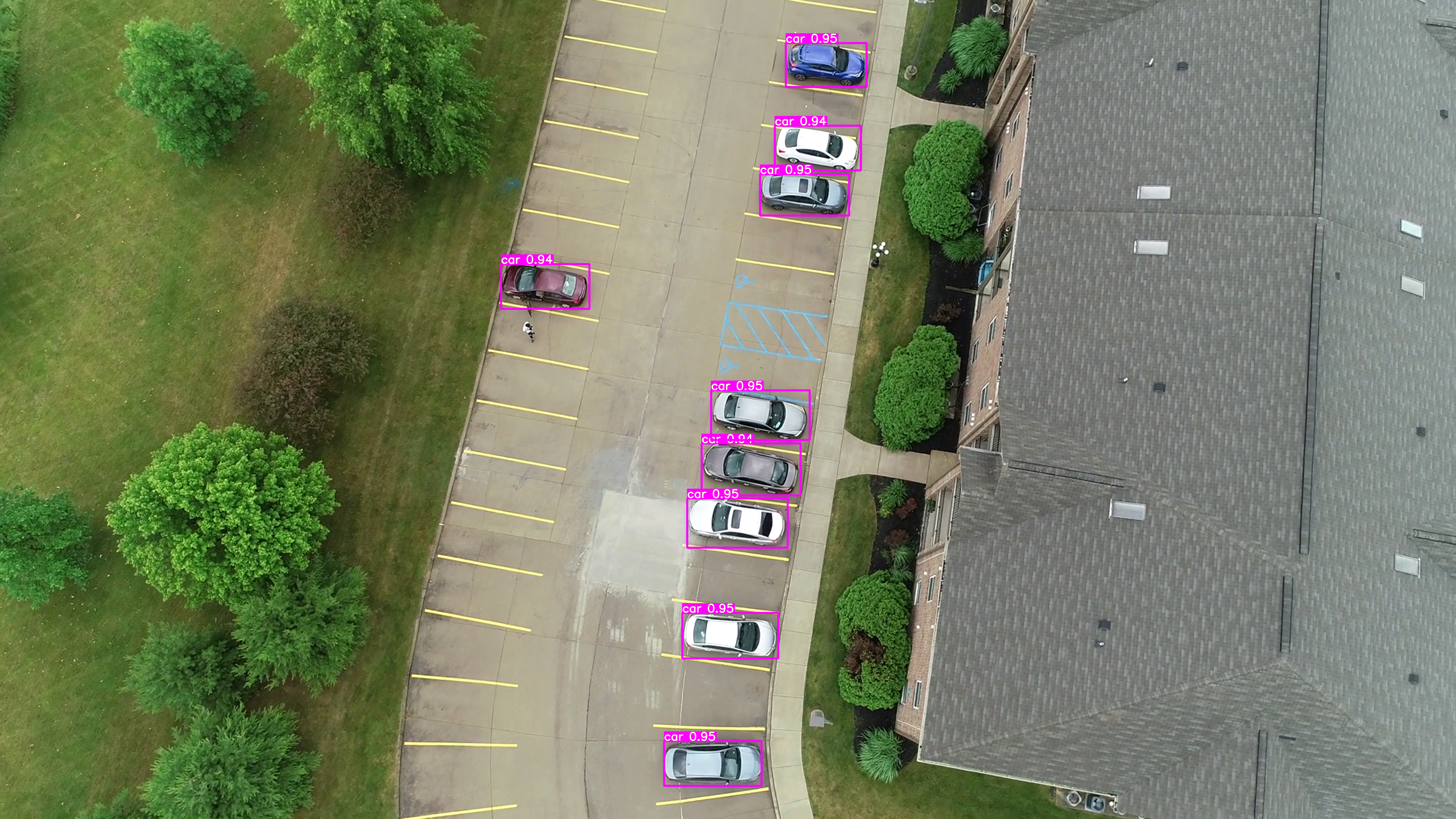}
\includegraphics[width=0.195\textwidth]{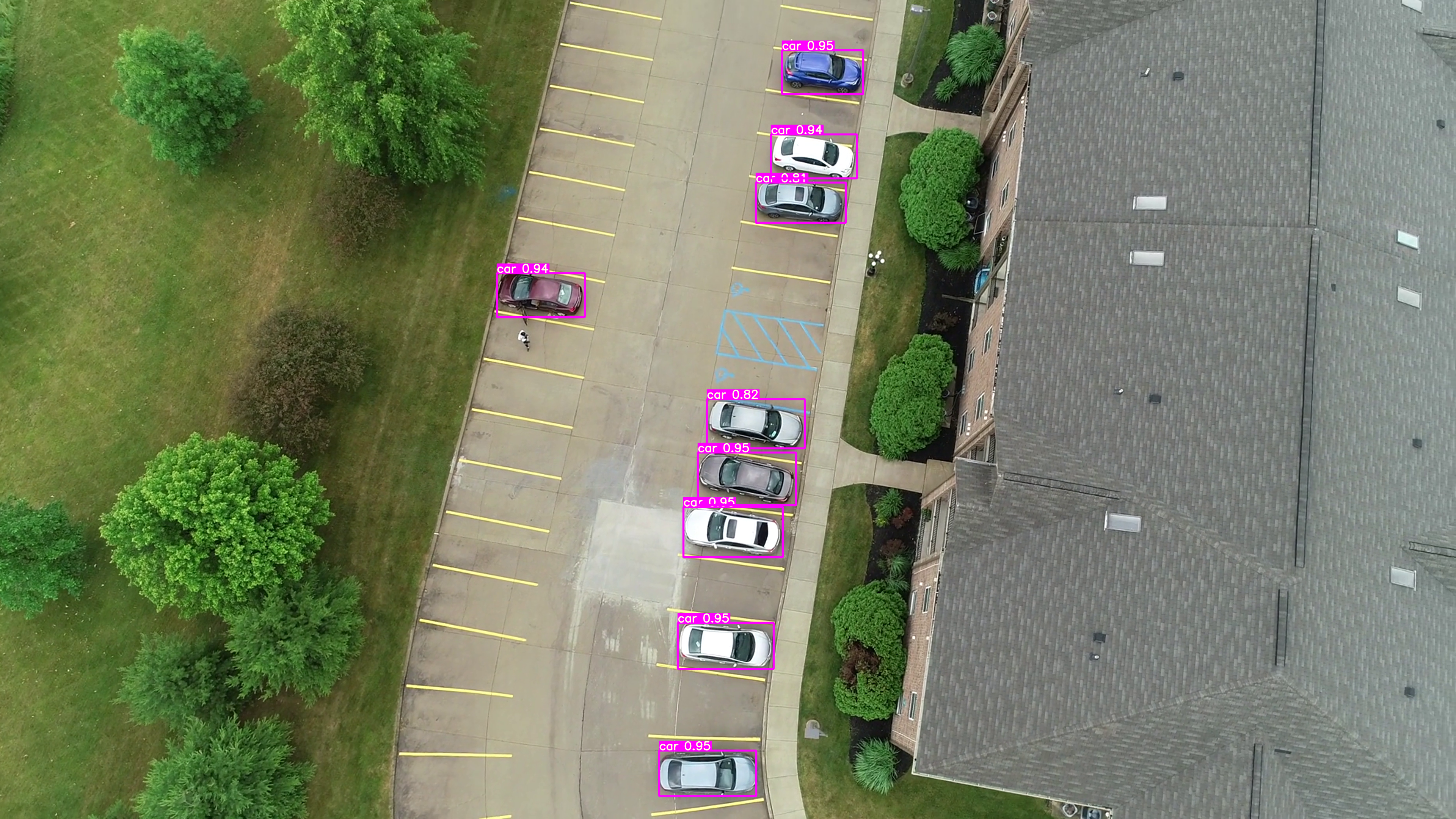}
\includegraphics[width=0.195\textwidth]{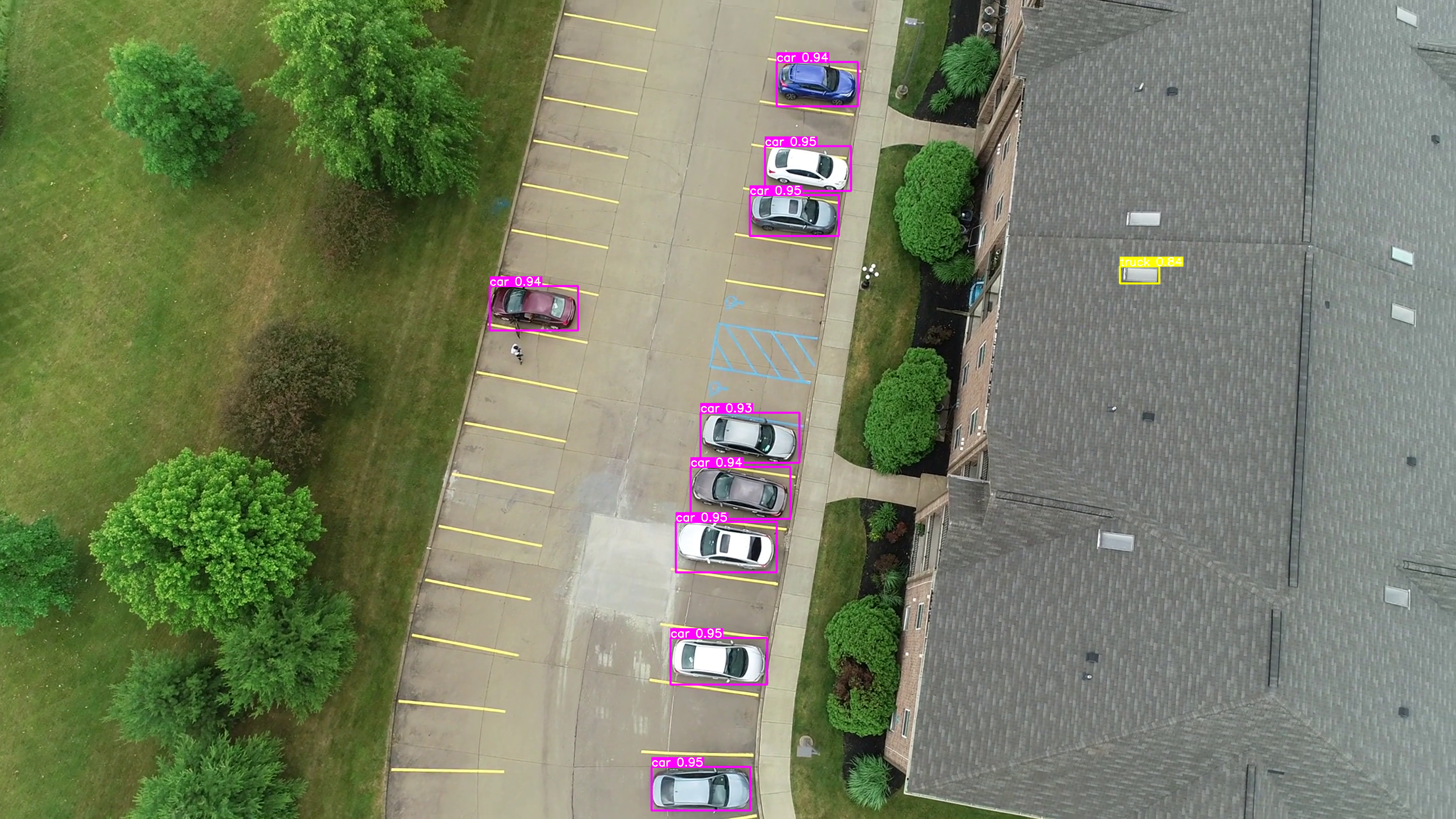}
\includegraphics[width=0.195\textwidth]{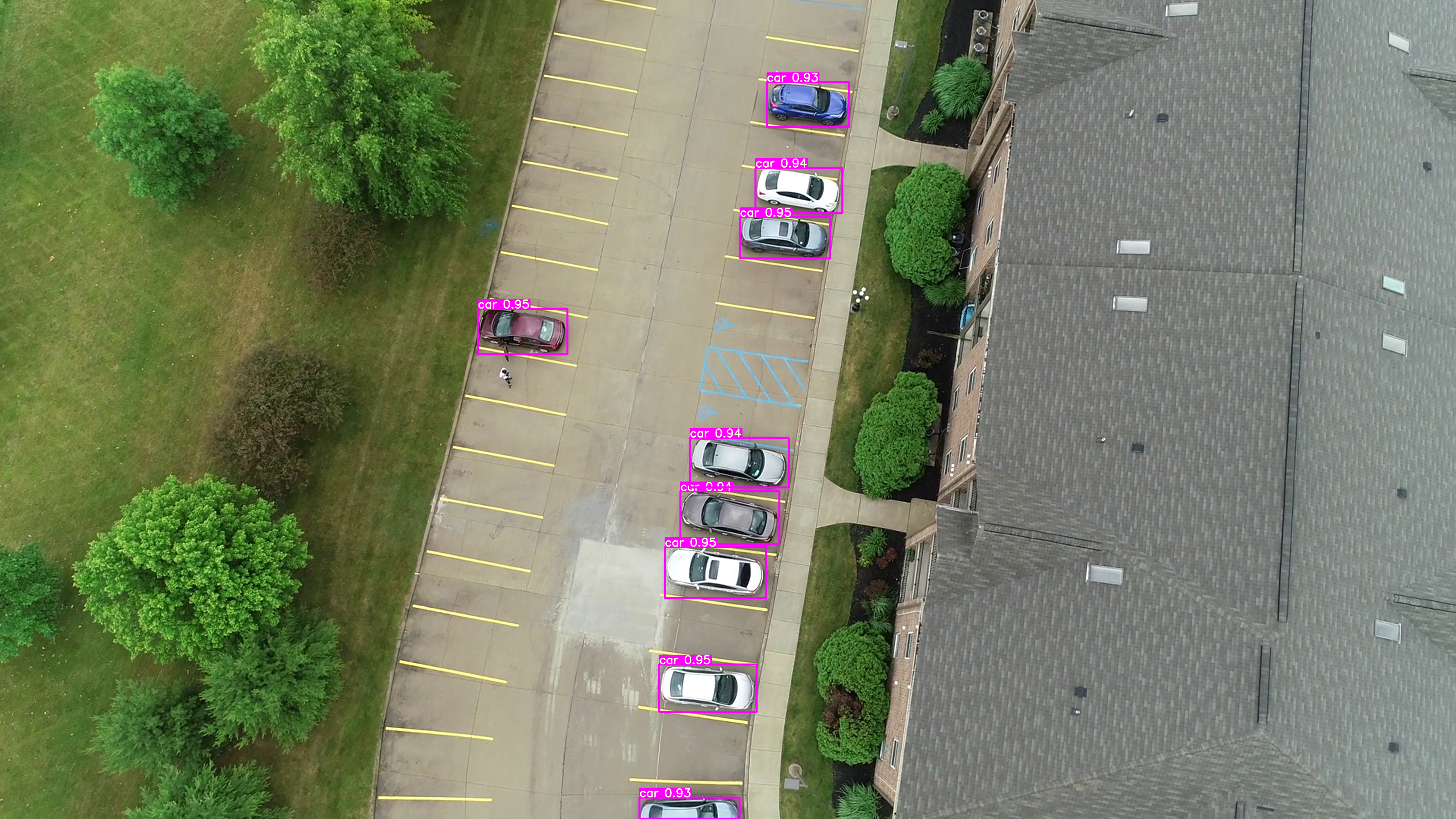}
\caption{YOLOv5s intermediate detection results for consecutive frames in case (a): top (original) and bottom (\name). 
In this example, all cars are successfully detected, although with some true negatives(regard roof as a vehicle).
}
\label{fig:ex-accuracy-a-yolov5}
\vspace{-3mm}
\end{figure*}
\begin{figure*}[t]
\centering

\includegraphics[width=0.195\textwidth]{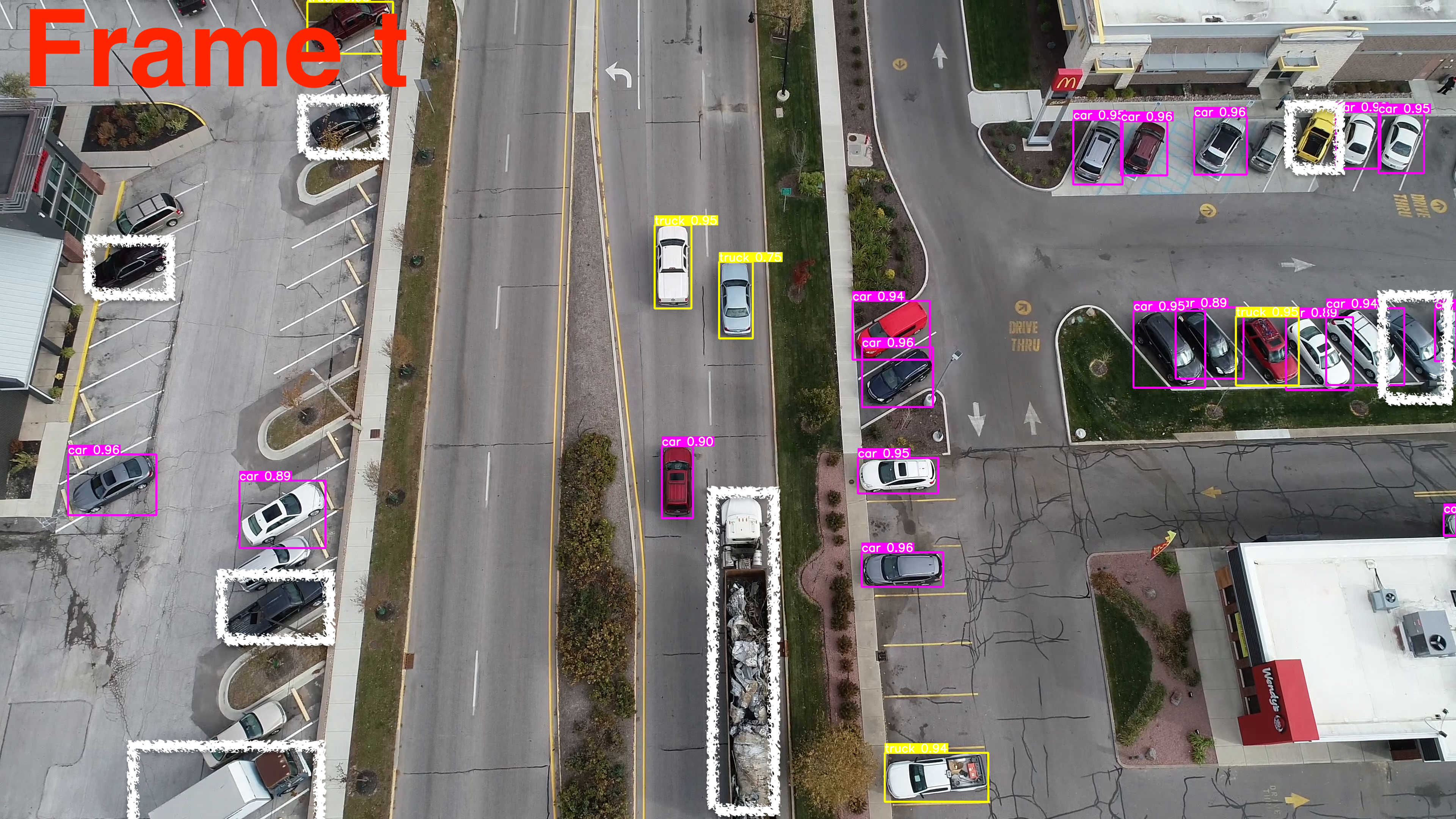}
\includegraphics[width=0.195\textwidth]{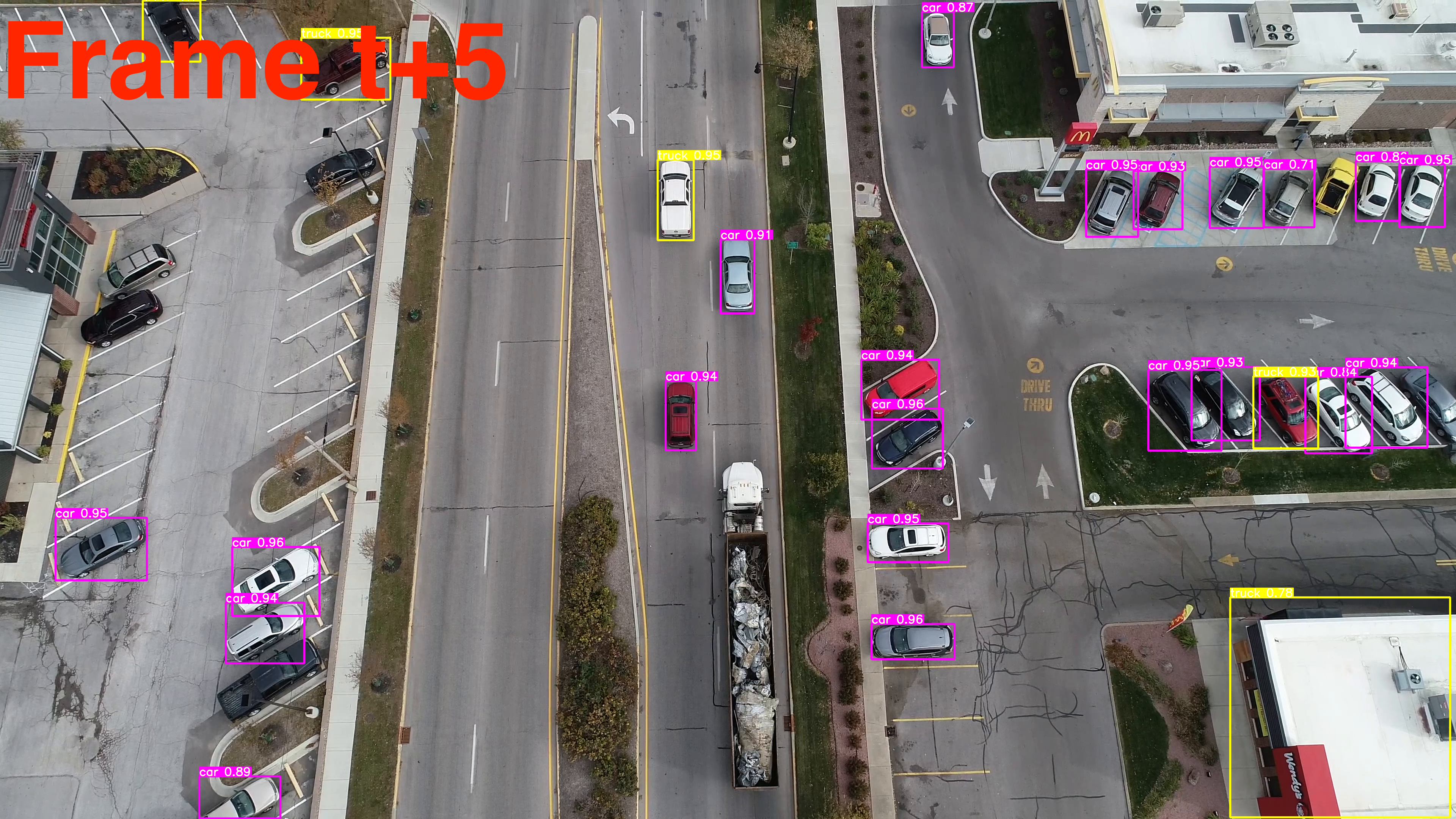}
\includegraphics[width=0.195\textwidth]{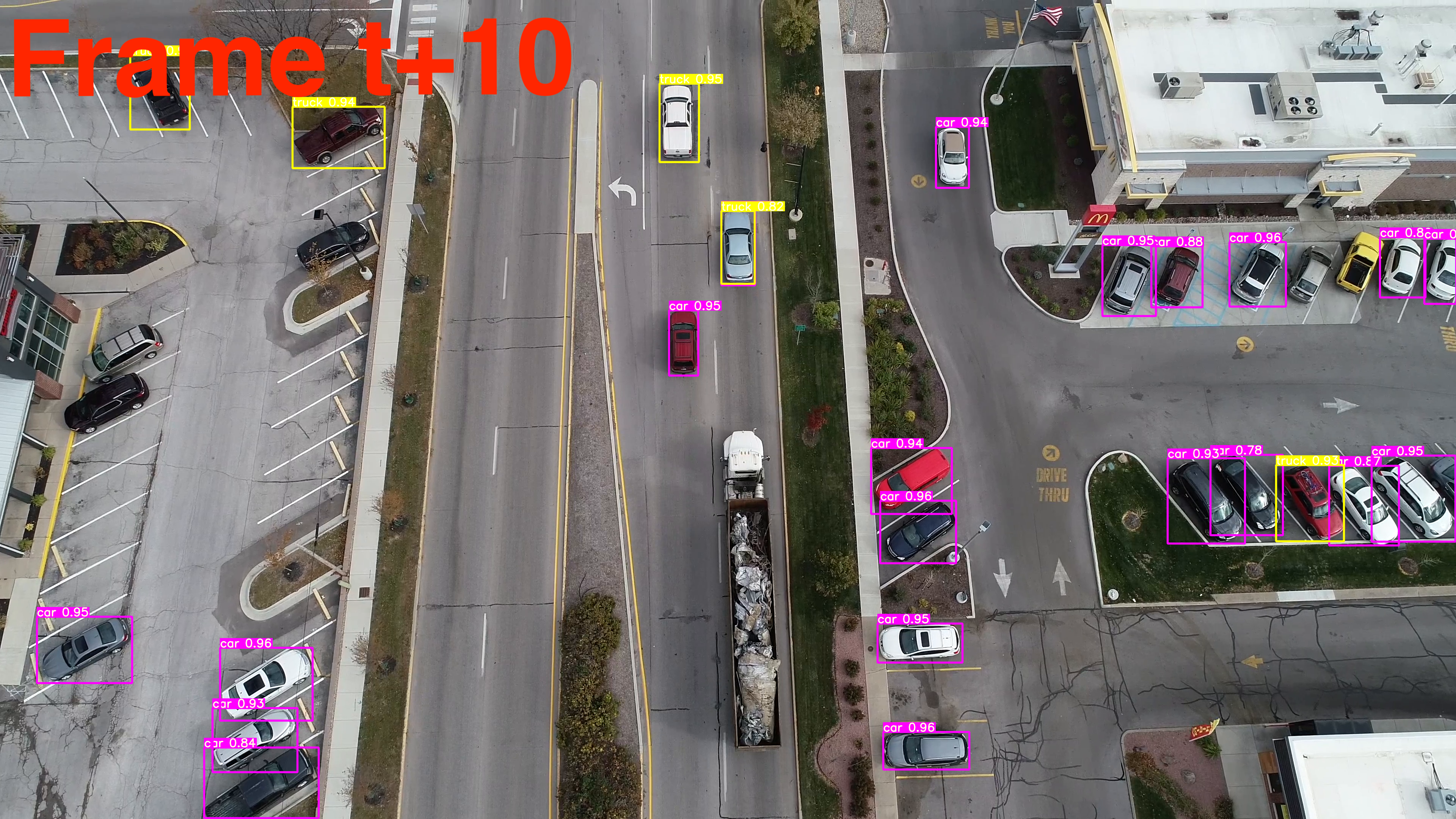}
\includegraphics[width=0.195\textwidth]{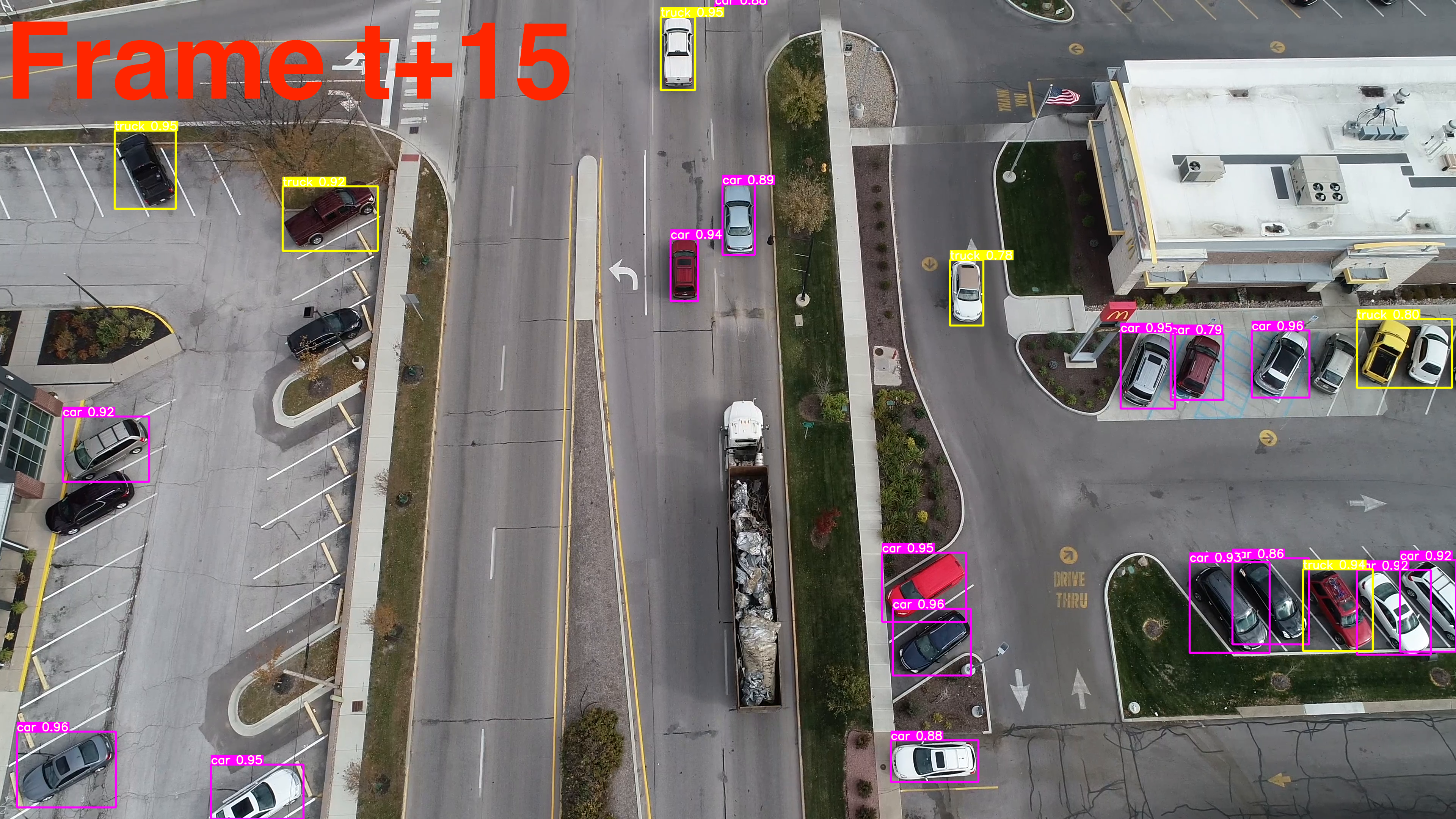}
\includegraphics[width=0.195\textwidth]{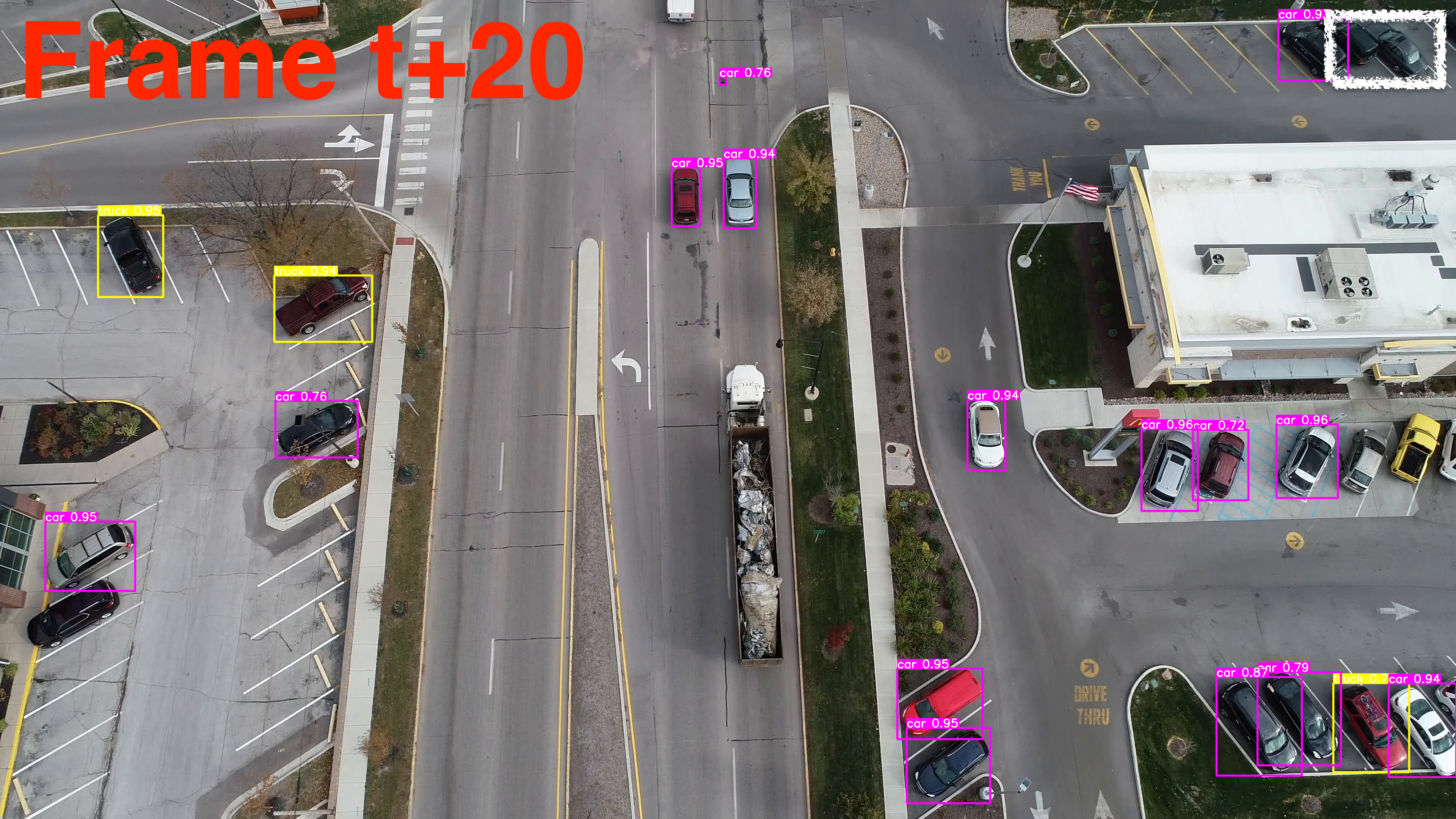}
\\
\includegraphics[width=0.195\textwidth]{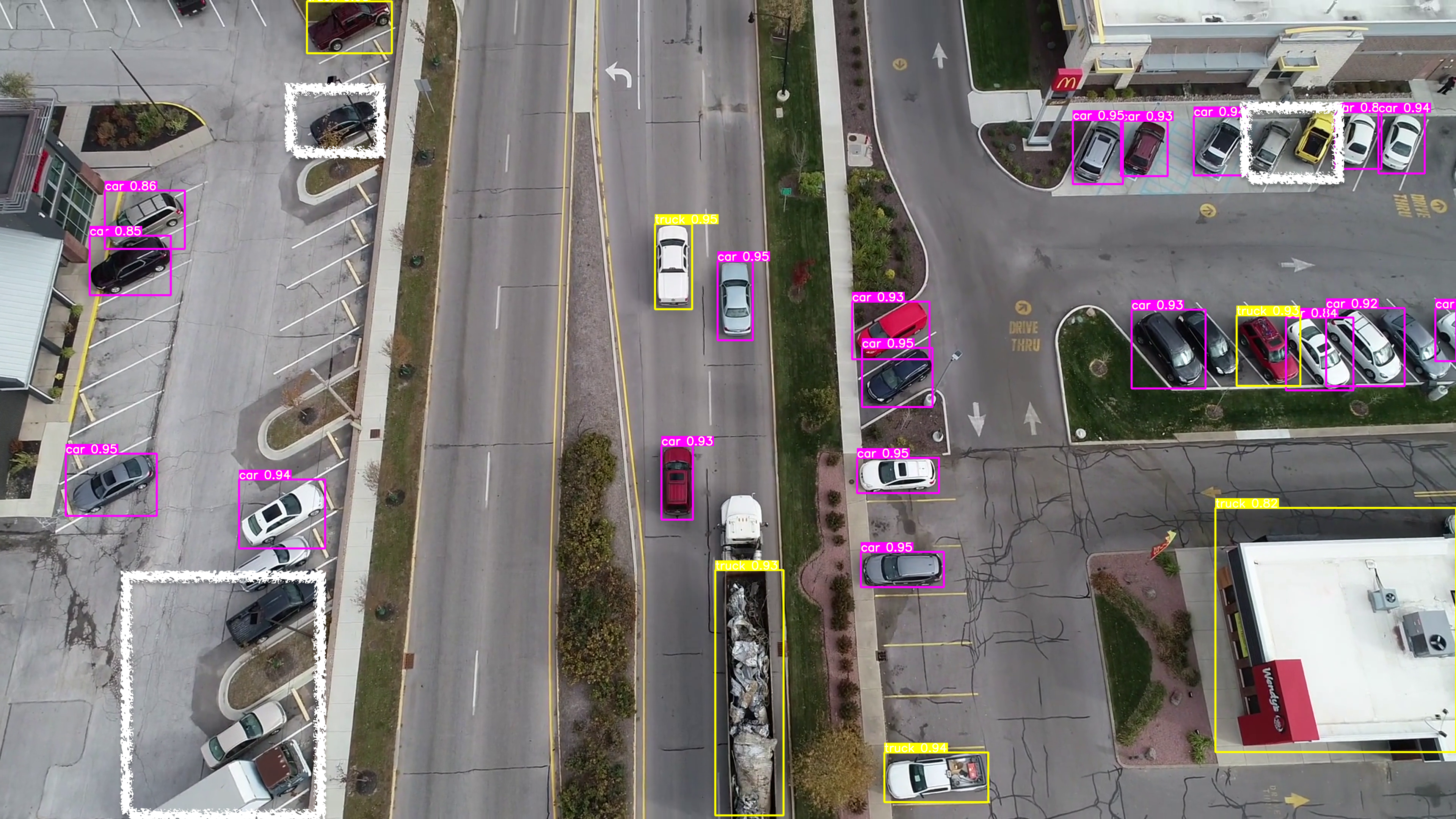}
\includegraphics[width=0.195\textwidth]{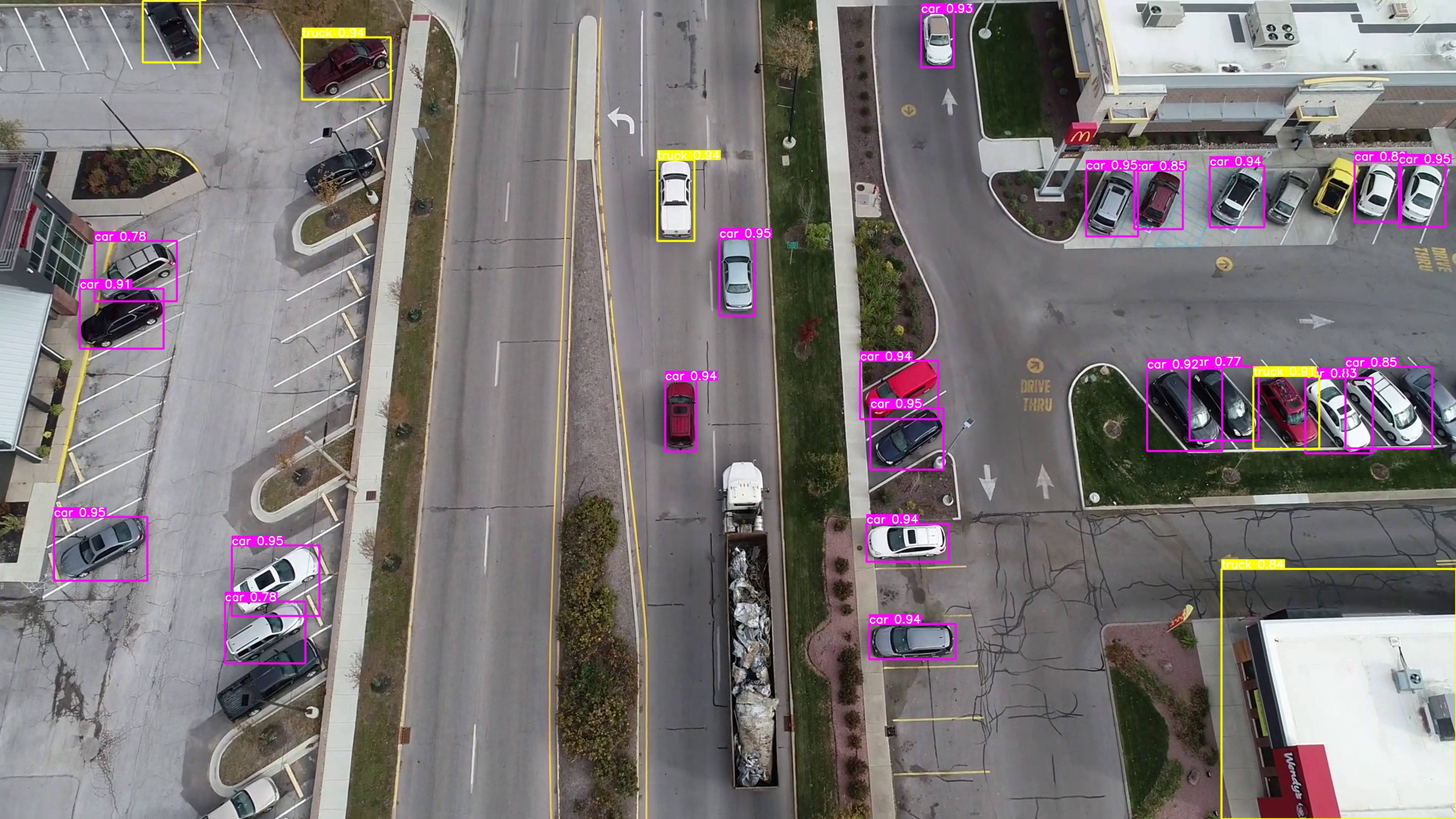}
\includegraphics[width=0.195\textwidth]{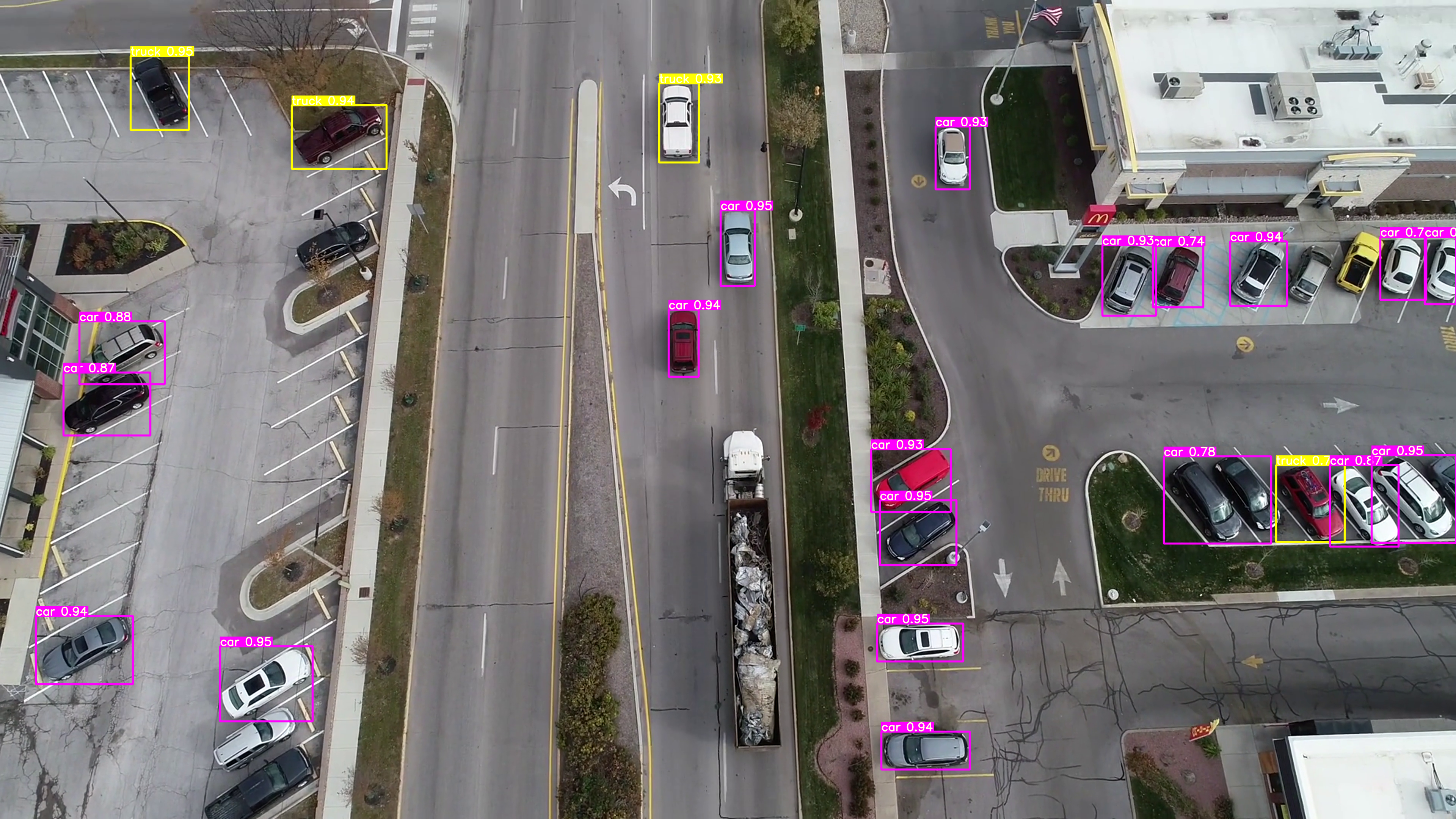}
\includegraphics[width=0.195\textwidth]{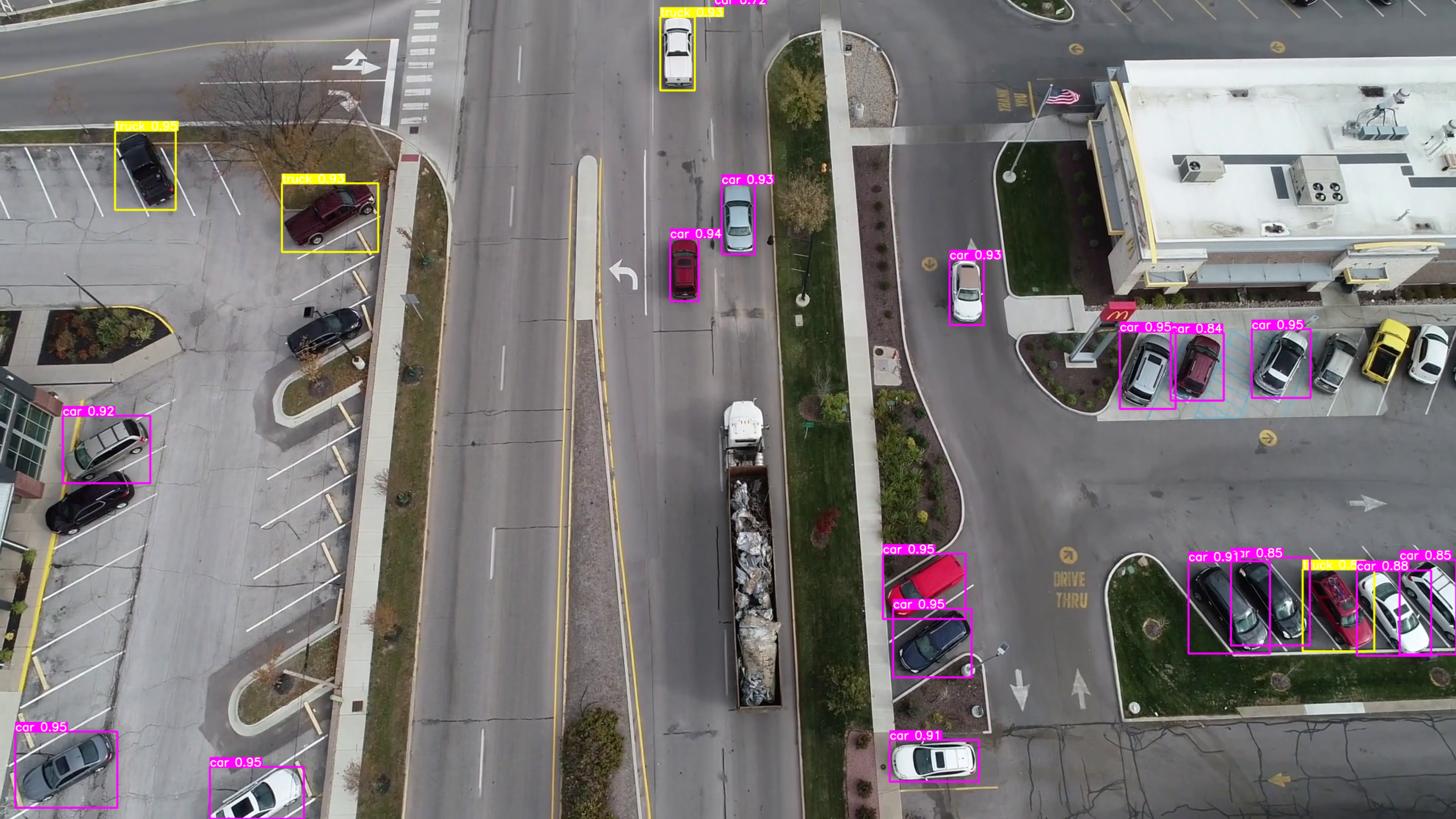}
\includegraphics[width=0.195\textwidth]{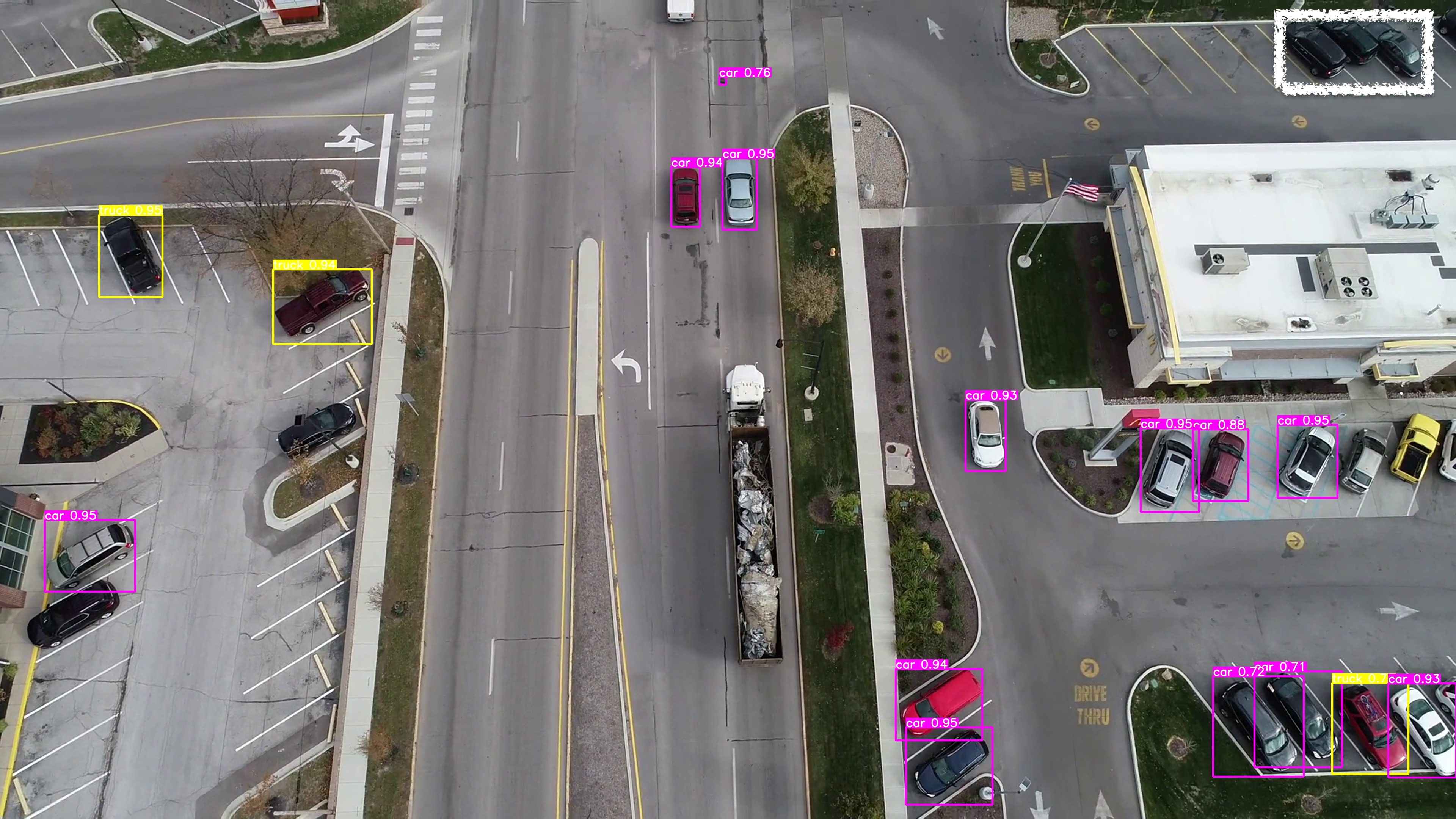}
\caption{YOLOv5s intermediate detection results for a 4-second clip in case (b): top (original), bottom (\name). 
In this example, not all the vehicles in the street parking lots (at both sides) are correctly detected, missed are labeled in white boundingbox. 
}
\label{fig:ex-accuracy-b}
\vspace{-3mm}
\end{figure*}

Second, the compression gains of \name's solution components vary in these six use scenarios; 
They are impacted by a number of factors like altitude, pitch angle, speed and area type. 
Adaptive resolution alone (\name-R) has already saved 77\% (25.4MB ), 80\% (26.5MB), 78\%(24.2MB), 76\%(36.4MB), 78\% (29.7MB), 53\%(12.6MB) of the baseline volume in case (a)-(f), respectively. Its compression contribution declines at longer working distance(see case(f)). When we take a close look at the assigned resolutions, we find that it is all 1K for the cases(a - e). As for case(f), 1K, 2K and 4K are assigned respectively from the near to far-end. Digging deeper, if the drone hovers at the height lower than 50m, only 720p needed to be assigned. Actually, in some instances, 720p is enough for 50m as well without scarifying detection accuracy. In this case, \name-R can save up to 92\%. This matches with our expectation, at a longer working distance, the objects become smaller with fewer pixels, which leaves less room to tolerate adaptive(smaller) resolution. 
In contrast, the rest components ({motion-compensated encoding and RoPI-aware quantization}) are more effective at longer working distance. 
In case(f), they further reduce the size from 11.3MB  to 3.6 MB. This is mainly because more blocks in the captured frame are RoNIs for a broader view and it opens up compression room for RoPI-aware quantization.
We also observe the impacts of speeds when the drone flies. High mobility reduces compression room with likely more inter-frame changes.
From \name-R to \name, we reduce extra 56\% volume (11.3MB $\rightarrow$ 5MB) in case (d) but more gain (8.1MB $\rightarrow$ 2.9MB) is observed in case (e).
Moreover, through the comparison of these cases, we can infer that distinct ground scenes may have a great influence on the compression. This is because the traffic density and background scene are closely related to the motion-compensated encoding and RoPI-aware quantization respectively. Like busy traffic may involve more mobilities and open environment may contain more RoNIs.
A thorough evaluation of impact factors over a larger dataset (\tab~\ref{tab:dataset}) is given next (\S\ref{sect:eval_sta}).

Third, \name lead to a small accuracy loss, this is because \name somehow downgrades video quality via lossy compression and downsampling. We examine how accuracy is sacrificed by \name by delving into intermediate detection results. Due to space limited, we use two illustrative examples in (a), and (b). More results will be released on \textit{https://github.com/JpKuo24/}.
As shown in \fig~\ref{fig:ex-accuracy-a-yolov3}, we observe that the detection is unstable across the frames, even for the original video it cannot detect all the vehicle for all the frame. 
The loss of information makes this instability worse and eventually misses it just as the vehicle labeled in white bounding box. Here, we also explain the benefit of aggregating detection results over consecutive frames. This is taking advantage of video to make the detection more robust. 
In terms of (b)  \fig~\ref{fig:ex-accuracy-b},  we do observe unsatisfactory accuracy for both precision and recall rate. Both of the original video and \name missed 6 vehicles out of 36. This is not the fault of \name; Instead, the object detection algorithm in use holds accountable. We will discuss on this problem in \S\ref{sect:disc} later. 

Forth, YOLOv5s can makes up for the missing in tiny-YOLOv3, especially for the video captured above residence. The recall loss decreases from 4.8\% to 0.4\%. As shown in \fig~\ref{fig:ex-accuracy-a-yolov5}, it can detect all the vehicles. Statistical results for these two detectors will be shown in  (\S\ref{sect:eval_sta}).

\begin{figure*}
\vspace{-2mm}
\begin{tabular}{rrrr}
\includegraphics[width=0.23\textwidth]{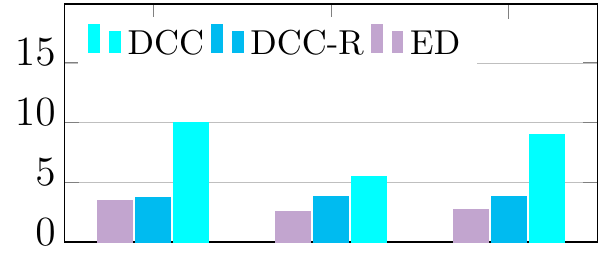} &
\includegraphics[width=0.23\textwidth]{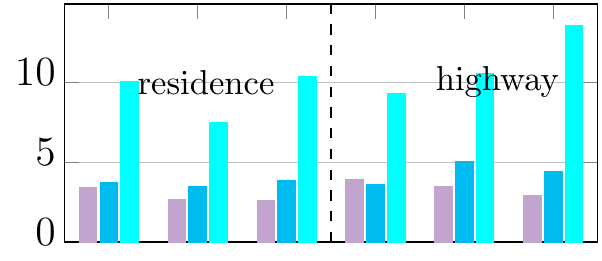} &
\includegraphics[width=0.23\textwidth]{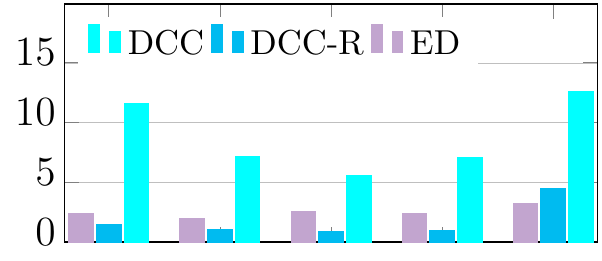} &
\includegraphics[width=0.23\textwidth]{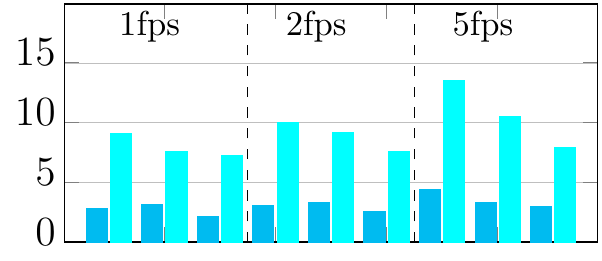} \\[-2mm]
%
\includegraphics[width=0.225\textwidth]{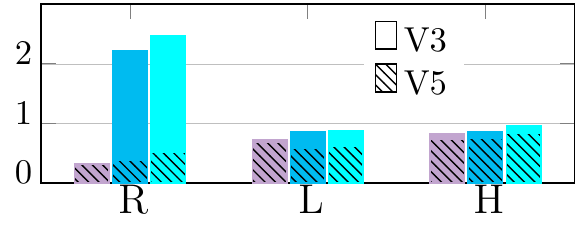} &
\includegraphics[width=0.225\textwidth]{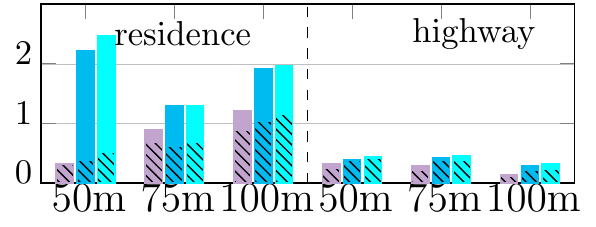} &
\includegraphics[width=0.225\textwidth]{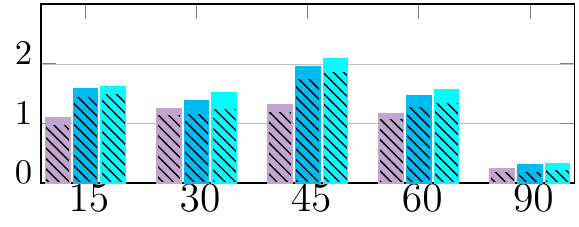} &
\includegraphics[width=0.225\textwidth]{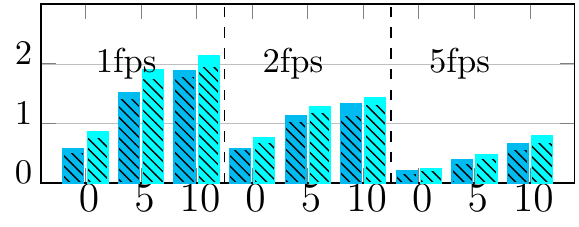} \\[-2mm]
%
\multicolumn{1}{c}{(a) area type} & 
\multicolumn{1}{c}{(b) altitude} & 
\multicolumn{1}{c}{(c) pitch angle}  &
\multicolumn{1}{c}{(d) drone speed and frame rate} \\
\end{tabular}
\caption{Impact of different factors. Top: compression gain $\gamma(V)$, bottom: accuracy loss $\gamma(F_1)$.}
\vspace{-3mm}
\label{fig:eval}
\end{figure*}

\subsection{Statistical Results}
\label{sect:eval_sta}


 
To assess factor impacts, we extend the above case study to a larger scale evaluation over more runs (\tab~\ref{tab:dataset}). We evaluate each method (\name, \name-R and \texttt{ED}) in terms of its average compression gain $\gamma(V)$ and accuracy loss $\gamma({F_1})$. For each setting, we randomly select 2 test instances(in total 20s) to calculate the accuracy since we need to manually aggregate the results frame-by-frame. In this section, we first talk about the overall performance and compare \name to \texttt{ED}, then break down to the analysis of different impact factors.

\paragraphb{Comparison to \texttt{ED}.}
Overall, \name outperforms \texttt{ED}(Baseline approach), with 19\% -- 683\%(402\% -- 3174\%) compression gain. Unsurprisingly, \texttt{ED} saves less in presence of more vehicles as every subframe needed to be transmitted. Note that the compression gain of \texttt{ED} is also limited by the number of tiles. To strike a balance between speed and accuracy, only a small number of tiles are considered in this comparison(2 by default in~\cite{wang2018bandwidth}). However, the vehicles \save{(target objects)} appear frequently in all cases since drones often fly on demand. The compression gain is thus quite limited. To save more, more small tiles should be considered but it will increase their computing load.

In addition, \texttt{ED} is much computing-heavier as it has to run a compressed CNN on device, which is slow($<$ 1 fps) for 4K video without hardware acceleration. In contrast, \name is able to operate at 10.1 fps on average. Digging deeper, the bulk of the processing overheads mainly come from the compression. Frame differencing and downsampling that run at 50fps, 32fps respectively. These newly introduced components are relatively lightweight. \save{Besides, all the sub-frames(output of {\bf resolution adaptation}) can be parallel processed and having no independence between different slices.}

Although accuracy loss for \texttt{ED} is better than \name, \name still can guarantee that the loss of $F_1$ score $<$  1.5\% in most cases by YOLOv5s. 

\paragraphb{YOLOv5s v.s. tiny-YOLOv3.}  
YOLOv5s outperforms tiny-YOLOv3 in accuracy for majority of cases, both recall and precision rate are increase as shown in \tab~\ref{tab:example}, which is constant with the performance on COCO dataset. In terms of accuracy loss, both detectors have similar performance for highway scenario, less gain is found for YOLOv5s over tiny-YOLOv3 since the detection is relatively stable. But for the rest scenarios, especially for residential area, YOLOv5s performs much better. \fig~\ref{fig:eval}(b) shows that YOLOv5s has reduced losses by 41.6\%--79.3\% compared to tiny-YOLOv3 across the three different altitudes above residence. We believe this is because better detectors can tolerate more noises, thereby reducing the impact of information loss caused by \name. Note that reduced losses are more for the case of 50m, this is because fewer vehicles are captured above lower altitudes. Correspondingly, one more successful detection can improve a lot of accuracy. The comparison between two detectors also indicate that \name can apply to different detectors. We expect better performance will be achieved when the CNN is training with the noise or training on compressed videos~\cite{wang2019fast}. They are orthogonal to our work. Then we move to impact factors analysis.

\paragraphb{Area types.}
The compression gain varies in different areas for two reasons. On the one hand, the diverse background may contains different number of RoNIs and influence on the RoPI-aware quantization accordingly; On the other hand, the density of vehicles and their speeds vary a lot, resulting in different compression room. 
We evaluate the impact of ground areas by flying drones above three areas (R, L, H) under the same setting with altitude = 50m, pitch  $\approx$ 90\textdegree, average speed = 5m/s. 
\fig~\ref{fig:eval}(a) shows the average compression gains and accuracy losses of \name, \name-R and \texttt{ED}. 
In terms of \name's compression gain, the residence (1001\%) $>$ highway (897\%) $>$ local (549\%). After a closer look, \name-R has the similar contribution(around 380\%) for all three cases as the flight altitude is the same and correspondingly same resolution(1K) is assigned. We confirm the difference of compression gain mainly comes from the number of RoNI. Relatively speaking, residential areas and highways have more background leaving big room to compression. We also observe a large variance(5.32) in residence. This is because we have two different types of residence in our dataset, one is to have more apartments, and the other is to houses. If it is a latter case, there are less cars parked outside and more greenery involves. As for local case, even the street scene is similar, there still have a variance of 0.97. We find that this difference mainly comes from the density of vehicles: Its compression gain drops at the busy hour (with many more cars) and grows at idle hours (with fewer cars).

\paragraphb{Altitude.}
We evaluate the impact of altitude in two typical settings: (1) residence, 5m/s and (2) highway, 0m/s (hovering). 
Pitch $\approx$ 90\textdegree and we test with three altitudes: 50m, 75m and 100m. 
The results are shown in \fig~\ref{fig:eval}(b). We have two new findings. 
First, there is no fixed or monotonic relation between the altitude and the compression gain realized by \name. 
Above the highway, the maximal compression gain is achieved at 100m. That is, it goes up from 50m  (925\%) to 75m (1055\%) and then inclines when the drone climbs up to 100m (1351\%). This is not the case above the residence.
Second, we find that such results are stemmed in distinct impacts of altitude on \name's components.
The contribution of resolution adaptation is similar due to all assigned 1K according to the look-up table. Note here if the shooting altitude decrease to 40m, smaller resolution can be assigned, which can bring additional compression room. In fact, some test case for 50m, 720p is enough. But to make it consistent, 1K is assigned by default at the beginning. But the assigned resolution may be changed based on the real-time detection results to meet different situations in specific cases.
If the recall is continuously over than 90\%, we will inform the drone to reassign a lower resolution(decrease by one) than the previous assignment. 
In the highway case, RoPI-aware quantization brings huger compression gains at higher altitudes.
This is because more backgrounds like trees and farms are seen along the highway when drone flies higher while the highway is almost covered with the entire frame at the height of 50m.
This is not the case in residential zones.  Broader FoV covers new roads and vehicles, thus there are comparable extra gain from \name-R to \name for the case of 50m and 100m.

\paragraphb{Pitch Angle.}
We use highway as the study case to assess the impact of pitch angles, the similar findings hold in other test cases. 
We test with 90\textdegree, 60\textdegree, 45\textdegree, 30\textdegree and 15\textdegree \, when the drone hovers at 100m above the highway. 
%
We see the best performance with a vertical shot (90\textdegree) due to shorter working distance. The compression gain gradually degrade with the decrease of the pitch angle since higher resolutions we need to be assigned accordingly. The worst performance is on 45\textdegree, which has the stronger far-near effect for the ground part. Then the performance increase due to the appearance of sky, which we set as RoNI and no need for transmission. 
In a nutshell, the smaller pitch angle, the larger chances with more blocks of no interests due to the broader view. Note that the growing contribution of RoPI-aware quantization is only in the highway case. In the local and residence case, broader view cover new roads and vehicles due to the dense layout. For the 45\textdegree case, the relatively high loss of accuracy is because of more ground being captured, which correspondingly increases the chance of vehicle loss due to the information loss.

\paragraphb{Drone's speed and frame rate.}
High mobility reduces compression room with likely more inter-frame changes. 
Drone's speed and frame rate jointly impact the movement between frames. 
We evaluate their impact together with three speeds  0, 5 and 10m/s at one highway setting (100m, pitch $\approx$ 90\textdegree). We consider three frame rates: 1fps, 2fps and 5fps. 
%
When the speed grows, global motion compensation performs worse. 
This is mainly because the newly introduced parts are large, requiring to encode everything with no reference frames. Note that there are still huge residuals needed to be encoded, even when P-frame is selected to use. This is why the degradation is more obvious when the frame rate drops.  At 1fps, more changes must be encoded as I-frames as there are huge offsets over consecutive frames(over a larger interval).
The accuracy loss increase for high mobility. This is because less temporal results of detection we can count for video object detection. 
\section{Conclusion and Future Work}
\label{sect:conc}
\label{sect:disc}

We present \name to realize adaptive and deeper video compression for edge-assisted drone-sourced video analytics. 
It reserves both good of well-established video codec technology and recent advances on efficient device-edge video analytics. By exploiting tolerance for lower-quality frames allowed by video analytical tasks, we leverage drone's runtime context to build \name on top of the existing codec and make it lightweight and efficient in computing and compression. We use vehicle detection as a showcase to demonstrate its significant compression gain
with no/negligible impact on object detection accuracy. It possesses more degrees of adaptation and develops a `{\em fit}' compression technique tailored to drone-source video analytics.



Despite its effectiveness in our study, there are many remaining and open issues as future work.

{\em Video frame rate.}  Clearly, a constant frame rate is unnecessary for video analytics, and on-demand delivery seems a better solution. However, it cannot be compatible with any existing video codec, and thus have to use images, not videos. It is hard to tell which compresses more. It depends on the frequency of target objects. In the use scenarios where objects occasionally appear, it can be more efficient to transmit the interesting frames (images) only, rather than streaming the video source at a constant frame rate. However, in our target drone-sourced scenarios, target objects often appear much more often as the drone's flight is on-demand. This is why compressing a video is a better solution over sending the selected frames. How to set a proper frame rate or even to implement an adaptive frame rate is a hard problem. A high frame rate is useless with similar views, while a low frame rate may miss the objects or fail to make power of video frames because they are less temporally correlated. One promising direction for our future work is to adjust the intervals of I-frames and P-frames to implement an adaptive frame rate over a constant source frame rate.

{\em New video codec for machines.} \name is still built on traditional video codec that are optimized for human viewing. Another promising direction is to compress the feature stream instead of the video signal stream. Descriptors for video analysis can be compact and greatly reduce the data volume needed to be encoded and transmitted. The key challenges lie in how to choose the descriptors(features) that enable multiple machine vision tasks and how to encode them. One possible idea is to encode the backbone intermediate results, which are low level features and shared by multi-tasks for machine vision. We believe that there are lot of rooms in collaborative operations between video and feature streams for video codec for machines and humans.


{\em Accuracy fluctuation.}  The classification result is not stable over frames even using the original video. This is a problem inherent in CNN-based detection. To improve the robustness, we consider aggregating the detection results within a detection window. But it will not cure the disease, essentially there are two problems need to be solved: First, occlusions, widely appears in local and residence such as trees and shadows of buildings, make it difficult for vehicle to be detected, the moving camera will increase such instability. Detecting partially occluded object is still a difficult task in CV domain, apply context-aware network (e.g.CompositionalNet ~\cite{wang2020robust}) might release this problem, but more efforts are needed to label the data. 
Second, there is a problem of class imbalance in our dataset. Specifically, the class car has most instances compared to others. Accordingly, the performance for trucks detection will not be stable as the car. Essentially, class rebalancing should be involved like creating importance re-weighting into loss function over label distribution. However, the label distribution is hard to be known ahead of time. Moreover, there are not much difference from the perspective of bird view, making it hard to differentiate pickup trucks(should be classified as truck) and SUVs(should be classified as car).

{\em Public Dataset.} No public dataset contains detailed drone information such as shooting altitude, angle, etc. However these information can be really helpful to assist in object detection. We expect large and diverse video datasets in the future since it is a very time-consuming task to label on the video. Besides, densely annotated dataset can benefit for delay evaluation. Recently, average delay(AD) ~\cite{mao2019delay} is presented as a new metric to measure the detection delay of video object detectors. We want to know whether AD is suitable for video taken by drones, since the low resolution and wide FoV caused by long distance may hinder the delay measurement.



%
%


\bibliographystyle{9pt,unsrt}

\bibliography{../bib/drone-video,../bib/all_career,../bib/chunyipub,../bib/standard}

\begin{thebibliography}{10}

\bibitem{wang2018bandwidth}
Junjue Wang, Ziqiang Feng, Zhuo Chen, Shilpa George, Mihir Bala, Padmanabhan
  Pillai, Shao-Wen Yang, and Mahadev Satyanarayanan.
\newblock Bandwidth-efficient live video analytics for drones via edge
  computing.
\newblock In {\em SEC}, pages 159--173. IEEE, 2018.

\bibitem{restas2015drone}
Agoston Restas.
\newblock Drone applications for supporting disaster management.
\newblock {\em World Journal of Engineering and Technology}, 3(03):316, 2015.

\bibitem{ham2016visual}
Youngjib Ham, Kevin~K Han, Jacob~J Lin, and Mani Golparvar-Fard.
\newblock Visual monitoring of civil infrastructure systems via camera-equipped
  unmanned aerial vehicles (uavs): a review of related works.
\newblock {\em Visualization in Engineering}, 4(1):1, 2016.

\bibitem{george2019towards}
Shilpa George, Junjue Wang, Mihir Bala, Thomas Eiszler, Padmanabhan Pillai, and
  Mahadev Satyanarayanan.
\newblock Towards drone-sourced live video analytics for the construction
  industry.
\newblock In {\em Proceedings of the 20th International Workshop on Mobile
  Computing Systems and Applications (HotMobile'19)}, 2019.

\bibitem{tsag19-drone-app-public-safety}
Transportation~Safety~Advancement~Group (TSAG).
\newblock White paper: The uses continue to emerge: Public safety drones and
  considerations, 2019.

\bibitem{sudhakar2020unmanned}
S~Sudhakar, V~Vijayakumar, C~Sathiya Kumar, V~Priya, Logesh Ravi, and
  V~Subramaniyaswamy.
\newblock Unmanned aerial vehicle (uav) based forest fire detection and
  monitoring for reducing false alarms in forest-fires.
\newblock {\em Computer Communications}, 149:1--16, 2020.

\bibitem{cai2018cascade}
Zhaowei Cai and Nuno Vasconcelos.
\newblock Cascade r-cnn: Delving into high quality object detection.
\newblock In {\em Proceedings of the IEEE conference on computer vision and
  pattern recognition}, pages 6154--6162, 2018.

\bibitem{zhang2017live}
Haoyu Zhang, Ganesh Ananthanarayanan, Peter Bodik, Matthai Philipose, Paramvir
  Bahl, and Michael~J Freedman.
\newblock Live video analytics at scale with approximation and delay-tolerance.
\newblock In {\em 14th $\{$USENIX$\}$ Symposium on Networked Systems Design and
  Implementation ($\{$NSDI$\}$ 17)}, pages 377--392, 2017.

\bibitem{Kristan2019a}
Matej Kristan, Jiri Matas, Ales Leonardis, Michael Felsberg, Roman Pflugfelder,
  Joni-Kristian Kamarainen, Luka \v{C}ehovin Zajc, Ondrej Drbohlav, Alan
  Lukezic, Amanda Berg, Abdelrahman Eldesokey, Jani Kapyla, and Gustavo
  Fernandez.
\newblock The seventh visual object tracking vot2019 challenge results, 2019.

\bibitem{redmon2018yolov3}
Joseph Redmon and Ali Farhadi.
\newblock Yolov3: An incremental improvement.
\newblock {\em arXiv preprint arXiv:1804.02767}, 2018.

\bibitem{bochkovskiy2020yolov4}
Alexey Bochkovskiy, Chien-Yao Wang, and Hong-Yuan~Mark Liao.
\newblock Yolov4: Optimal speed and accuracy of object detection.
\newblock {\em arXiv preprint arXiv:2004.10934}, 2020.

\bibitem{jocher2020yolov5}
Glenn Jocher et~al.
\newblock Yolov5.
\newblock {\em Code repository https://github. com/ultralytics/yolov5}, 2020.

\bibitem{ren2015faster}
Shaoqing Ren, Kaiming He, Ross Girshick, and Jian Sun.
\newblock Faster r-cnn: Towards real-time object detection with region proposal
  networks.
\newblock In {\em Advances in neural information processing systems}, pages
  91--99, 2015.

\bibitem{denton2017unsupervised}
Emily Denton and Vighnesh Birodkar.
\newblock Unsupervised learning of disentangled representations from video.
\newblock {\em arXiv preprint arXiv:1705.10915}, 2017.

\bibitem{majid2018shuffledet}
Seyed Majid~Azimi.
\newblock Shuffledet: Real-time vehicle detection network in on-board embedded
  uav imagery.
\newblock In {\em ECCV}, pages 0--0, 2018.

\bibitem{TR36.777}
3GPP.
\newblock {TR36.777: Enhanced LTE support for aerial vehicles}, Jan. 2018.
\newblock (Release 15).

\bibitem{hayat2019experimental}
Samira Hayat, Christian Bettstetter, Aymen Fakhreddine, Raheeb Muzaffar, and
  Driton Emini.
\newblock An experimental evaluation of lte-a throughput for drones.
\newblock In {\em DroNet}, pages 3--8. ACM, 2019.

\bibitem{lin2019mobile}
Xingqin Lin, Richard Wiren, Sebastian Euler, Arvi Sadam, Helka-Liina Maattanen,
  Siva Muruganathan, Shiwei Gao, Y-P~Eric Wang, Juhani Kauppi, Zhenhua Zou,
  et~al.
\newblock Mobile network-connected drones: Field trials, simulations, and
  design insights.
\newblock {\em IEEE Vehicular Technology Magazine}, 14(3):115--125, 2019.

\bibitem{guo2021towards}
Junpeng Guo and Chunyi Peng.
\newblock Towards drone-sourced live video analytics via
  adaptive-yet-compatible compression.
\newblock In {\em Proceedings of the 22nd International Workshop on Mobile
  Computing Systems and Applications}, pages 172--173, 2021.

\bibitem{wiki-codec}
Wikipedia.
\newblock Video coding format.
\newblock \url{https://en.wikipedia.org/wiki/Video_coding_format}, 2019.

\bibitem{han2016mcdnn}
Seungyeop Han, Haichen Shen, Matthai Philipose, Sharad Agarwal, Alec Wolman,
  and Arvind Krishnamurthy.
\newblock Mcdnn: An approximation-based execution framework for deep stream
  processing under resource constraints.
\newblock In {\em Proceedings of the 14th Annual International Conference on
  Mobile Systems, Applications, and Services}, pages 123--136, 2016.

\bibitem{ananthanarayanan2020project}
Ganesh Ananthanarayanan, Yuanchao Shu, Landon Cox, and Victor Bahl.
\newblock Project rocket platform—designed for easy, customizable live video
  analytics—is open source.
\newblock Microsoft Research Blog, January 2020.

\bibitem{rocket}
Microsoft rocket for live video analytics.
\newblock
  \url{https://www.microsoft.com/en-us/research/project/live-video-analytics/},
  January 2020.

\bibitem{zhu2017deep}
Xizhou Zhu, Yuwen Xiong, Jifeng Dai, Lu~Yuan, and Yichen Wei.
\newblock Deep feature flow for video recognition.
\newblock In {\em Proceedings of the IEEE conference on computer vision and
  pattern recognition}, pages 2349--2358, 2017.

\bibitem{wu2019sequence}
Haiping Wu, Yuntao Chen, Naiyan Wang, and Zhaoxiang Zhang.
\newblock Sequence level semantics aggregation for video object detection.
\newblock In {\em Proceedings of the IEEE/CVF International Conference on
  Computer Vision}, pages 9217--9225, 2019.

\bibitem{lu2017online}
Yongyi Lu, Cewu Lu, and Chi-Keung Tang.
\newblock Online video object detection using association lstm.
\newblock In {\em Proceedings of the IEEE International Conference on Computer
  Vision}, pages 2344--2352, 2017.

\bibitem{kang2017object}
Kai Kang, Hongsheng Li, Tong Xiao, Wanli Ouyang, Junjie Yan, Xihui Liu, and
  Xiaogang Wang.
\newblock Object detection in videos with tubelet proposal networks.
\newblock In {\em Proceedings of the IEEE Conference on Computer Vision and
  Pattern Recognition}, pages 727--735, 2017.

\bibitem{wu2019delving}
Zhenyu Wu, Karthik Suresh, Priya Narayanan, Hongyu Xu, Heesung Kwon, and
  Zhangyang Wang.
\newblock Delving into robust object detection from unmanned aerial vehicles: A
  deep nuisance disentanglement approach.
\newblock In {\em ICCV}, 2019.

\bibitem{benjdira2019car}
Bilel Benjdira, Taha Khursheed, Anis Koubaa, Adel Ammar, and Kais Ouni.
\newblock Car detection using unmanned aerial vehicles: Comparison between
  faster r-cnn and yolov3.
\newblock In {\em 2019 1st International Conference on Unmanned Vehicle
  Systems-Oman (UVS)}, pages 1--6. IEEE, 2019.

\bibitem{liu2019edge}
Luyang Liu, Hongyu Li, and Marco Gruteser.
\newblock Edge assisted real-time object detection for mobile augmented
  reality.
\newblock In {\em MobiCom}. ACM, 2019.

\bibitem{ma2020video}
Di~Ma, Fan Zhang, and David~R Bull.
\newblock Video compression with low complexity cnn-based spatial resolution
  adaptation.
\newblock In {\em Applications of Digital Image Processing XLIII}, volume
  11510, page 115100D. International Society for Optics and Photonics, 2020.

\bibitem{duan2020video}
Ling-Yu Duan, Jiaying Liu, Wenhan Yang, Tiejun Huang, and Wen Gao.
\newblock Video coding for machines: A paradigm of collaborative compression
  and intelligent analytics.
\newblock {\em arXiv preprint arXiv:2001.03569}, 2020.

\bibitem{duan2018compact}
Ling-Yu Duan, Yihang Lou, Yan Bai, Tiejun Huang, Wen Gao, Vijay Chandrasekhar,
  Jie Lin, Shiqi Wang, and Alex~Chichung Kot.
\newblock Compact descriptors for video analysis: The emerging mpeg standard.
\newblock {\em IEEE MultiMedia}, 26(2):44--54, 2018.

\bibitem{chen2015glimpse}
Tiffany Yu-Han Chen, Lenin Ravindranath, Shuo Deng, Paramvir Bahl, and Hari
  Balakrishnan.
\newblock Glimpse: Continuous, real-time object recognition on mobile devices.
\newblock In {\em SenSys}, pages 155--168, 2015.

\bibitem{meuel2018region}
Holger Meuel, Florian Kluger, and J{\"o}rn Ostermann.
\newblock Region of interest (roi) coding for aerial surveillance video using
  avc \& hevc.
\newblock {\em arXiv preprint arXiv:1801.06442}, 2018.

\bibitem{zhang2015design}
Tan Zhang, Aakanksha Chowdhery, Paramvir Bahl, Kyle Jamieson, and Suman
  Banerjee.
\newblock The design and implementation of a wireless video surveillance
  system.
\newblock In {\em Proceedings of the 21st Annual International Conference on
  Mobile Computing and Networking}, pages 426--438, 2015.

\bibitem{cheng2018vitrack}
Linsong Cheng and Jiliang Wang.
\newblock Vitrack: Efficient tracking on the edge for commodity video
  surveillance systems.
\newblock In {\em IEEE INFOCOM 2018-IEEE Conference on Computer
  Communications}, pages 1052--1060. IEEE, 2018.

\bibitem{jain2018rexcam}
Samvit Jain, Junchen Jiang, Yuanchao Shu, Ganesh Ananthanarayanan, and Joseph
  Gonzalez.
\newblock Rexcam: Resource-efficient, cross-camera video analytics at
  enterprise scale.
\newblock {\em arXiv preprint arXiv:1811.01268}, 2018.

\bibitem{liu2019fusioneye}
Hansi Liu, Pengfei Ren, Shubham Jain, Mohannad Murad, Marco Gruteser, and Fan
  Bai.
\newblock Fusioneye: Perception sharing for connected vehicles and its
  bandwidth-accuracy trade-offs.
\newblock In {\em 2019 16th Annual IEEE International Conference on Sensing,
  Communication, and Networking (SECON)}, pages 1--9. IEEE, 2019.

\bibitem{greengard2019drones}
Samuel Greengard.
\newblock When drones fly.
\newblock {\em Communications of the ACM}, 62(11):16--18, 2019.

\bibitem{szeliski2010computer}
Richard Szeliski.
\newblock {\em Computer vision: algorithms and applications}.
\newblock Springer Science \& Business Media, 2010.

\bibitem{mi2019sensor}
Yang Mi, Chunbo Luo, Geyong Min, Wang Miao, Liang Wu, and Tianxiao Zhao.
\newblock Sensor-assisted global motion estimation for efficient uav video
  coding.
\newblock In {\em ICASSP}, pages 2237--2241, 2019.

\bibitem{wang2019fast}
Shiyao Wang, Hongchao Lu, and Zhidong Deng.
\newblock Fast object detection in compressed video.
\newblock In {\em Proceedings of the IEEE/CVF International Conference on
  Computer Vision}, pages 7104--7113, 2019.

\bibitem{wang2020robust}
Angtian Wang, Yihong Sun, Adam Kortylewski, and Alan~L Yuille.
\newblock Robust object detection under occlusion with context-aware
  compositionalnets.
\newblock In {\em Proceedings of the IEEE/CVF Conference on Computer Vision and
  Pattern Recognition}, pages 12645--12654, 2020.

\bibitem{mao2019delay}
Huizi Mao, Xiaodong Yang, and William~J Dally.
\newblock A delay metric for video object detection: What average precision
  fails to tell.
\newblock In {\em Proceedings of the IEEE/CVF International Conference on
  Computer Vision}, pages 573--582, 2019.

\end{thebibliography}

\end{document}